\documentstyle[prb,aps,eqsecnum,epsf,multicol]{revtex}
\begin{document}
\draft
\tighten

\title{Nonequilibrium Steady States of Driven Periodic Media}

\author{Leon Balents}
\address{Institute for Theoretical Physics, University of California, 
Santa Barbara, CA 93106--4030}
\author{M. Cristina Marchetti}
\address{Physics Department, Syracuse University, Syracuse, NY 13244--1130}
\author{Leo Radzihovsky}
\address{Physics Department, University of Colorado, Boulder, CO 80309--0390}

\date{\today}
\maketitle

\begin{abstract}
  
  We study a periodic medium driven over a random or periodic
  substrate, characterizing the non-equilibrium phases which occur by
  dynamic order parameters and their correlations.  Starting with a
  microscopic lattice Hamiltonian, we perform a careful
  coarse-graining procedure to derive continuum hydrodynamic equations
  of motion in the laboratory frame.  This procedure induces
  nonequilibrium effects (e.g. convective terms, KPZ nonlinearities,
  and non-conservative forces) which cannot be derived by a naive
  Galileian boost.  Rather than attempting a general analysis of these
  equations of motion, we argue that in the random case instabilities
  will always destroy the LRO of the lattice.  We suggest that the
  only periodicity that can survive in the driven state is that of a
  transverse smectic, with ordering wavevector perpendicular to the
  direction of motion. This conjecture is supported by an analysis of
  the linearized equations of motion showing that the induced
  nonequilibrium component of the force leads to displacements
  parallel to the mean velocity that diverge with the system size.  In
  two dimensions, this divergence is extremely strong and can drive a
  melting of the crystal along the direction of motion.  The resulting
  driven smectic phase should also occur in three dimensions at
  intermediate driving.  It consists of a periodic array of flowing
  liquid channels, with transverse displacements and density
  (``permeation mode'') as hydrodynamic variables.  We study the
  hydrodynamics of the driven smectic within the dynamic functional
  renormalization group in two and three dimensions.  The finite
  temperature behavior is much less glassy than in equilibrium, owing
  to a disorder-driven effective ``heating'' (allowed by the absence
  of the fluctuation-dissipation theorem).  This, in conjunction with
  the permeation mode, leads to a fundamentally {\sl analytic}
  transverse response for $T>0$.

\end{abstract}


\begin{multicols}{2}

\section{Introduction}
\label{intro}

Non-equilibrium driven solids and liquids arise in a wide variety of
different physical contexts.  A common means of driving is to apply a
constant or low frequency spatially uniform shear (either a constant
shear rate or shear stress), which has been studied extensively in
colloidal and polymeric systems\cite{Clark}.  Such a driving is in a
sense severe, since it is incompatible with a macroscopically ordered
solid, requiring the continual breaking of a non-zero density of bonds
per unit time.  If translational symmetry is broken explicitly by the
presence of, e.g. a periodic substrate or quenched impurities fixed in
space, a gentler sort of driving is possible.  In this case, even a
uniform translation of the system is nontrivial, and it can be driven
out of equilibrium simply by applying a constant force or pulling at a
constant velocity.

A considerable number of such systems have been subjects of recent
investigation.  These include flux lattices in type-II
superconductors\cite{blatter,huse_radzihovsky}, charge density waves
(CDWs) in anisotropic conductors\cite{gruner}, magnetic bubble
arrays\cite{seshadri}, and the magnetically-induced Wigner crystal in
a two-dimensional electron gas\cite{andrei,willet}.  In all these
systems the relevant degrees of freedom - be they vortices in
superconductors or electrons in metals - form a lattice inside a solid
matrix, provided by the superconducting or conducting material.  Both
a periodic potential (due to the underlying crystal lattice) and a
quenched random one (due to material impurities and defects) are
generally present, though their relative importance can vary from
system to system.  Closely related problems also arise in microscopic
models of friction and lubrication, in which a surface or monolayer is
brought into contact with another surface and forced to slide relative
to it. Some recent simple models of earthquakes\cite{quakes}, in which
two elastic half-spaces are slowly driven past each other, also fall
into the general class of driven disordered periodic elastic systems.

Much of the recent focus in these {\sl pinned elastic media} has been
on {\em equilibrium} behavior, since, unlike their colloidal and
polymeric counterparts studied in the shear geometry, these systems
exhibit a nontrivial competition between the external (substrate or
disorder-induced) potential and the tendency for local order.  In the
random case, this was argued by Larkin\cite{larkin}\ to generate
long-range elastic distortions.  More recent works have reinvestigated
this problem in some detail, suggesting the existence of a novel
``Bragg glass'' phase in three and possibly two
dimensions.\cite{glstatic,gingras,DSFrecent}\ In the periodic case
(known as intrinsic pinning in the vortex community), the potential
can either lock-in commensurate phases, generate finite (but
qualitatively unimportant) incommensurate distortions of the lattice,
or stabilize anisotropic {\sl liquid-crystalline} states.  Though for
both types of matrix some detailed questions remain unanswered (in
particular, the stability of the studied elastic glassy phases to
proliferation of topological defects), the {\em equilibrium} phases and
transitions in these systems have been extensively studied and are
fairly well understood.

Once the elastic medium is driven, however, a host of new questions
arises: Under a uniform applied external force $f_{\rm ext.}$, what is
the mean velocity $v(f_{\rm ext.})$ (the IV curve, in the context of
superconductors and CDWs), and is this uniquely determined or
dependent upon history or other variables?  Do ordered solid phases
exist at low temperatures or weak disorder, and if so how are they
characterized?  Can one develop non-equilibrium phase diagrams, with
phases classified by order parameters and symmetry analogous to
equilibrium problems? What are the properties of the resulting
non-equilibrium dynamic phase transitions and the nature of the
fluctuations?  Under what conditions does such a system reach a
non-equilibrium {\sl steady state}, and what are the hydrodynamic
modes in this case?

Along with these new questions come a number of new physical variables
which play important roles.  While equilibrium behavior is relatively
insensitive to detailed dynamical laws, there is no reason to expect
this for a driven system.  In particular, the nature of dissipation
surely plays a crucial role.  This distinguishes, e.g. friction and
lubrication, in which (to a good approximation) energy can only be
redistributed among vibrational modes of the solid, from
charge-density-wave or vortex solids, in which energy can be
transfered out of the collective modes to dissipate into electronic
degrees of freedom.  In the latter case, which we focus on here, it is
appropriate to consider overdamped dynamics, while for the former
inertial effects may be significant.  Also crucial are conservation
laws, which generally give rise to additional hydrodynamic modes.  In
equilibrium, the interactions of these with elastic degrees of freedom
are constrained by the fluctuation-dissipation theorem (FDT) not to
modify static correlations.  For the nonequilibrium lattice, however,
they must be treated explicitly.  Finally, a driven system may exhibit
complex dynamics even in the absence of external ``thermal'', or
time-dependent, noise.  When the external noise level is low, this
deterministic dynamics can give rise to steady, periodic,
quasiperiodic, or (spatio-temporally) chaotic solutions.  While in the
latter case, dynamical chaos likely gives rise to an effective
``temperature'' and restoration of ergodicity, more regular solutions
need not explore the full phase space and can in principle behave with
almost arbitrary complexity!  Of course, factors such as the nature of
the external potential, dimensionality, and range of inter-particle
interactions, which control the equilibrium state, influence the
driven dynamics as well.

A complete answer to these questions for all possible cases is beyond
the scope of a single paper (or researcher!).  Instead, we will focus
here on systems with overdamped, dissipative dynamics which reach
statistically steady states.  This can occur either due to the
presence of external time-dependent noise, which forces the system to
explore the available phase space, or due to intrinsic chaotic
dynamics.  Note that this condition is violated by overdamped
phase-only models (e.g. Fukuyama-Lee-Rice\cite{FLR}) of CDWs, in which
there is a unique periodic long-time attractor, as shown by
Middleton.\cite{Middleton}\ It may also be violated in
zero-temperature simulations of vortex dynamics under some
circumstances, as observed recently by Nori in some regimes for a
strong periodic pinning potential.\cite{Nori}\ We expect, however,
that in the presence of a random potential chaotic dynamics is much
more germane, and furthermore, that most systems of experimental
interest contain appreciable external thermal noise.  Nevertheless,
where possible, we will comment on the extensions of our conclusions
to the noiseless case.  

Our approach to the problem is first to devise a means of classifying
the non-equilibrium phases, then to determine the dynamical equations
of motion governing these phases.  Finally, using these equations, we
can begin to calculate the phase diagram and the properties thereof.
We summarize the main results below, reserving a comparison with
previous work for the discussion section of the paper.  Some
preliminary versions of some of the results of this paper appeared in
Ref.~\onlinecite{us}.

\subsection{Order parameters and correlation functions}

A framework for classifying phases was introduced in
Ref.~\onlinecite{bf}\ in the context of CDWs, and we generalize it
here to more complex periodic media.  As in equilibrium, phases are
characterized by broken symmetry.  An ordinary crystal is globally
periodic, and hence has broken translational and rotational symmetry.
Because the periodicity is sustained over long distances, we say that
it has translational long range order (LRO).  To quantify this notion,
we must introduce suitable order parameters.

We consider initially a general model where the constituents of the
driven lattice may be ``oriented manifolds'', which are extended in
some direction, such as magnetic flux lines in a three-dimensional
superconductor.  The configuration of such a lattice is described by
labeling each manifold by its undisplaced equilibrium transverse
position, ${\bf x}$. The internal coordinates parallel to the
oriented manifold (e.g., along the magnetic field direction in the
vortex lattice) are parameterized by a $d_l$-dimensional vector
${\bf z}$, and ${\bf r}=({\bf x},{\bf z})$.  The number of
transverse dimensions is denoted by $d_t$ and the dimensionality
of space is $d=d_t+d_l$.  The lattice may be described by
a density field smoothed out on the scale of the lattice spacing,
$\rho({\bf x},{\bf z})$, which in the ordered phase becomes
periodic in the transverse direction.  For the vortex lattice the
density is the local magnetic induction, while for the CDW and
the Wigner crystal it is the electronic charge density.  In the
ordered phase the collection of oriented manifolds acquires long-range
periodicity, defined by a reciprocal lattice with basis vectors
$\{{\bf Q}\}$. The density field is then written as
\begin{equation}
\label{density}
\rho({\bf x},{\bf z})=\rho_0+{1\over 2}\sum_{\bf Q}
       \tilde\rho_{\bf Q}({\bf r}) 
        e^{i{\bf Q}\cdot{\bf x}},
\end{equation}
where $\rho_0$ is the mean density.  The complex Fourier components
$\tilde\rho_{\bf Q}({\bf r})$ satisfy $\tilde\rho_{-{\bf
    Q}}=\tilde\rho^*_{\bf Q}$.  All the amplitudes
transform as $\tilde\rho_{\bf Q}\rightarrow \tilde\rho_{\bf Q}
e^{i{\bf Q}\cdot{\bf a}}$ under a lattice translation ${\bf
  x}\rightarrow {\bf x}+{\bf a}$ \cite{notea}.  A nonzero
expectation value of the $\tilde\rho_{\bf Q}$'s thereby indicates
broken translational symmetry.  As a result of the broken
translational symmetry, long-wavelength fluctuations in the ordered
phase can be described in terms of long-wavelength distortions of the
Fourier modes, which can in general be written in terms of an
amplitude and a phase as
\begin{equation}
\label{phonon}
\tilde\rho_{\bf Q}({\bf r})=\tilde\rho_{{\bf Q},0}({\bf r}) 
e^{i{\bf Q}\cdot\tilde{\bbox{\phi}}({\bf r})}.
\end{equation}
Note that the phase in each Fourier amplitude is not an independently
fluctuating variable in the ordered state.  It is constrained by, e.g.
cubic interactions of the $\tilde\rho_{\bf Q}$, which leave only the
vector of phases $\tilde{\bbox{\phi}}$.  They correspond to distortions
of the lattice, and indeed $\tilde{\bbox{\phi}}$ can be interpreted as a
sort of displacement field (but see below).

So far, this discussion applies equally well to both equilibrium and
driven solids.  Consider now the {\sl time-dependence} of the local
density.  In a moving solid, we expect $\tilde{\bbox{\phi}} = {\bf v} t +
\bbox{\phi}$, so that
\begin{equation}
\label{moving_density}
\rho({\bf x},{\bf z},t)=\rho_0+{1\over 2}\sum_{\bf Q}
       \rho_{\bf Q}({\bf r},t) 
        e^{i{\bf Q}\cdot\left[{\bf x}+ {\bf v} t\right]},
\end{equation}
where
\begin{equation}
  \rho_{\bf Q}({\bf r},t) = \tilde\rho_{\bf Q}({\bf r},t)
  e^{-i{\bf Q}\cdot {\bf v} t}.
\end{equation}
Physically, the oscillations in the density simply reflect the fact
that individual constituents of the lattice pass any given point in a
regular periodic fashion.  The set of $\rho_{\bf Q}$ fields comprises
the order parameters for the non-equilibrium system.  Neglecting
topological defects (i.e.  dislocations) amounts to assuming that the
amplitude $\rho_{{\bf Q},0}$ is constant.  Elastic deformations of the
solid are then described entirely by phase fluctuations in terms of a
single-valued coarse-grained displacement field $\bbox{\phi}({\bf
  r})$.  It is important to distinguish between the displacement field
defined in the moving or crystal frame (where the conventional phonon
field is defined) and the displacement field in the laboratory frame.
Throughout this paper we denote by $\bbox{\phi}({\bf r})$ the
displacement field in the laboratory frame, while we reserve the
symbol ${\bf u}({\bf r})$ for the conventional phonon field.  Many of
the experimental systems of interest form triangular lattices.  In
this case, $\langle\rho_{\bf Q} \rangle \neq 0$ for three shortest
wavevectors in the ordered phase (the brackets indicate a time average
or average over thermal noise).

A quantity of immediate experimental interest is the static structure
factor, which is the Fourier transform of the equal time
density-density correlation function and is given by
\begin{eqnarray}
  \label{structure}
  S({\bf q}) &=& \langle|\delta\rho({\bf q},t)|^2\rangle \nonumber\\
  &=&\sum_{\bf Q}\int_{\bf r} e^{i({\bf q}-{\bf Q})\cdot{\bf r}}
  \langle\rho_{\bf Q}({\bf r},t)\rho^*_{\bf Q}({\bf 0},t)\rangle.
\end{eqnarray}
Here ${\bf q}=({\bf q}_t,{\bf q}_z)$ is the full wavevector.
The order parameter correlator, 
\begin{equation}
  \label{opcorr}
  C_{\bf Q}({\bf r},t)= \big[\langle\rho_{\bf Q}({\bf r},t)\rho^*_{\bf 
    Q}({\bf 0},0)\rangle\big]_{\rm ens.},
\end{equation}
governs the behavior of the structure factor. We used angular (square)
brackets to denote thermal (disorder) averages. If $C_{\bf Q}({\bf
r},0) \rightarrow {\rm const.}$ for large $|r|$, $S(\bbox{q})$ has a
sharp (resolution limited) Bragg peak at ${\bf q}_t={\bf Q}$.  If
$C_{\bf Q}({\bf r},0) \sim |r|^{-\eta}$, these broaden to power-law
form.  If $C_{\bf Q}({\bf r},0)$ decays exponentially for large $|r|$,
only diffuse peaks if any exist.  In these three cases, the system is
said to possess long-range, quasi-long-range, and short-range
translational order (at wavevector ${\bf Q}$), respectively.  These
will be abbreviated LRO, QLRO, and SRO, respectively, in the remainder
of the paper.

$C_{\bf Q}({\bf r},t)$ also describes the extent of temporal
periodicity.  By analogy to the translational characterization, we say
that the system possesses long-range, quasi-long-range, or short-range
temporal order (at frequency $\omega = {\bf Q}\cdot {\bf v}$) if
$C_{\bf Q}({\bf 0},t)$ goes to a constant value, decays algebraically,
or decays exponentially, respectively.  This temporal order can be
experimentally probed through the {\em dynamic} structure function
\begin{eqnarray}
  \label{structure_dyn} 
  S({\bf q},\omega) &=& \langle|\delta\rho({\bf q},\omega)|^2\rangle\\ 
  &=&\sum_{\bf Q}\int_{{\bf r},t} e^{i({\bf q}-{\bf Q})\cdot{\bf r}
  -i(\omega-{\bf Q}\cdot{\bf v})t} \langle\rho_{\bf Q}({\bf
  r},t)\rho^*_{\bf Q}({\bf 0},0)\rangle.\nonumber
\end{eqnarray}
Systems with temporal LRO will display sharp peaks at multiples of
the washboard frequency $\omega={\bf Q}\cdot{\bf v}$ in the $S({\bf
q},\omega)$ and in the related power spectrum of velocity
fluctuations.  Such system should also exhibit ``complete mode
locking'' to an arbitrarily weak external periodic driving at these
frequencies.

In a random system, temporal order is generally more robust than the
translational order.  Physically, the difference arises because
impurities inhomogeneously stress the system.  The responding
non-uniform distortion, however, can have very little fluctuation in
time, and thereby can leave the temporal ordered relatively unaffected.
It will, of course, have some effect on the dynamics, because
disturbances propagate differently atop the non-uniform background,
and because the local strains lower the barriers to defect nucleation.
We cannot exclude the possibility of a phase in which dislocations are 
unbound, but frozen in the laboratory frame. Such a phase 
would exhibit translational SRO but temporal LRO, and
would be the driven analog of the vortex glass phase originally
proposed by Fisher, Fisher and Huse \cite{FFH}.  Given the relative 
instability of this phase in equilibrium, we think this scenario is,
however, somewhat unlikely.
\begin{figure}[bth]
{\centering
\setlength{\unitlength}{1mm}
\begin{picture}(150,90)(0,0)
\put(-3,-20){\begin{picture}(150,90)(0,0) 
\includegraphics{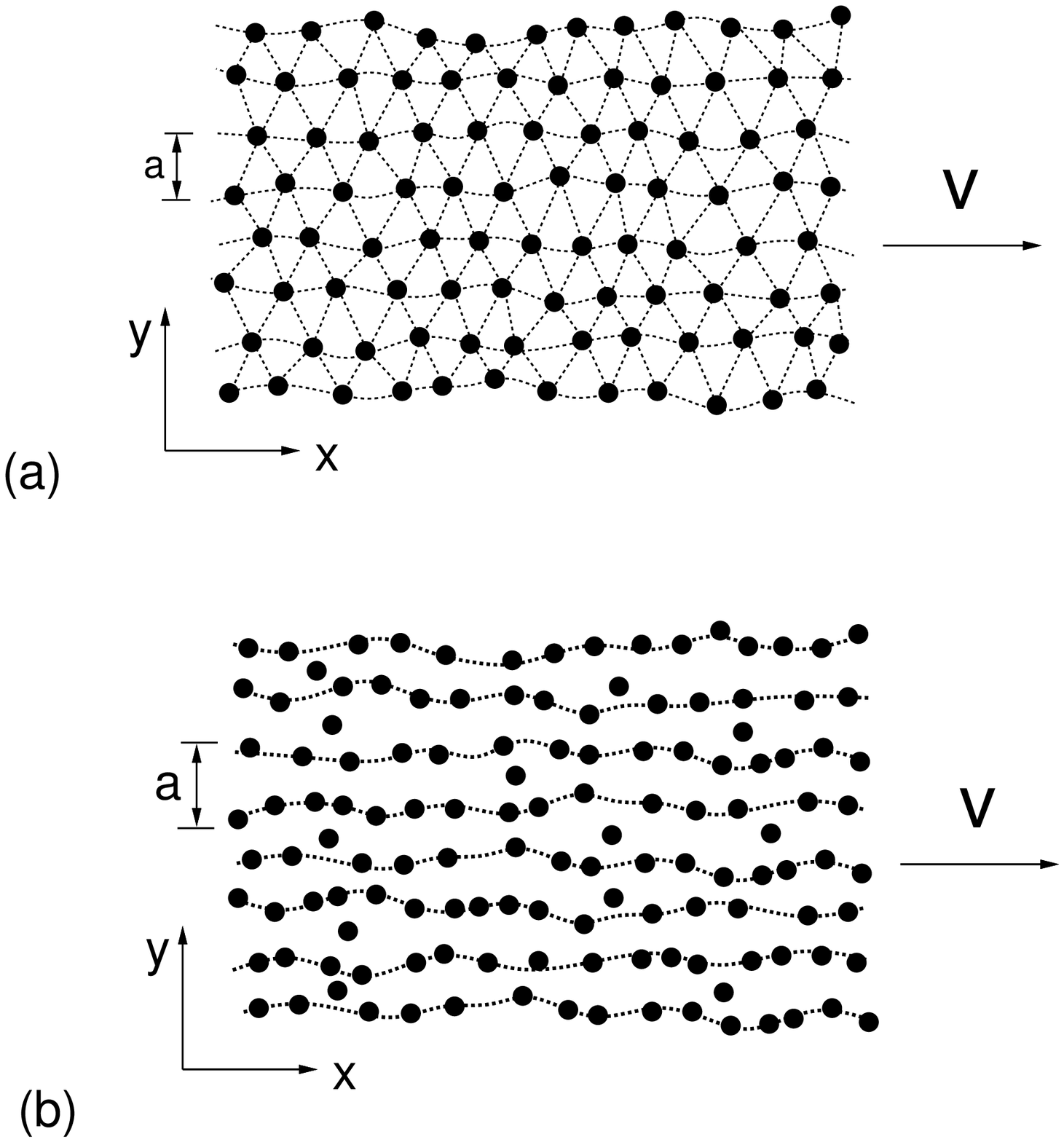}
\end{picture}}
\end{picture}}
Fig.1:{(a) Schematic illustration of a moving glass, which becomes
  unstable to the proliferation of dislocations with Burgers vectors
  along the direction of motion.  At large velocities, we expect that
  unbound dislocations will be widely separated, with spacings $x_{\rm
    sm}$ and $y_{\rm sm}$ in the directions longitudinal and
  transverse to the motion, respectively.  A naive estimate based on
  the linearized theory probably gives an upper bound for these
  scales: $x_{\rm sm} \lesssim c_{66}a^8(\gamma v)^7/\Delta^4$ and
  $y_{\rm sm} \lesssim c_{66}a^4(\gamma v)^3/\Delta^2$.  Nonlinear
  (KPZ) effects very likely shorten these lengths considerably.  On
  larger scales, the ``lattice'' crosses over to the driven smectic
  state displayed in (b), with only transverse periodicity and
  liquid-like order along the direction of motion.  The dislocations
  with Burgers vectors along the direction of motion allow the weakly
  coupled liquid channels to move at different velocities.}
\label{real_space}
\end{figure}

What particular types of translational order can in principle arise in
a driven system?  Of course, a disordered liquid state is possible
(and may be the only stable phase in low dimensions).  When
translational order is present, it can occur at a variety of
wavevectors ${\bf Q}$.  For weak disorder at low temperatures, it is
natural to expect that a full reciprocal lattice of wavevectors
characterizing a crystal should be important (i.e. have appreciable
$C_{\bf Q}$).  In two dimensions, for instance, the smallest of these
would typically be arranged into a hexagon.  If, as seems natural, and
can be shown in some simple models, reflection symmetry perpendicular
to the applied force is not broken, this can take one of two
orientations, with a diagonal oriented either parallel or
perpendicular to the force.  Since the applied force breaks rotational
symmetry, however, there is no reason for the correlations at all six
points to be identical.  Instead, if the system evolves continuously
with increasing temperature or disorder, it is natural to expect that
LRO will be lost first at some subset of these wavevectors.  The
surviving state has a lesser periodicity, with only a single line of
Bragg peaks (symmetry requires that the other solid peaks disappear
pairwise).  It represents therefore a layered liquid or {\sl smectic}
state.  This can be either a longitudinal smectic, with ordering
wavevector parallel to the velocity, or a transverse smectic, with
periodicity in the direction perpendicular to the velocity.  We argue
below that only the case of a transverse smectic shown in Figure 1(b)
is stable, and indeed, that this is likely to be a more generic state
than the true solid when the system is driven.  

\subsection{Hydrodynamics and elasticity}

There are at least two analytical approaches to calculating $C_{\bf
Q}({\bf r},t)$.  One is to construct a density-functional or
Landau-like theory for the order parameters $\rho_{\bf Q}$.  For the
non-equilibrium driven system considered here, this would be a set of
stochastic partial differential equations.  This has the advantage of
allowing large amplitude fluctuations, and hence including
dislocations in a natural way.  The disadvantage of this approach is
its intractability.  It is often difficult to recover relatively
simple properties in the ordered phase.  A second approach is to
assume a large degree of local order, so that the amplitude of the
order parameters fluctuates very little.  In this case, a phase-only
or elastic approximation is natural, and only the $\bbox{\phi}$ fields
remain in the description.  This method has the disadvantage of
excluding topological defects, which must be reintroduced by hand to
complete the description, as has been successfully done for
equilibrium systems, such as for example Kosterlitz and Thouless
treatment of 2d superfluids and 2d melting\cite{kt}.  Determining
the relevance or irrelevance of such defects upon the elastic
description in disordered systems is a difficult and unsolved problem.
If, however, the phase-only approximation predicts only small
displacements in the ordered phase, it can be expected to be a
self-consistent approximation out to a rather long length scale.  If,
by contrast, large distortions are found, the original assumption of
order is inconsistent, and we have discovered an instability of the
solid phase.

To determine the nature and stability of the possible ordered
non-equilibrium phases, this paper focuses on the elastic approach.
Assuming constant amplitude of the order parameters, $C_{\bf Q}$ can be
written in terms of the displacement field correlation as
\begin{equation}
  \label{dw} C_{\bf Q}({\bf r},t) = |\rho_{\bf Q,0}|^2 \left[\langle
  e^{i{\bf Q}\cdot [\bbox{\phi}({\bf r},t)-\bbox{\phi}({\bf
  0},0)]}\rangle\right]_{\rm ens.},
\end{equation} 
where we have pulled out the amplitude factor. If $\bbox{\phi}$ is a
Gaussian random variable,
\begin{equation}
  C_{\bf Q}({\bf r},t) = |\rho_{\bf Q,0}|^2 \exp\{-Q_{i}Q_{j} 
   B_{ij}({\bf r},t)\}, 
\end{equation}
where $i$ and $j$ denote Cartesian components and $B_{ij}({\bf r})$
are the components of a mean-square displacement tensor,
\begin{equation}
  \label{Gaussian_approx}
  B_{ij}({\bf r},t)= \left[\langle[\phi_i({\bf r},t)-\phi_i({\bf 0},0)]
  [\phi_j({\bf r},t)-\phi_j({\bf 0},0)]\rangle\right]_{\rm ens.}.
\end{equation}
Generally, the displacements fluctuate in a non-Gaussian manner, so
that Eq.~\ref{Gaussian_approx}\ is not strictly correct.  However, we
expect its qualitative implications to hold.  If $B_{ij}({\bf r},t)$
goes to a constant at long distances (times), then so does $C_{\bf
Q}({\bf r},t)$.  If $B_{ij}({\bf r},t)$ grows algebraically, then
$C_{\bf Q}({\bf r},t)$ has stretched exponential form (though not
necessarily with the naive stretching exponent), suggesting that in a
physical system topological defects might proliferate and lead to
short-range order, and if $B_{ij}({\bf r},t)$ grows logarithmically,
then $C_{\bf Q}({\bf r},t)$ decays as or slower than a power-law.

To calculate $B_{ij}({\bf r},t)$, we employ the analog of hydrodynamic
equations of motion.  In general, the hydrodynamics of systems far
from equilibrium is far more complex than its equilibrium counterpart.
In particular, fluctuations about the nonequilibrium steady state do
not satisfy a fluctuation-dissipation theorem and the external driving
force breaks both the rotational and reflection (parallel to the
force) invariance of the equilibrium system.  As a result, the
hydrodynamic equations contain many more parameters than in
equilibrium.  A general construction based only on symmetry
constraints is thus not very useful, and a concrete derivation, which
provides precise values for these parameters, is desirable.  We
perform such a derivation  in
Section~\ref{sec:hydro_derivation}.  Our first main result is the
complete non-equilibrium hydrodynamic equation of motion for the
driven lattice, 
\begin{eqnarray}
  \label{maineq}
  \gamma_{ij}\partial_t\phi_j & = &
  A_{i\alpha j}\partial_\alpha\phi_i
  +B_{i\alpha\beta j}\partial_\alpha\partial_\beta\phi_j\\ \nonumber
  & & +C_{i\alpha j\beta k}\partial_\alpha\phi_j\partial_\beta\phi_k
  + \tilde{F}_i[{\bf r},\bbox{\phi},t] + \eta_i({\bf r},t),
\end{eqnarray}
where the coefficients $A$,$B$, and $C$ are non-zero only when the
number of subscripts taking values of axes perpendicular to the
driving force is even.  Explicit expressions for these are given in
Section~\ref{sec:hydro_derivation}. $\tilde{F}$ and $\eta$ represent quenched
random and external ``thermal'' forces, respectively.
Eq.~\ref{maineq}\ remedies the deficiencies of various ad-hoc
equations of motion proposed in previous works \cite{scheidl}.

One noteworthy feature of Eq.~\ref{maineq}\  is the
proliferation of gradient terms beyond the usual equilibrium elastic
ones (contained in the $B$ term).  These represent convective effects
and dependence of the substrate-lattice interaction upon the local
deformation of the lattice.  To properly account for them, it is
crucial to treat the phonon modes near the zone boundary, not
considered in previous calculations.  Especially important are the $A$
terms, which lead to propagating modes at low frequencies and
wavevectors. 

Another key feature is the random force $\tilde{F}_i[{\bf r},
\bbox{\phi},t]$.  It is distinguished from the equilibrium form in two
ways.  First, it contains non-trivial time-dependence, as can be seen from
\begin{equation}
\label{pinforce}
\tilde{F}_i[{\bf r},\bbox{\phi},t]=\sum_{{\bf Q}}
  e^{i{\bf Q}\cdot({\bf x}-{\bf v}t-\bbox{\phi}({\bf r},t))}
  F_i({\bf r}),
\end{equation}
where $F_i({\bf r})$ is a time-{\sl independent} quenched random
variable, which we will refer to as a {\it static} pinning force (the
exponential factor in Eq.~\ref{pinforce}\ comes from the transformation
from the moving to the laboratory frame).  The terms with ${\bf
  Q}\cdot{\bf v} \neq 0$ therefore oscillate in the sliding
solid.  The second distinction is seen from the decomposition
\begin{equation}
  \label{pftwo}
  F_i({\bf r})=F_i^{\rm eq}({\bf r})+F_i^{\rm neq}({\bf r}).
\end{equation}
The first term on the right hand side
of Eq.~\ref{pftwo}\ is the {\it equilibrium} component of the static
pinning force and can be written as the gradient of a potential, as
required by the fluctuation-dissipation theorem.  Its correlations are
approximately given by
\begin{equation}
  \label{pfeq}
  [F_i^{\rm eq}({\bf r})F_j^{\rm eq}({\bf r'})]_{\rm ens.}=
  -\partial_i\partial_j\Gamma({\bf r}-{\bf r'}),
\end{equation}
where $\Gamma({\bf r})$ is the correlator of the Gaussian random potential.
The second term in Eq.~\ref{pftwo}\ is the {\it nonequilibrium} part
of the static pinning force, with correlator
\begin{equation}
\left [F_i^{\rm neq}({\bf r})F_j^{\rm neq}({\bf r'})\right ]_{\rm ens.}=
  g_{ij}\delta({\bf r-r'}).
\end{equation}
The variance $g_{ij}$ of this static force is given in 
Eq.~\ref{gij_result} and
vanishes in the absence of external drive.  A non-zero $g_{ij}$
implies that $F_i^{\rm neq}$ is non-conservative, violating the
fluctuation-dissipation theorem.

\subsection{Analysis}

A general analysis of Eq.~\ref{maineq}\ is quite difficult.  
In principle, the stability and behavior of the putative moving glass can be
determined by an RG analysis of the full equations of motion 
(Eqs.~1.11), including {\it both} the KPZ nonlinearities 
($C_{i\alpha j\beta k}$) and the random forces ($F_i({\bf r},\bbox{\phi},t)$).
Past experience with other nonequilibrium KPZ-like equations
\cite{bf,chen}\ suggests that instabilities are quite ubiquitous
in low dimensions. We expect such instabilities will occur also in this case,
at least in two dimensions and quite possibly in three, and leave an 
analytical check of this hypothesis to younger researchers who may
have enough career years ahead of them to complete the calculation.
Provided such an instability occurs, can any residual order survive
in the driven state? Two physical realizations of partially ordered
moving states have already been suggested: the longitudinal and the 
transverse smectic. A longitudinal smectic is equivalent to a conventional
driven CDW, studied earlier by Chen et al.~\cite{chen}
and by Balents and Fisher~\cite{bf}. These authors concluded that this
phase is unstable in two dimensions, but may exist at large velocities in
three dimensions (although the role of dislocations and KPZ nonlinearities
in three dimensions deserves further study).
\begin{figure}[bth]
{\centering
\setlength{\unitlength}{1mm}
\begin{picture}(150,95)(0,0)
\put(-3,-18){\begin{picture}(150,95)(0,0) 
\includegraphics{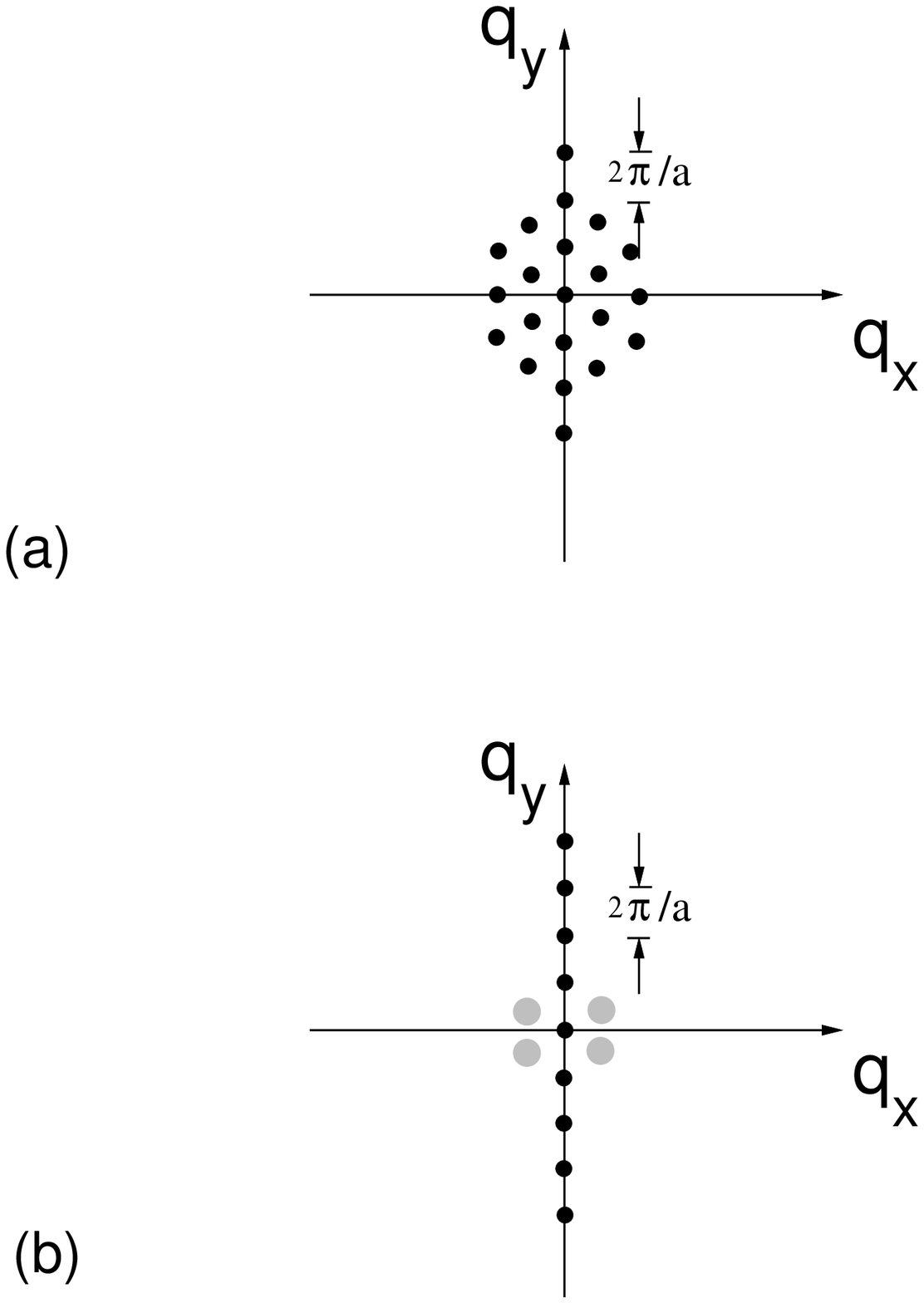}
\end{picture}}
\end{picture}}
Fig.2:{(a) The static structure factor for the moving glass sketched
in Fig.1(a), with sharp peaks at all the reciprocal lattice vectors of
a triangular lattice. The static structure factor of the driven
transverse smectic is shown in (b).  Sharp peaks (dark spots) only
appear along the $q_y$ axis; some remnant of short-range local
triangular lattice order can appear in the form of weak and diffuse
peaks at finite $q_x$ (light spots).}

\label{lattice}
\end{figure}
The only possible ordered phase in $d=2$ is thus a transverse smectic
(discussed in the previous subsection and illustrated in Fig.~1(b))
with some degree of order at a periodic set of ${\bf Q}$ perpendicular
to the velocity.  This is of course also a possibility in $d=3$,
regardless of the stability of the lattice. The marginal stability of
the driven lattice in three dimensions allows for a dynamic phase
transition between the driven smectic (at intermediate velocities) to
a moving lattice (at high velocities).  We illustrate the
corresponding dynamic phase diagrams in Fig.~3.

The latter part of the paper is devoted to an analysis of this
possibility.  In Section~\ref{sec:smectic}, we present the
hydrodynamic equations of motion for the smectic.  These include a
simplified version of Eq.~\ref{maineq}, supplemented by an additional
one for the conserved particle density.  Unlike in a
vacancy/interstitial-free solid, this is not generally slaved to the
smectic displacement, and constitutes a separate hydrodynamic
mode.\cite{vacancy_note}\ This is the well-known permeation mode in
smectic liquid crystals.

Even these equations are somewhat intractable, so in
Sections~\ref{sec:RGa}--\ref{sec:RGb}\ we study the ``toy model'' in
which the permeation mode is decoupled from the smectic displacement
$\phi_y \equiv \phi$.  This is best done using renormalization group
(RG) techniques, which are controlled in two limits.  At and near
$d=3$, the RG is controlled by a low-temperature fixed point, which is
analyzed using a {\sl functional} RG in Section~\ref{sec:RGa}.  The
fixed-point temperature $T^*$ increases with decreasing dimension
until in $d=2$, more conventional sine-Gordon RG techniques can be
applied.  Directly in three dimensions, there is a slow asymptotic
approach to zero temperature.

From these RG calculations come several concrete predictions.  In a
three-dimensional smectic, the structure factor has power-law Bragg
peaks along the axis in momentum space perpendicular to the velocity,
as illustrated in Fig.~2(b).  Because these peaks are entirely
transverse, the smectic exhibits neither narrow-band-noise nor
complete mode locking (although incomplete mode locking is of course
possible for sizeable ac drives).  The response to a force ${\bf
f}_\perp$ transverse to the mean direction of motion is a
superposition of two effects.  First, the permeation mode provides a
non-vanishing linear component of the velocity ${\bf v}_\perp^{\rm
perm.} = \mu_{\rm perm.}{\bf f}_\perp$.  Secondly, the smectic
responds in a non-linear manner, {\it resembling} a threshold at low
temperatures but crossing over to linear response at very low forces
for $T\neq 0$.

In two dimensions, the RG study of the elastic model also predicts
only short-range {\em asymptotic transverse} translational
correlations and quasi-long-range temporal correlations.  These
calculations actually indicate that the system is outside the regime
of validity of the elastic approximation. We expect that in $d=2$ an
eventual instability of even the smectic state to unbinding of
transverse dislocations may occur at larger length scales.
Nevertheless, the RG results should hold up to this length, and, as in
three dimensions, also predict superimposed linear and non-linear
transverse velocity responses.

We conclude the Introduction with a summary of the remainder of the
text.  Section II describes the derivation of hydrodynamic equations
of motion for the driven lattice, which we study perturbatively in
Section III.  Sections IV presents the equations of motion for the
transverse smectic, which are analyzed using renormalization group
techniques in Sections V and VI.  The specific predictions of the RG
for correlation and response functions are given in section VII.
Section VIII summarizes our results, makes comparisons with other
work, and gives a synopsis of remaining open questions and future
applications of these ideas.  Finally, six Appendices give technical
details unsuitable for the body of the paper.
\begin{figure}[bth]
{\centering
\setlength{\unitlength}{1mm}
\begin{picture}(150,140)(0,0)
\put(-25,-15){\begin{picture}(150,60)(0,0) 
\includegraphics{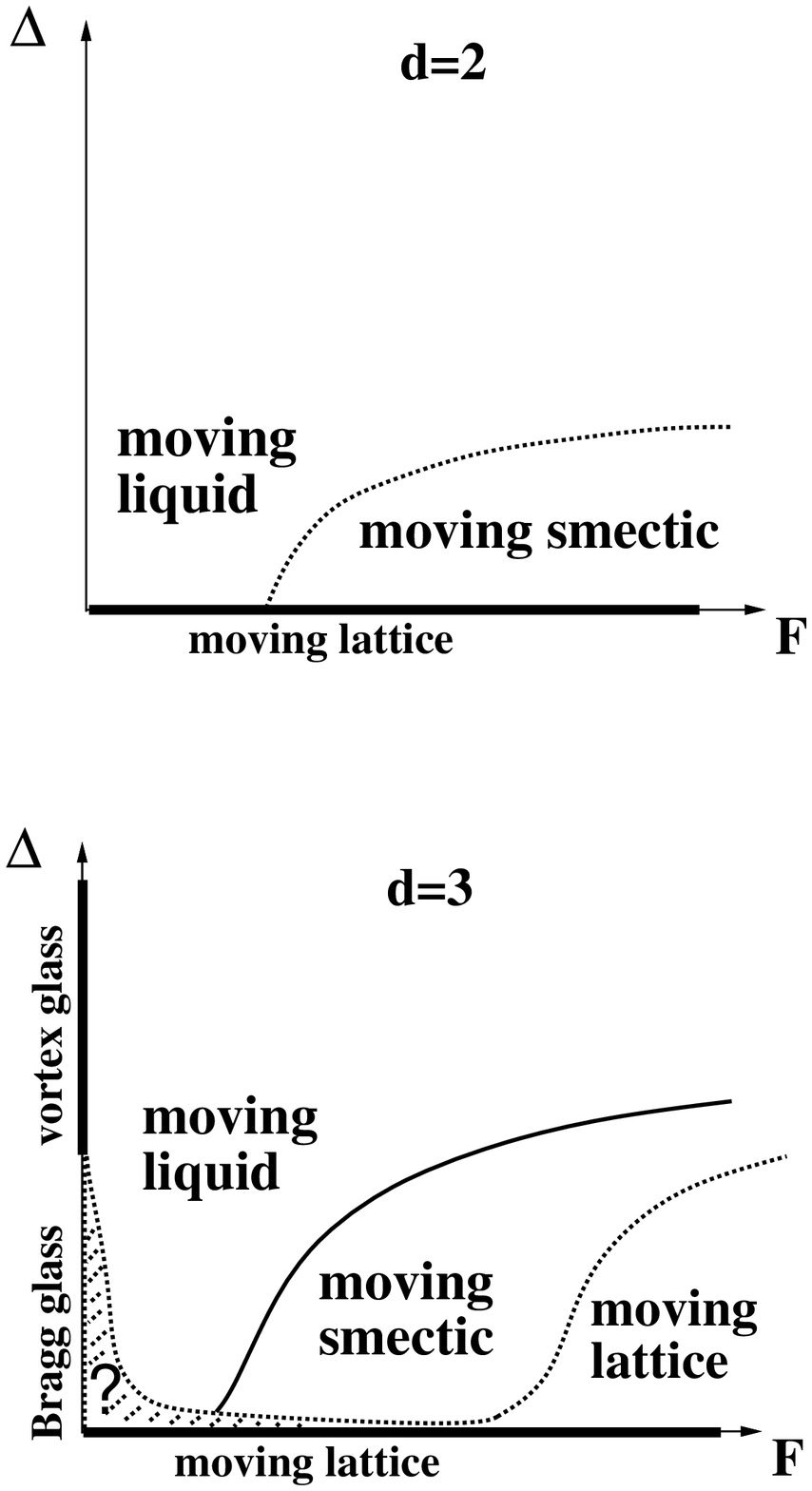}
\end{picture}}
\end{picture}}
Fig.3:{Schematic phase diagram for a finite temperature driven solid
in (a) $d=2$ and (b) $d=3$ dimensions.  In (a) we have used a dashed
line as a boundary between the moving liquid and the transverse
smectic to emphasize that the smectic might be unstable in 2d and
therefore asymptotically indistinguishable from a moving liquid. In
(b) we have similarly used a dashed boundary between the moving
smectic and lattice, to emphasize that it is likely that the moving
lattice is unstable even in 3d. In (b) the hatched ``?'' region
indicates the interesting possibility of a moving Bragg glass in 3d,
at low drives.}
\label{phase_diagram}
\end{figure}

\section{Derivation of Hydrodynamics}
\label{sec:hydro_derivation}

As discussed above, the goal of this paper is to study the
non-equilibrium steady states that arise in driven periodic media.  We
will focus on the low-energy and long-wavelength properties of these
steady states, in cases where these are uniquely defined.  This should
be the case provided ergodicity is achieved, either through a small
non-zero ``thermal'' noise or via intrinsic chaotic dynamics of the
system in the absence of external noise.  We will not discuss
zero-temperature systems in which relatively simple global attractors
exist with non-chaotic (e.g. periodic) dynamics.

In the limit of interest, then, we expect a sort of hydrodynamic
description to hold.  Such hydrodynamics is particularly successful in
equilibrium because it is highly constrained.  It must respect both
detailed balance (to reduce to equilibrium statistical mechanics) and
the symmetries of the system -- in this case translations, rotations,
and reflections.  Out of equilibrium, a putative steady state equation
of motion can be much more general.  The fluctuations around the
driven state need not satisfy any fluctuation--dissipation relation,
and the external driving force breaks both rotational and reflection
invariance.  

For this reason, there are many more parameters in the hydrodynamic
description.  In the absence of further input, it is therefore
considerably less powerful than equilibrium hydrodynamics.  To make it
useful, we need a means of calculating these parameters
``microscopically''.  This is possible for weak external potentials.

The calculation proceeds in two stages.  Beginning with a microscopic
lattice Hamiltonian, we first explicitly coarse-grain the equations of
motion.  This is a mode-elimination procedure in momentum space,
reminiscent of a single step of rescaling in a renormalization group
calculation.  It is this stage which requires a weak external
potential, since the elimination can only be performed perturbatively
in these mode-coupling terms.  The end result is an effective equation
of motion for the phonon modes with momenta within a small sphere of
radius $\Lambda \ll 2\pi/a$ in momentum space ($a$ is the lattice
spacing).

The second step is to transform the equation of motion from the moving
frame (in which the conventional phonon coordinate or displacement
field is defined) to the laboratory frame (more relevant for physical
measurements).  While technically much simpler than the previous
mode-elimination step, it is possible only at this stage, since the
transformation can be carried out only as a gradient expansion, with
the small parameter $\Lambda a \ll 1$.  

\subsection{Formulation}

The formulation of the problem begins with some microscopic
Hamiltonian, describing particles connected by springs.  The natural
degrees of freedom are thus the displacements of these particles,
whose gradients are expected to be small provided the potential acting
on the particles is weak.  The high-momentum modes being eliminated
here therefore describe {\sl small} relative displacements of nearby
(say, neighboring) particles.  Small though these are, they are
crucial to the physics of the {\sl sliding} solid.

Furthermore, the effects of the fixed external potential are expected
to be most pronounced when it is has strong periodic components
commensurate with the driven lattice.  This is intuitively reasonable,
and indeed comes out of our more detailed calculations.  This physics,
however, comes precisely from these high-momentum modes at the scale
of the lattice spacing.  Even if the lattice is not pinned into a
static configuration, these modes are the most strongly coupled to the
static matrix, and thereby give rise to the seeds of interesting
non-linear dynamics.

We consider a general model in which the constituents of the driven
lattice may be ``oriented manifolds'', which are extended in some
directions.  This allows for systems including, for example, vortex
lines in a three-dimensional superconductor.  
The conformation of a manifold with undisplaced equilibrium transverse 
position ${\bf x}$ in a distorted lattice is described
by the displacement vector ${\bf u({\bf x,z})}$, such that 
the true position is
\begin{equation}
  {\bf X}({\bf x,z}) = {\bf x + u(x,z)},
\end{equation}
where ${\bf u}$ and ${\bf
  x}$ are $d_t$-component vectors, while ${\bf z}$ has
$d_l$ components.  

A natural microscopic Hamiltonian, valid in most cases, is
\begin{eqnarray}
  H & = & \sum_{\bf x x'} \int_{\bf z z'} V[ {\bf
    x-x' + u(x,z) - u(x',z'),z-z'}] \nonumber
  \\
  & & +
  \sum_{\bf x} \int_z U[{\bf x + u(x,z),z}].
\end{eqnarray}
Here $V[{\bf x,z}]$ is a two-body manifold-manifold interaction,
and $U[{\bf x,z}]$ is the external potential of the static
medium.

Expanding the interaction potential in the usual way gives the elastic
energy
\begin{eqnarray}
  H & \approx & {1 \over 2}\sum_{\bf x,x'}\int_{\bf zz'}
  K_{ij}({\bf x -x',z-z'}) u^i({\bf
  x,z}) u^j({\bf x',z'}) \nonumber \\
& & + \sum_{\bf x}\int_z  U[{\bf x + u(x,z),z}], 
\end{eqnarray}
where the elastic matrix
\begin{equation}
  K_{ij}({\bf r}) = \sum_{{\bf x'}}\int_{\bf z'} \partial_i\partial_j V({\bf 
r'})
\delta_{\bf x,0}\delta({\bf z}) - \partial_i\partial_j V[{\bf
  r}]. 
\end{equation}

At this point, it pays to establish some notation.  In the previous
equation, we have already adopted a convention for indices.  Latin
indices alphabetically following $i$ denote transverse (${\bf x}$)
coordinates, while $a,b,\ldots,h$ denote longitudinal (${\bf z}$)
ones.  If an index may range over the full space, a Greek index
$\alpha,\beta,\ldots$ will be used.  We will attempt to use ${\bf x}$
and ${\bf z}$ (with primes, etc.) exclusively as transverse and
longitudinal coordinate vectors, with corresponding momenta ${\bf q}_t
$ and ${\bf q}_z$.  These will often be assembled into vectors in the
full $d$-dimensional space denoted ${\bf r = (x,z)}$ and ${\bf q} =
({\bf q}_t,{\bf q}_z)$.

In Fourier space, using reflection symmetry, the elastic matrix is
\begin{equation}
  K_{ij}({\bf q}) \equiv \sum_{\bf x} \int_z K_{ij}({\bf r}) e^{i{\bf
      q \cdot r}} = \sum_{\bf x} \int_z \partial_i\partial_j V({\bf r}) 
[1-e^{i{\bf q\cdot
      r}} ].
\end{equation}

Overdamped relaxational dynamics is defined by
\begin{equation}
  \gamma \partial_t u_i({\bf r},t) = - {{\delta H} \over {\delta
    u_i({\bf r},t)}} + f_i,
\end{equation}
where ${\bf f}$ is an external force.  This is equivalent to the
equation of motion
\begin{eqnarray}
  \gamma \partial_t u_i({\bf r},t) & = & - \sum_{\bf x'}\int_{{\bf
      z}'} K_{ij}({\bf r-r'}) u_j({\bf r'},t) \nonumber \\ 
      & & +\tilde{F}_i[{\bf x} + {\bf u}({\bf r},t) + {\bf v}t,{\bf z}],
      \label{eom0}
\end{eqnarray}
where we have shifted ${\bf u}$ to remove the external force,  
${\bf v} = {\bf f}/\gamma$, and
\begin{equation}
 \tilde{F}_i({\bf r}) = -\partial_i U({\bf r}).
\end{equation}
In general, ${\bf v} = {\bf f}/\gamma$ is not the true steady-state
velocity for a given force ${\bf f}$, since interactions with the
impurities intermittently pin the lattice and thereby reduce the
sliding speed.  This finite renormalization (which is quantitatively
small at large velocities) can, however, be neglected for the current
purpose of studying the properties of the steady state.  This is
analogous to ignoring the mass renormalization in field theory (or
$T_c$ shift in critical phenomena). Of course from the experimental
point of view, the velocity curve $v(f)$ (like the true T$_c$) is an
important and measurable quantity.

\subsection{Mode elimination}
\label{mode_elimination}

Having set up the dynamics, we are now in a position to perform the
coarse-graining.  We will use the Martin-Siggia-Rose (MSR) formalism\cite{MSR},
which allows the mode-elimination to be performed by functional
integration.  The object of interest in the MSR method is the
statistical weight
\begin{equation}
  W[\hat{\bf u},{\bf u}] = e^{-\tilde{S}},
\end{equation}
where the action $\tilde{S}=S_0 + \tilde{S}_1$, with
\begin{equation}
  S_0 = \sum_{\bf x} \int_{{\bf z}t} \hat{u}_i({\bf r},t) \bigg[
  \gamma \partial_t 
  u_i({\bf r},t) + \sum_{\bf x'} \int_{\bf z'} K_{ij}({\bf
    r-r'}) u_j({\bf r'},t)\bigg]
\end{equation}
and
\begin{equation}
 \tilde{S}_1 =  - \sum_{\bf x} \int_{{\bf z} t} \hat{u}_i({\bf r},t)
   \tilde{F}_i[{\bf x}+{\bf u}({\bf r},t)+{\bf v} t,{\bf z}].
\end{equation}

With this choice of weight, correlation functions are given by the
functional integral
\begin{equation}
  \langle u_i({\bf r},t) u_j({\bf r'},t')\cdots\rangle = \int
  [d\hat{\bf u}][d{\bf u}] u_i({\bf r},t) u_j({\bf r'},t') \cdots e^{-\tilde{S}}.
\end{equation}
In this expression, the left-hand side should be interpreted as simply
the product of the specified $u_i$ evaluated along the solution of
Eq.~\ref{eom0}. The right-hand side is the result of functional
integration over {\sl all} functions ${\bf u},\hat{\bf u}$.  Because
of this identity, we will freely employ angular brackets in either
context.  We note that this equality relies crucially on proper
regularization of the equal-time correlators in the field theory.  In
particular, we will choose the causal convention 
\begin{equation}
  \langle u_i({\bf r},t) \hat{u}_j({\bf r'},t) \rangle = 0.
\end{equation}
Correlation functions involving the field $\hat{\bf u}$ have the
physical interpretation of {\sl response} functions, as can be seen by
simply differentiating a correlation function with respect to an
applied force.  This convention then simply implies that there is no
instantaneous (and hence discontinuous) response to a perturbation.

Within the MSR formalism, we can now readily carry out the mode
elimination.  The fields are first separated
into two parts via
\begin{eqnarray}
  {\bf u}({\bf r},t) & = & {\bf u}_<({\bf r},t) + {\bf u}_>({\bf
    r},t), \nonumber \\
  \hat{\bf u}({\bf r},t) & = & \hat{\bf u}_<({\bf r},t) + \hat{\bf
    u}_>({\bf 
    r},t), \label{splitting}
\end{eqnarray}
Here the slow and fast fields are defined by
\begin{eqnarray}
  {\bf u}_<({\bf r},t) & \equiv & \int_{|{\bf q}_t|<\Lambda} \int_{{\bf
      q}_z\omega} {\bf u}({\bf q},\omega) e^{i{\bf q\cdot r}-i\omega
    t}, \\
  \hat{\bf u}_<({\bf r},t) & \equiv & \int_{|{\bf q}_t|<\Lambda} \int_{{\bf
      q}_z\omega} \hat{\bf u}({\bf q},\omega) e^{-i{\bf q\cdot r}+i\omega
    t}, \\
  {\bf u}_>({\bf r},t) & \equiv & \int_{|{\bf q}_t|>\Lambda} \int_{{\bf
      q}_z\omega} {\bf u}({\bf q},\omega) e^{i{\bf q\cdot r}-i\omega
    t}, \\
  \hat{\bf u}_>({\bf r},t) & \equiv & \int_{|{\bf q}_t|>\Lambda} \int_{{\bf
      q}_z\omega} \hat{\bf u}({\bf q},\omega) e^{-i{\bf q\cdot r}+i\omega
    t},
\end{eqnarray}
where we have adopted opposite Fourier sign conventions for the
displacement and response fields, and all momentum integrations are
restricted to within the Brillouin zone.  Correlation functions of the
slow fields describe all the long-wavelength (i.e. hydrodynamic and
elastic) behaviors of the system, and can be obtained from the
effective weight
\begin{equation}
  W_{\rm eff.}[{\bf u}_<,\hat{\bf u}_<] = e^{-\tilde{S}_{\rm eff}} = \int
  [d{\bf u}_>][d\hat{\bf u}_>] e^{-\tilde{S}}.
\end{equation}

For simplicity, let us consider the model at zero temperature -- i.e.,
in the absence of any external time-dependent noise.  Because the
potential couples the slow and fast modes, we will
nevertheless obtain non-trivial renormalizations of the slow dynamics
from the mode elimination.  

We proceed by inserting the decomposition in Eq.~\ref{splitting}
into the action $\tilde{S}$.  Furthermore assuming that the potential
$U$ is weak, we may expand the exponential and perform the functional
integrations over the fast modes order by order, re-exponentiating the
resulting expressions, which then depend only upon the slow fields.
The elastic part of the Hamiltonian is diagonal in momentum space,
\begin{equation}
   S_0 = \int_{{\bf q},\omega} \hat{u}_i({\bf q},\omega) \bigg[
  i \gamma \omega + K_{ij}({\bf q},\omega) \bigg]
  u_i({\bf q},\omega),
\end{equation}
so that upon decomposition the slow and fast fields are decoupled in
this term: 
\begin{equation}
  S_0[{\bf u},\hat{\bf u}] = S_0[{\bf u}_<,\hat{\bf u}_<] + S_0[{\bf
    u}_>,\hat{\bf u}_>].
\end{equation}
The effective potential is therefore
\begin{equation}
  \tilde{S}_{\rm eff} = S_0[{\bf u}_<,\hat{\bf u}_<] - \ln \left\langle
  e^{-\tilde{S}_1} \right\rangle_{0>},
\label{effectiveS}
\end{equation}
where the angular bracket with the subscripts indicates integration
over the fast modes ($>$) with the additional weight factor $e^{-S_0}$
($0$).

From this point onward, the treatment differs for the periodic and
random potential, so we divide the remainder of this subsection into
two parts.  Each involves the conceptually
straightforward perturbative calculation of the average in
Eq.~\ref{effectiveS}.  This is somewhat tedious technically, so
details will be given in Appendices A and B.

\subsubsection{Periodic potential}

We can specify a periodic potential by the Fourier decomposition,
\begin{equation}
  U({\bf x}) = \sum_{\bf Q} e^{i{\bf Q \cdot x}}
  U_{\bf Q}.
  \label{Fourier}
\end{equation}
Note that we have taken the potential to be independent of the
continuous longitudinal coordinates ${\bf z}$.  

\end{multicols}

Inserting the above Fourier decomposition and evaluating the
expectation value in Eq.~\ref{effectiveS}\ to second order in $U$
gives corrections to $S_0$ which correspond to a renormalized equation
of motion (see Appendix A).  Since for small $\Lambda$, the remaining
``slow'' fields have small gradients, the corrected equation of motion
may be written in the continuum approximation.  It takes the general
form
\begin{equation}
  \tilde\gamma_{ij} \partial_t u_j = \tilde{A}_{i\alpha j}
  \partial_\alpha u_j + 
  \tilde{B}_{i\alpha\beta j}\partial_\alpha\partial_\beta u_j +
  \tilde{C}_{i\alpha 
    j\beta k} \partial_\alpha u_j \partial_\beta u_k +F^s_i[{\bf x}
  + {\bf u}({\bf r},t) + {\bf v} t,{\bf z}]. 
  \label{eom}
\end{equation}
Here the gradients may be understood either as lattice differences or
as the corresponding expressions in momentum space.  The coefficients
are
\begin{eqnarray}
  \tilde\gamma_{ij} & = & \gamma \delta_{ij} - \sum_{\bf Q} Q_{ i}
  Q_{ j} 
  Q_{ k} Q_{ l} |U_{\bf Q}|^2 i {\partial \over
    {\partial \omega}} G_{kl}({\bf Q},{\bf v} \cdot {\bf
    Q}), \\
  \tilde{A}_{i\alpha j} & = & \sum_{\bf Q} Q_{ i} Q_{ j}
  Q_{ k} Q_{ l} |U_{\bf Q}|^2 i {\partial \over
    {\partial q^\alpha}} G_{kl}({\bf Q},{\bf v} \cdot {\bf
    Q}), \\
  \tilde{B}_{i\alpha\beta j} & = &  B^0_{i\alpha\beta j} -{1 \over
    2}\sum_{\bf Q} 
  Q_{ i} Q_{ j} 
  Q_{ k} Q_{ l} |U_{\bf Q}|^2 {\partial \over
    {\partial q^\alpha}} {\partial \over
    {\partial q^\beta}} G_{kl}({\bf Q},{\bf v} \cdot {\bf
    Q}), \\
  \tilde{C}_{i\alpha j\beta k} & = & {i \over 2}\sum_{\bf Q} Q_{ i}
  Q_{ j} Q_{ k} Q_{ l} Q_{ m} |U_{\bf Q}|^2
  {\partial \over {\partial q^\alpha}} {\partial \over
    {\partial q^\beta}} G_{lm}({\bf Q},{\bf v} \cdot {\bf
    Q}),
\end{eqnarray}
where
\begin{equation}
  B^0_{i\alpha\beta j} \equiv {\partial \over
    {\partial q^\alpha}} {\partial \over
    {\partial q^\beta}} K_{ij}({\bf q=0}) 
\end{equation}
is the bare linearized elastic matrix.  To leading order, the force is
unrenormalized,
\begin{equation}
  F_i^s[{\bf r}] = \tilde{F}_i[{\bf r}].
\end{equation}
Here we have written the expressions in terms of the Fourier transform
of the Green's function, $G({\bf q},\omega)$, which is defined  in the
extended zone scheme (i.e. it is periodically repeated in each
translated copy of the fundamental Brillouin zone).  It is
\begin{equation}  
  G({\bf q},\omega) = \bigg[ i\omega I + K({\bf q}) \bigg]^{-1}
  \theta(|q| - \Lambda). \label{fast_response}
\end{equation}
The $\theta$-function is present because only a partial mode
elimination has been performed, so that the slow modes remain as
dynamical variables in the coarse-grained theory.  Note that these
expressions {\sl diverge} in the limit of zero velocity and identical
periodicities, since in this case all the ${\bf Q}$ are
equivalent to the origin in momentum space.

\begin{multicols}{2}

\subsubsection{Random potential}

In the case of the random potential, we adopt the approach of disorder
pre-averaging the MSR functional, thereby working directly with the
variance of the random force.  This is done purely for technical
reasons: identical results are obtained by coarse-graining for a fixed
realization of disorder and eventually averaging only physical
quantities.

We take the simplest model for disorder, in which $U({\bf r})$ is
Gaussian distributed and statistically translationally invariant, with
zero mean and second cumulant
\begin{equation}
  \left[ U({\bf r}) U({\bf r'}) \right]_{\rm ens.} = \tilde\Gamma({\bf r-r'}).
\end{equation}
The force-force correlator is thus
\begin{equation}
  \bigg[ \tilde{F}_i({\bf r}) \tilde{F}_j({\bf r'})\bigg]_{\rm ens.} =
  -\partial_i\partial_j \tilde\Gamma({\bf r-r'}).
\end{equation}
The averaged statistical weight is readily computed:
\begin{equation}
  \left[ W[\hat{\bf u},{\bf u}] \right]_{\rm ens.} = e^{-S},
\end{equation}
with $S= S_0 + S_1$, and
\begin{eqnarray}
  S_1 & = & {1 \over 2} \sum_{\bf r,r'} \int_{{\bf zz'}tt'}
  \hat{u}^i({\bf r},t) 
  \hat{u}^j({\bf r}',t') \nonumber \\
  & & \times \partial_i\partial_j \tilde\Gamma[{\bf
    x-x'+{\bf u-u'} +{\bf v}}(t-t'),{\bf z-z'}].
\end{eqnarray} 

As before, we can coarse-grain by integrating out the fast ($>$)
modes.  The analog of Eq.~\ref{effectiveS}\ is
\begin{equation}
  S_{\rm eff} = S_0[{\bf u}_<,\hat{\bf u}_<] - \ln \left\langle
  e^{-S_1} \right\rangle_{0>}.
\label{effectiveStilde}
\end{equation}
In Appendix B, we compute this average to second order in $S_1$.  This
again gives an equation of motion of the form, Eq.~\ref{eom}, but with
\begin{eqnarray}
  \tilde{A}_{i\alpha j} & = & - \sum_{\bf x} \int_{{\bf z}t}
  \tilde\Gamma_{ijkl}[{\bf x} + {\bf v} t,{\bf z}] r^\alpha
  G_{kl}({\bf r},t), \\
  \tilde\gamma_{ij} & = & \gamma\delta_{ij}+\sum_{\bf x} \int_{{\bf z}t}
  \tilde\Gamma_{ijkl}[{\bf x} + {\bf v} t,{\bf z}] t
  G_{kl}({\bf r},t), \\
  \tilde{B}_{i\alpha\beta j} & = & B^0_{i\alpha\beta j} \nonumber \\
  &  & \hspace{-0.5in} + {1 \over 2} \sum_{\bf x} \int_{{\bf z}t}
  \tilde\Gamma_{ijkl}[{\bf x} + {\bf v} t,{\bf z}] r^\alpha r^\beta
  G_{kl}({\bf r},t), \\
  \tilde{C}_{i\alpha j\beta k} & & \nonumber \\
   &  & \hspace{-0.5in}= -{1 \over 2} \sum_{\bf x} \int_{{\bf z}t}
  \tilde\Gamma_{ijklm}[{\bf x} + {\bf v} t,{\bf z}] r^\alpha r^\beta
  G_{lm}({\bf r},t),
\end{eqnarray}
where we have abbreviated $\tilde{\Gamma}_{ij\cdots} \equiv
\partial_i\partial_j \cdots \tilde\Gamma$.
Also useful are the corresponding expressions in momentum space,
\begin{eqnarray}
  \tilde{A}_{i\alpha j} & = & \int_{\bf q} q_i q_j q_k q_l
  \tilde\Gamma({\bf q}) i 
  {\partial \over {\partial q^\alpha}} G_{kl}({\bf q}, {\bf v} \cdot
  {\bf q}_t ), \\
  \tilde\gamma_{ij} & = & \gamma\delta_{ij}-\int_{\bf q} q_i q_j q_k q_l 
\tilde\Gamma({\bf q}) i
  {\partial \over {\partial \omega}} G_{kl}({\bf q}, {\bf v} \cdot
  {\bf q}_t ), \\
  \tilde{B}_{i\alpha\beta j} & = & B^0_{i\alpha\beta j} \nonumber \\
   & & \hspace{-0.5in} -  {1 \over 2}  \int_{\bf q} q_i q_j q_k q_l
  \tilde\Gamma({\bf q})  
  {\partial \over {\partial q^\alpha}} {\partial \over {\partial
      q^\beta}} G_{kl}({\bf q}, {\bf v} \cdot 
  {\bf q}_t ), \\
  \tilde{C}_{i\alpha j\beta k} & & \nonumber \\
  & & \hspace{-0.5in}=  {i \over 2}   \int_{\bf q} q_i q_j q_k q_l
  q_m \tilde\Gamma({\bf q})  
  {\partial \over {\partial q^\alpha}} {\partial \over {\partial
      q^\beta}} G_{kl}({\bf q}, {\bf v} \cdot 
  {\bf q}_t ).
\end{eqnarray}

A correction $\delta\Gamma_{ij}({\bf r})$ is also obtained to the
random force correlator.  Because it is unpardonably ugly, we quote it
only in Appendix B.  It is, however, straightforward to show that the
renormalized force-force correlator {\sl cannot} be written in terms of
a random potential correlator, i.e.
\begin{equation}
  \Gamma_{ij}({\bf r}) \equiv \tilde\Gamma_{ij}({\bf r}) +
  \delta\Gamma_{ij}({\bf r}) \neq -\partial_i \partial_j
  \Gamma({\bf r}),
\end{equation}
for {\sl any} function $\Gamma({\bf r})$.  The difference from
the equilibrium form can be accounted for by separating the force into
two components:
\begin{equation}
  F^s_i({\bf r}) = F_i^{\rm eq.}({\bf r}) + F_i^{\rm neq.}({\bf r}),
  \label{force_decomposition}
\end{equation}
where the equilibrium component $F_i^{\rm eq.}$ is the gradient of a
potential, so that
\begin{equation}
  \left[ F_i^{\rm eq.}({\bf r}) F_j^{\rm eq.}({\bf r}')
  \right]_{\rm. ens.} = -\partial_i \partial_j \Gamma^{\rm
    eq.}({\bf r-r'}). 
\end{equation}
The other component we denote a (${\bf u}$-independent) {\sl static
  force}, with the correlator
\begin{equation}
  \left[ F_i^{\rm neq.}({\bf r}) F_j^{\rm neq.}({\bf r}')
  \right]_{\rm. ens.} = g_{ij} \delta({\bf r-r'}).
\end{equation}
Clearly, since the correction term $\delta\Gamma_{ij}$ is small, we
have
\begin{equation}
  \Gamma^{\rm eq.}({\bf r}) \approx \tilde\Gamma({\bf r}).
\end{equation}
The static force variance is determined, however, entirely by the
correction.  It can be obtained by integrating 
\begin{equation}
  g_{ij} = \int_{\bf r} \delta\Gamma_{ij}({\bf r}).
\end{equation}
Substituting in Eq.~(\ref{toodamnlong}) from Appendix B, all but the
first and fourth terms are total derivatives and hence integrate to
zero.  After a certain amount of manipulation, we find
\begin{eqnarray}
  g_{ij} & = & \int_{\bf q} q_i q_j q_k q_l q_m q_n |\tilde\Gamma({\bf q})|^2
  G_{km}({\bf q},{\bf q_t\!\cdot\! v}) \nonumber \\
  & & \times \left[ G_{ln}({\bf q},-{\bf q_t\!\cdot\! v})\!  -
    \! G_{ln}({\bf q},{\bf q_t\!\cdot\! v}) \right].
  \label{gij_result}
\end{eqnarray}
As expected, this expression vanishes for ${\bf v} = 0$,
i.e. in order to satisfy the fluctuation-dissipation relation in
equilibrium, the random force {\sl must} be the derivative of a random
potential.

\subsection{Transformation to Laboratory Frame}

Up to this point, we have worked with conventional displacement
fields, defined in the crystal frame.  This means that each particle
in the lattice is labeled by its equilibrium position ${\bf r} =
({\bf x},{\bf z})$, and that its actual transverse position is
given by
\begin{equation}
  {\bf X} = {\bf x} + {\bf v} t + {\bf u}({\bf
    x},{\bf z},t).
\end{equation}
Most measurements in the systems of interest are, however, conducted
in the laboratory frame.  It is therefore advantageous to adopt a
description based directly in this frame.  To do so, we define a new
field $\bbox{\phi}({\bf r},t)$ to be the displacement of the particle
which is located at position ${\bf r}$ {\sl at time $t$}.  Formally,
this is described by the implicit equation
\begin{equation}
  \phi_i({\bf X},{\bf z},t) = \phi_i({\bf x} + {\bf v} t
  +{\bf u}({\bf r},t),{\bf z},t) = 
  \tilde{u}_i({\bf r},t). \label{phi_def}
\end{equation}
Here, because ${\bf u}$ is defined only at a discrete set of points
$\{{\bf x}\}$, we have written the previous equation in terms of
a smoothed continuum field $\tilde{\bf u}$ defined at all space
points.  For many purposes this distinction is insignificant, but it
will return at one important juncture.

At this point the need for coarse-graining {\em prior} to the frame
transformation is clear.  Eq.~(\ref{phi_def}) has an unambiguous
solution defining $\bbox{\phi}$ only when $\partial_\alpha u_i \ll
1$.  In this case, we can obtain the transformation rules for
gradients by simply differentiating.  They are
\begin{eqnarray}
  \partial_t \tilde{u}_i & = & \left(\delta_{ij} +\partial_j \phi_i
  \right) \left( \partial_t + {\bf v} \cdot
    \bbox{\nabla}\right)\phi_j + \ldots, \\
  \partial_\alpha \tilde{u}_i & = & \partial_\alpha \phi_i + \partial_\alpha
  \phi_j \partial_j \phi_i + \ldots, \\
  \partial_\alpha \partial_\beta \tilde{u}_i & = &
  \partial_\alpha\partial_\beta\phi_i + \ldots. 
\end{eqnarray}

We are now in a position to transform the equation of motion,
Eq.~(\ref{eom}).  The first step is change from ${\bf u}$ field to the
$\tilde{\bf u}$ field.  This is done by multiplying by a smoothing
function and summing, using
\begin{equation}
  \tilde{\bf u}({\bf x},{\bf z},t) = a^{d_t}\sum_{{\bf x}'}
  D({\bf x} - {\bf x'}) {\bf u}({\bf x'},{\bf z},t),
\end{equation}
where $D({\bf x})$ is a delta-like function smoothed-out on the
scale of the lattice spacing $a$, i.e. $\int_{{\bf x}'} D({\bf
  x}'- {\bf x}) f({\bf x}') = f({\bf x})$ and
$D({\bf 0}) = 1$.

Carrying out this procedure, the gradient and time-derivative terms
are essentially unchanged, with ${\bf u} \rightarrow \tilde{\bf u}$ to
a good approximation.  The discreteness of the lattice sum is
important for the force term, however.  It becomes
\begin{eqnarray}
  a^{d_t}\sum_{{\bf x'}} && D({\bf x} - {\bf
    x'}) F^s_i[{\bf x'} + {\bf v} t + {\bf u}({\bf
    x'},{\bf z},t),z] 
  \nonumber \\
  & & \hspace{-0.25in} = \sum_{\bf Q} \int_{{\bf x'}}
  \!\!\!\! e^{i{\bf
      Q}_\perp\cdot{\bf x'}} D({\bf x}\! -\! {\bf
    x}'\!) F^s_i[{\bf x'}\! +\! {\bf v} t\! +\! {\bf u}({\bf
    x}',{\bf z},t),z] \nonumber \\
  & & \approx \sum_{\bf Q} e^{i{\bf
      Q}_\perp\cdot{\bf x}} F^s_i[{\bf x} + {\bf v} t +
  \tilde{\bf u}({\bf x},{\bf z},t),z].
\end{eqnarray}

Making the final transformation from $\tilde{\bf u}$ to $\bbox{\phi}$,
the force term becomes
\begin{equation}
  F_i[{\bf X},{\bf z};\bbox{\phi},t] = \sum_{\bf
    Q} e^{i{\bf 
      Q}\cdot \left( {\bf X} - {\bf v} t -
      \bbox{\phi}({\bf X},{\bf z},t)\right)} F^s_i[{\bf
    X},{\bf z}].
  \label{random_force_form}
\end{equation}

Putting this together with the gradient transformations,
Eq.~(\ref{eom}) becomes
\begin{eqnarray}
  \gamma_{ij} \partial_t \phi_j & & ({\bf X},{\bf z},t)  = 
  A_{i\alpha j} 
  \partial_\alpha \phi_j + 
  B_{i\alpha\beta j}\partial_\alpha\partial_\beta \phi_j
  \nonumber \\
  & & + 
  C_{i\alpha  
    j\beta k} \partial_\alpha \phi_j \partial_\beta \phi_k 
  + F_i[{\bf
    X},{\bf z};\bbox{\phi},t].
\label{eom2}
\end{eqnarray}
At this point, we can regard ${\bf X}$ as a dummy variable and
treat Eq.~\ref{eom2}\ as simply a continuum partial differential
equation.  The transformed gradient coefficients are
\begin{eqnarray}
  \gamma_{ij} & = & \tilde\gamma_{ij}, \\
  A_{i\alpha j} & = & \tilde{A}_{i\alpha j} - v^\alpha \tilde\gamma_{ij},
  \\
  B_{i\alpha\beta j} & = & \tilde{B}_{i\alpha\beta j}, \\
  C_{i\alpha j \beta k} & = & \tilde{C}_{i\alpha j\beta k} -
  \tilde\gamma_{ij}\delta_{\alpha k} v^\beta + \tilde{A}_{i\alpha k}
  \delta_{j\beta}. 
\end{eqnarray}

\section{Linearized theory}
\label{sec:linearized}

In the previous section we derived the proper form of the hydrodynamic
equations, Eqs.~\ref{maineq}, describing the low-energy,
long-wavelength properties of a periodic medium driven through either
a periodic or quenched random potential.  These equations are distinct
from their equilibrium counterpart in three respects. First, in
addition to the well-known convective term arising when transforming
from the crystal to the laboratory frame, the coupling of the driven
system to the external potential yields several other terms linear in
the gradients of the displacement field; these are collectively
described by non-zero coefficients $A_{i\alpha j}$. Second, the
equations contain nonequilibrium KPZ-type nonlinearities that can be
thought of as corrections to linear elasticity.  Third, in the random
case there is a {\it nonequilibrium} component of the static (i.e.,
${\bf \phi}$-independent) pinning force.  This force is a genuine
nonequilibrium effect, as it cannot be represented as the gradient of
a potential and vanishes in the absence of external drive.  

The remainder of this paper focuses on the random case, which is of
more immediate experimental relevance.  Although, as discussed in the
Introduction, a general analysis of Eqs.~\ref{maineq}, is beyond the
scope of this paper, we take this section to discuss the weak-disorder
limit, in which only the leading (in $\tilde\Gamma$) contributions to
each term are kept.  In this approximation the damping $\gamma_{ij} =
\gamma\delta_{ij}$ is diagonal, and the leading linear gradient term
is simply the convective derivative, $A_{i\alpha j} \approx
-\gamma\delta_{ij}v^\alpha$.  Similarly keeping only the bare elastic
matrix, we obtain the simplified set of nonequilibrium
elastic-hydrodynamic equations for the driven lattice,
\begin{eqnarray}
\label{modelh}
\gamma\big(\partial_t+ & &{\bf v}\cdot\bbox{\nabla}_t\big)\phi_i=
B^0_{i\alpha\beta j}\partial_\alpha\partial_\beta\phi_j\nonumber\\
 & & +F_i[{\bf r},\bbox{\phi},t]+\eta_i({\bf r},t),
\end{eqnarray}
where $B^0$ is the usual elastic matrix, $\eta_i$ represents
``thermal'' noise (which as we will see in Sec.\ref{sec:RGa} can
appear even at zero physical temperature), and the random force $F_i$
is given by Eqs.\ref{pinforce} and \ref{pftwo} and contains all
nonlinearities.  Note that we have kept both the equilibrium and
non-equilibrium components contained in $F_i$, since, although these
are of differing order in $\tilde\Gamma$, they cannot be regarded as
corrections to any zeroth order terms.  One important ingredient which
has been left out of Eq.~\ref{modelh}\ is the set of KPZ
non-linearities ($C_{i\alpha j\beta k}$ terms).  While these
coefficients are probably small, they may very well modify the
asymptotic long-distance behavior.  Based on previous work on somewhat
simpler (but instructive!)  models,\cite{chen,unpublished}\ we expect
that these effects will only {\sl increase} the distortions of the
moving lattice, and furthermore, that the increase in displacements
will be most pronounced in the longitudinal direction.

We therefore proceed with the analysis of Eq.~\ref{modelh}, expecting
that our results {\sl underestimate} the roughness in the moving
lattice, and therefore provide a necessary (but not sufficient)
condition for its stability.  To proceed, we note that, as pointed out
by Giamarchi and Le Doussal \cite{gl}, many of the terms in the random
force, Eq.~\ref{random_force_form}, are oscillatory in time.  Such
oscillatory terms average out at large sliding velocities, when the
``washboard'' frequencies $\omega_{\bf Q} = {\bf Q}\cdot{\bf v}$ are
large.  Even when they are not large, they generate only finite
renormalizations of the other parameters in the model.  We therefore
drop this oscillatory part of ${\bf F}$, keeping only the Fourier
components of the force orthogonal to the mean velocity,
\begin{equation}
F_i[{\bf r},\bbox{\phi}]\rightarrow
  \sum_{{\bf Q}\cdot{\bf v}=0}
  e^{i{\bf Q}\cdot({\bf x}-\bbox{\phi})}
  F^s_i({\bf r}),\label{fnosc}
\end{equation}
where we have dropped the explicit time dependence $t$ inside $F_i$,
since it has disappeared in this approximation.  Recall that the
$\phi$-independent force $F^s_i({\bf r})$ contains both
equilibrium and nonequilibrium contributions, with the corresponding
correlators given in Eqs.~\ref{force_decomposition}--\ref{gij_result}.

In this section we will further simplify the problem and linearize the
nonequilibrium hydrodynamic equations (\ref{modelh}).  We will then
discuss the predictions of the linearized theory for the decay of
translational and temporal order in the driven lattice.

We recall that we have used the labels longitudinal ($l$) and
transverse ($t$) to denote the directions along the oriented manifolds
and transverse to them, respectively. The $d$-dimensional position
vector was then written in terms of a transverse coordinate ${\bf x}$
and a longitudinal coordinate vector ${\bf z}$ as ${\bf r}=({\bf x,
z})$.  We now choose the $x$ direction (one of the $d_t$ ${\bf x}$
coordinates) along the direction of the driving force (${\bf v} =
v\hat{\bf x}$) and denote by ${\bf r}_\perp$ the $d-1$ directions
transverse to the external drive ($d=d_l+d_t$).  These $d-1$
directions can be further broken up into components transverse and
longitudinal to the oriented manifolds, ${\bf r}_\perp = ({\bf y,
z})$, with ${\bf y}$ a $d_t-1$-dimensional vectors denoting the
components of the transverse displacement that are also perpendicular
to the external driving force.  The $d_t$-dimensional transverse
coordinate space is then described by a 1-dimensional coordinate $x$
(not to be confused with $\bf x$) and $\bf y$, respectively parallel
and perpendicular to the direction of motion, with ${\bf x}=(x,{\bf
y})$. The $d$-dimensional position vector is written as ${\bf r}=({\bf
x, z})=(x,{\bf y, z})$.  We will also use the labels parallel
($\parallel$) and perpendicular ($\perp$) to denote the directions
parallel and perpendicular to the external drive.  For concreteness we
specialize to a lattice where the $d_l$ directions along the manifolds
that compose the lattice are isotropic and the longitudinal elastic
properties are described by a single elastic constant, denoted by
$c_{44}$.  We assume the $d_t$-dimensional lattice is described by
isotropic elasticity, with two elastic constants, a compressional
modulus $c_{11}$ and a shear modulus $c_{66}$.  Our model hydrodynamic
equations for the driven lattice are then given by,
\begin{eqnarray}
\label{eoms}
\gamma\Big(\partial_t+& & v\partial_x \Big)\phi_i=
 [c_{66} \nabla_t^2 +
 c_{44} \nabla_z^2]\phi_i\\ \nonumber 
 & & + c_{11}\partial_i \bbox{\nabla}_t\cdot\bbox{\phi} 
 +F_i[{\bf r},\bbox{\phi}],
\end{eqnarray}
where for the purposes of this section we have dropped thermal noise,
thereby ignoring the {\it subdominant} thermal fluctuations. These
equations describe the important physical case of the lattice of
magnetic flux lines in a three-dimensional superconductors ($d_l=1$,
$d_t=2$).  For $d_l=0$ and $d_t=2$ the equations describe the elastic
properties of driven vortex lattices in superconducting films, driven
magnetic bubble arrays, or driven Wigner crystals.

The displacement field can be split into components parallel and perpendicular
to the mean motion, $\bbox{\phi}=(\phi_x,\bbox{\phi}_{\bf y})$.  The
transverse displacement $\bbox{\phi}_{\bf y}$ is a $(d_t-1)$-dimensional
vector.  The pinning force is then seen to be independent of $\phi_x$: 
\begin{eqnarray} 
F_i[{\bf r},\bbox{\phi}_{\bf y}]= \sum_{{\bf Q}\cdot{\bf
v}=0}& & e^{i{\bf Q}\cdot({\bf x} -\bbox{\phi}_{\bf y})}\nonumber\\ & & \times
[F_i^{\rm eq}({\bf r})+F_i^{\rm neq}({\bf r})], 
\label{pfnot} 
\end{eqnarray} 
The independence of the static pinning force on $\phi_x$ is a
consequence of the precise time-translational invariance of the
system.  This requires that the equation of motion be unchanged upon
transforming $t \rightarrow t+\tau$, $\phi_x \rightarrow \phi_x + v
\tau$.  As argued above, all explicitly time dependent terms are
irrelevant at low frequencies, thereby implying independence of
$\phi_x$ in the same limit.  Eq.~\ref{eoms} is thus {\it linear} in
the longitudinal displacement $\phi_x$, which can therefore be treated
exactly.

The only remaining nonlinearity in the equation of motion,
Eq.\ref{eoms} is in $\bbox{\phi}_{\bf y}$, entering through the random
force $F_i[{\bf r},\bbox{\phi}_{\bf y}]$, which, in this section, we
treat perturbatively. We stress that the validity of this perturbative
calculation requires the displacements $\bbox{\phi}_{\bf y}$
transverse to the mean motion to be small, but places no constraints
on the size of the displacements along the direction of motion.  The
perturbation theory in $F_i[{\bf r},\bbox{\phi}_{\bf y}]$ thus gives a
complete description of the behavior of positional correlations in the
$x$ direction out to asymptotic length scales for the model of
Eq.~\ref{modelh}.  In contrast, the predictions of the perturbation
theory for the correlations of the {\em transverse} displacement
$\bbox{\phi}_{\bf y}$ presented below are only valid in the Larkin
regime; RG methods of Sections~\ref{sec:RGa},\ref{sec:RGb} must be
used to go beyond the Larkin length scale.

To lowest order in the random force, we simply neglect the
displacement field $\bbox{\phi}_{\bf y}$ in $F_i[{\bf
r},\bbox{\phi}_{\bf y}]$, Eq.~(\ref{pfnot}). The Fourier components of
the pinning force are then given by
\begin{equation}
  \label{staticzero}
  F_i^{(0)}({\bf q})=\sum_{{\bf Q}\cdot{\bf v}=0}
    F^s_i({\bf q}_t+{\bf Q},{\bf q}_z),
\end{equation}
with correlations 
\begin{eqnarray}
\label{staticorr}
\big[& & F^{(0)}_i({\bf q})F^{(0)}_j 
   ({\bf q'})\big]_{\rm ens.}\nonumber \\
 & & =(2\pi)^{d_l}\delta^{(d_l)}({\bf q}_z+{\bf q}_z')
      (2\pi)^{d_t}\delta^{(d_t)}({\bf q}_t+{\bf q}_t')
  \Delta_{ij},
\end{eqnarray}
and
\begin{eqnarray}
\Delta_{ij}& =& g_{ij} + \sum_{{\bf Q}\cdot{\bf v}=0} Q_{ i}Q_{ j}
      \tilde\Gamma(Q)\nonumber\\
&=& (\Delta+g_0)(\delta_{ij}-\delta_{ix}\delta_{jx})
    +g_1\delta_{ix}\delta_{jx}.
\end{eqnarray}
The disorder strength $\Delta$
is the variance of the equilibrium part of the static pinning force,
\begin{equation}
  \Delta={1\over d_t}\sum_{{\bf Q}\cdot{\bf v}=0} Q^2
  \tilde\Gamma(Q).
\label{delta_eq}
\end{equation}
The coefficients $g_0$ and $g_1$ 
are determined by the 
correlations $g_{ij}$ of the nonequilibrium part of the static pinning
force, according to
\begin{equation}
g_{ij}=g_0
    (\delta_{ij}-\delta_{ix}\delta_{jx})
    +g_1\delta_{ix}\delta_{jx}.
\end{equation}
They are evaluated in Appendix C, where it is shown that for large sliding
velocities and short-ranged pinning potential ($\xi\ll a$, with $\xi$ the
range of the pinning potential),
\begin{equation}
g_{0,1}\sim {\Delta^2\over 
   v^{(2d_t-d_l)/2}}.
\end{equation}

The components of the mean square displacement tensor are given by
\begin{eqnarray}
  \label{invert}
  B_{ij}({\bf r})=2\int_{\bf q}& &\int_{\omega}
  [1-\cos({\bf q}\cdot{\bf r})]\nonumber\\
  & &\times \big[\langle\phi_i({\bf q},\omega)\rangle
  \langle\phi^*_j({\bf q},\omega)\rangle\big]_{\rm ens.}.
\end{eqnarray}
The correlation functions of the displacement field are easily
calculated in Fourier space. The linear equation of motion for the
disorder induced displacement yields
\begin{eqnarray}
\langle\phi_i({\bf q},\omega)\rangle
=\Big[G_L({\bf q},\omega) & &P^L_{ij}({\bf q}_t)+
     G_T({\bf q},\omega)P^T_{ij}({\bf q}_t)\Big]\nonumber \\
    & & \times 2\pi\delta(\omega)F_i^{(0)}({\bf q}),
\end{eqnarray}
where $G_L({\bf q},\omega)$ and $G_T({\bf q},\omega)$ are the
longitudinal and transverse elastic propagators, respectively,
\begin{eqnarray}
& & G_L({\bf q},\omega)={1\over i\gamma(\omega-vq_x)+
     (c_{11}+c_{66})q_t^2+c_{44}q_z^2},\label{propl}\\
& & G_T({\bf q},\omega)={1\over i\gamma(\omega-vq_x)+
     c_{66}q_t^2+c_{44}q_z^2},
\label{propt}
\end{eqnarray}
and 
$P^L_{ij}({\bf q}_t)=\hat{q}_{t i}\hat{q}_{t j}$
and $P^T_{ij}({\bf q}_t)=\delta_{ij}-\hat{q}_{t i}\hat{q}_{t j}$
are the familiar longitudinal and transverse projection operators,
with $\hat{\bf q}_t={\bf q}_t/q_t$.
The correlation functions of the displacement field are given by
\begin{eqnarray}
  \label{uxx}
  \big[|\langle\phi_x({\bf q},\omega)\rangle|^2\big]_{\rm ens.}
     =& &2\pi\delta(\omega)\bigg[
  {g_1\over (\gamma vq_x)^2+[c_{66}q_t^2
    +c_{44}q_z^2]^2}\nonumber\\
  & &+{\cal O}\Big(\Delta{q_x^2\over q_y^2}|G_L|^2\Big)\bigg]\delta(0),
\end{eqnarray}
\begin{eqnarray}
  \label{uyy}
\hspace{-.5cm}\big[|\langle\bbox{\phi}_{\bf y}({\bf q},\omega)\rangle|^2\big]_{\rm
   ens.} =&
\hspace{-.5cm}  &2\pi\delta(\omega)\bigg[ 
  {\Delta+g_0\over (\gamma vq_x)^2+
    [(c_{11}+c_{66})q_t^2+c_{44}q_z^2]^2}\nonumber\\
  & & +{\cal O}\Big(g_1{q_x^2\over q_y^2}|G_T|^2\Big)\bigg]\delta(0),
\end{eqnarray}
where $\delta(0)\equiv (2\pi)^{d+1}\delta(\omega=0)\delta^d({\bf
q}={\bf 0})$.  Because the wavevector integral in Eq. \ref{invert} is
dominated by $q_{x}\sim (cq_y^2/\gamma v)$, terms containing
$(q_x^2/q_y^2)|G_T|^2$ or $(q_x^2/q_y^2)|G_L|^2$ yield less divergent
(generally bounded) contributions to the fluctuations compared to the
terms without this angular factor.  The behavior in real space is
displayed here for the case $d_l=0$, relevant to vortex lattices in
thin superconducting films. The corresponding expressions for $d_l
=1$, describing flux-line arrays in three-dimensional superconductors,
are given in Appendix D.  For $d_l=0$ we obtain,
\begin{equation}
  \label{scalingx}
  B_{xx}(x,{\bf y})\sim g_1{|y|^{3-d_t}\over\gamma vc_{66}} 
  {\cal F}_0^{(d_t)}\bigg({|x|c_{66}\over\gamma v y^2}\bigg),
\end{equation}
\begin{eqnarray}
  \label{scalingy}
  B_{\bf \perp\perp}(x,{\bf y}) & \sim & 
  (\Delta+g_0){|y|^{3-d_t}
    \over\gamma v(c_{11}+c_{66})} \nonumber \\
  & & \times {\cal F}_0^{(d_t)}\bigg({|x|(c_{11}+c_{66})\over\gamma v
    y^2}\bigg), 
\end{eqnarray}
where 
\begin{equation}
{\cal F}_0^{(d_t)}(s)=\int{{d^{d_t-1}{\bf u}}\over (2\pi)^{d_t -1}}
  {1\over u^2}\Big[1-\cos({\bf \hat{y}}\cdot{\bf u})e^{-su^2}\Big].
\end{equation}
The asymptotic behavior of the scaling function is given by
${\cal F}_0^{(d_t)}(0)={\rm constant}$ and 
${\cal F}_0^{(d_t)}(s)\sim s^{(3-d_t)/2}$ 
for $s\gg 1$.

Fluctuations in the direction of the driving force are dominated by
the {\it nonequilibrium} part of the random drag and by {\it shear}\ 
modes.  As discussed earlier, the corresponding power-law scaling of
$\phi_x$ holds out to arbitrary length scales. Since
\begin{eqnarray}
B_{xx}&\sim&|y|^{3-d_t}\;,\;\;\;\mbox {for}\;\; y^2 \gg (c_{66}/\gamma
v)|x|\;,\\
B_{xx}&\sim& |x|^{(3-d_t)/2}\;,\;\;\;\mbox {for}\;\; 
y^2 \ll (c_{66}/\gamma v)|x|\;,
\end{eqnarray}
longitudinal density correlations are short-ranged in $d_t<3$, with a
stretched-exponential decay.  Since, as argued above, this behavior
will persist even once the full nonlinear form for $F_i[{\bf
r},\bbox{\phi}_{\bf y}]$ is taken into account, the validity of the
elastic model itself is in doubt.  In particular, for the physical
case of $d_t = 2$, the spatial correlations are in fact exponentially
decaying, and it is natural to expect that this drives the unbinding
of dislocations with Burger's vectors parallel to the drive.  If this
is the case, both translational and temporal order for ${\bf Q}\cdot
{\bf v} \neq 0$ are short range in the driven steady state.

In contrast, both the equilibrium and nonequilibrium parts of the
static pinning force contribute to fluctuations in the transverse
direction, which are controlled by the {\it compressional} modes of
the system.  For large sliding velocity or weak disorder and
intermediate length scales, the equilibrium part of the static pinning
force will dominate as $g_0\sim(\Delta/v)^2 \ll\Delta$ in this limit.
The perturbation theory employed here breaks down, however, at lengths
larger than the Larkin lengths, $R_c^y$ and $R_c^x$ defined by
$B_{\perp\perp}(R_c^x,R_c^\perp)\sim\xi^2$, where $\xi$ is the range
of the pinning potential.  The Larkin volume is defined for
deformations of the lattice in the transverse direction
($\phi_y\sim\xi$) and it is anisotropic, with
\begin{eqnarray}
  & & R_c^y\sim\Big({\xi^2vc_{11}\over\Delta}\Big)^{1/(3-d_t)}\\
  & & R_c^x\sim v(R_c^y)^2/c_{11}.
\end{eqnarray}
It is elongated in the longitudinal direction at large sliding
velocities.  Longitudinal displacements ($\phi_x$) grow without bound
on {\it all} length scales in $d_t<3$ and a Larkin domain cannot be
defined in this case.  As usual, the nonlinearities in the static
pinning force must be incorporated non-perturbatively to describe the
asymptotic decay of the correlations beyond the Larkin length scales.
We will consider this problem in more detail in the following
sections.  A natural suspicion is, however, that at asymptotically
large distances at a finite temperature in two dimensions, the
nonequilibrium random drag, $g_0$, will dominate, destroying also the
transverse periodicity.

We stress that longitudinal fluctuations - that grow without bound -
are induced by shear modes of the lattice. This is consistent with the
intuition that shear modes (rather than compressional ones) play the
dominant role in melting a crystal. In contrast, transverse
fluctuations are controlled by compressional modes. Compressional
modes are present in both solids and liquids and are generally not
expected to generate the unbounded strains needed to yield dislocation
unbinding.
\begin{figure}[bth]
{\centering
\setlength{\unitlength}{1mm}
\begin{picture}(150,85)(0,0)
\put(-25,-65){\begin{picture}(150,60)(0,0) 
\includegraphics{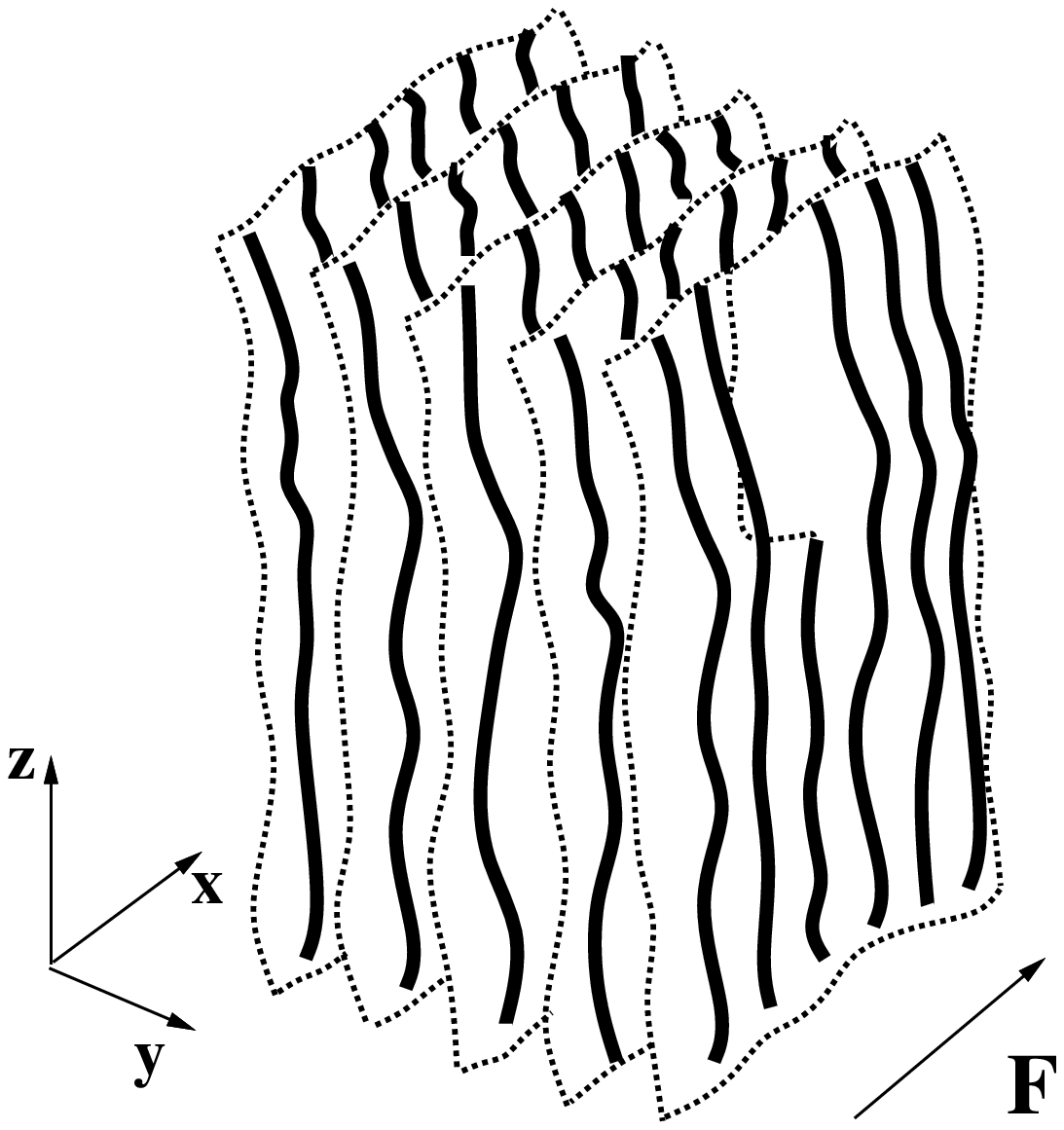}
\end{picture}}
\end{picture}}
Fig.4:{Schematic illustration of a driven line smectic, i.e. smectic
in three dimensions, with $d_t=2$, $d_l=1$.}
\label{line_smectic}
\end{figure}

\section{Driven Smectic Dynamics}
\label{sec:smectic}

It was shown in the previous section that disorder-induced
fluctuations parallel to the direction of the mean motion grow
algebraically in $d_t<3$.  This algebraic growth of fluctuations
yields short-range positional correlations along the direction of the
external drive, and it is our belief that this implies the breakdown
of the elastic description along the direction of motion ($x$).  While
we have not considered explicitly the role of dislocations here,
analysis of similar equations of motion {\sl including
non-linearities}\cite{chen,unpublished}\ suggests that this should
occur (for $d_t \leq 2$) via the unbinding of dislocations with
Burgers vectors {\sl parallel} to the driving force \cite{halp_ostl}.
This mechanism should convert the longitudinal translation and
temporal correlations to the exponential (or stretched exponential)
form typical of a liquid.  This is also in agreement with real space
images of driven two-dimensional vortex lattices ($d_t=2$, $d_l=0$)
obtained via numerical simulations.\cite{moon}\ Therefore for $d_t=2$
and $d_l=0$ the driven lattice can only retain periodicity at
reciprocal lattice vectors {\it transverse} to the direction of
motion.  At best, therefore, it consists of a stack of one-dimensional
liquid-like channels, sliding parallel to the direction of
motion\cite{us}\ and has the spatial symmetry of a smectic liquid
crystal. It is, however, important to stress that in real, {\it
equilibrium} smectic liquid crystals the underlying rotational
symmetry (which is broken {\em spontaneously}) enforces soft {\it
Laplacian} in-layer elasticity.\cite{GP} In contrast, the rotational
symmetry of the nonequilibrium ``driven smectic'' discussed here is
{\em explicitly} broken by the external drive. These systems are
therefore characterized by conventional {\it gradient} elasticity.

For the case $d_t=2$ and $d_l=1$, corresponding to magnetic flux-line
arrays in three-dimensional superconductors, the driven crystal is
potentially stable as translational correlations decay logarithmically
also in the direction of the applied force. We recall, however, that
the results obtained here by dropping the KPZ terms are expected to
strongly underestimate the growth of longitudinal fluctuations of a
moving {\it crystal}. Since in $d=3$ the transverse line smectic state
is equally stable it is possible to have a phase transition between
the transverse smectic (expected at intermediate velocities) and a
moving glass (stabilized at high velocities). The nonequilibrium
smectic state (illustrated in Fig.~4 then consists of liquid-like
sheets of flux lines lying in planes parallel to the $zx$ plane. These
sheets are periodically spaced in the $y$ direction, normal to the
external drive. Within each sheet, however, there is no positional
order of the flux lines and the correlations are liquid-like.

In a smectic liquid crystal the density field $\rho$ becomes an
independent variable, as it is no longer slaved to the displacement
field. The density is a conserved quantity and therefore a density
fluctuation relaxes at a rate that vanishes in the long-wavelength
limit, i.e., it is a hydrodynamic variable.  In addition, the system
still has broken translational symmetry in the direction perpendicular
to that of the mean motion.  {\sl Both} the Goldstone modes of the
broken translation symmetry (the displacement $\phi_{\bf y}$) and the
conserved density field ($\rho$) must be retained in a hydrodynamic
description.  In the remainder of this section we consider the case
$d_t=2$ and $d_l=0$ of a two-dimensional lattice driven over a
disordered substrate. The hydrodynamic variables of the driven smectic
are then a conserved density field $\rho$ and the one-dimensional
displacement vector $\phi_\perp({\bf x},z)\rightarrow \phi({\bf
x},z)$, describing displacements of the layers in the directions
normal to the external drive and to the layers themselves.  As we are
only interested in overdamped systems here, the momentum and energy
are not conserved and therefore need not be included explicitly in our
hydrodynamic description.

The continuum hydrodynamic free energy for the overdamped smectic
is given by
\begin{eqnarray}
\label{fsmec}
{\cal F}_s={1\over 2}\int_{{\bf x}}\Big\{& &c_L\Big({\delta\rho\over\rho_
0}\Big)
^2
  +c_{11}^y (\partial_y\phi)^2 \nonumber\\
& &  +K_1(\partial_x\phi)^2
  +2 K_2(\partial_y\phi){\delta\rho\over\rho_0}\Big\},
\end{eqnarray}
where $\delta\rho=\rho-\rho_0$, with $\rho_0$ the equilibrium density.
Here $c_L$ is the smectic bulk modulus, $c_{11}^y$ the in-layer
compressibility and $K_1$ the layer bending stiffness. The coupling
constant $K_2$ also has dimensions of an elastic constant.  The
hydrodynamic equations of the driven smectic contain additional
nonequilibrium terms, as compared to their equilibrium counterpart.
The nonequilibrium terms can be constructed by preserving the
invariance under inversions about the direction of the external drive
($y\rightarrow -y$, $\phi\rightarrow -\phi$) and the fact that
physical properties are invariant under the ``phase shift'' $\phi
\rightarrow \phi + a$.

Density conservation requires that $\rho$ satisfies a continuity
equation,
\begin{equation}
\label{continuity}
\partial_t\delta\rho+\bbox{\nabla}\cdot{\bf j}=0,
\end{equation}
where ${\bf j}$ is the number current density.
The equation for the layer displacement is given by
\begin{equation}
\label{smecphi}
(\partial_t+v\partial_x)\phi={j_y\over \rho}
  -{\Gamma_0\over\rho_0}\Big({\delta {\cal F}_s\over\delta\phi}
   -F_y[{\bf x},\phi]\Big),
\end{equation}
with $\Gamma_0$ a kinetic coefficient and $F_y$ the $y$-component of
the non-oscillating part of the pinning force given in Eq.~\ref{fnosc}.  
We neglect here the oscillatory contributions to the
pinning force that only give small corrections to the non-oscillatory
part. We also neglect other contributions to the pinning force that
couple to the density.  These would be important for a full systematic
RG treatment, which we do not attempt here.  A discussion of more general
equations for the driven smectic that incorporate these terms is left
for future work.  

To close the equations, a constitutive relation for
the current flux ${\bf j}$ is needed. This is given by
\begin{eqnarray}
& & j_x=v\delta\rho+v_1\delta\rho+\rho_0v_2\partial_y\phi
   -\rho_0\Gamma_1\partial_x{\delta {\cal F}_s\over\delta\rho},
 \label{currentx}\\
& & j_y=\rho_0v_3\partial_x\phi
   -\rho_0\Gamma_2\partial_y{\delta {\cal F}_s\over\delta\rho}.
\label{currenty}
\end{eqnarray}
The first term on the right hand side of Eq.~(\ref{currentx}) arises
from the transformation to the laboratory frame. The terms
proportional to the coefficients $v_1$, $v_2$ and $v_3$ are
nonequilibrium terms that can be generated upon coarse-graining, by
the method described in Section II for the the driven lattice.  Other
nonequilibrium terms that yield contributions that are higher order in
the gradients, and therefore subdominant, have been neglected here.
By inserting the constitutive equation for the current in
Eqs.~(\ref{continuity}) and (\ref{smecphi}), one obtains,
\begin{equation}
(\partial_t+\tilde{v}_1\partial_x)\delta\rho  =
   -\rho_0\tilde{v}_2\partial_x\partial_y\phi
   +[D_1\partial_x^2+D_2\partial_y^2]\delta\rho,  
\label{smdens}
\end{equation}
and
\begin{equation}
(\partial_t+ \tilde{v}_3\partial_x)\phi= 
    D_5\partial_y{\delta\rho\over\rho_0}
 +[D_3\partial_x^2+D_4\partial_y^2]\phi
   +{\Gamma_0\over\rho_0}F_y[{\bf x},\phi].
\label{smdis}
\end{equation}
Additional ``velocities'' $\tilde{v}_1$, $\tilde{v}_2$, $\tilde{v}_3$
have been defined as
\begin{mathletters}
\begin{eqnarray}
& & \tilde{v}_1=v+v_1,\\
& & \tilde{v}_2=v_2+v_3,\\
& & \tilde{v}_3=v-v_3
\end{eqnarray}
\end{mathletters}
The coefficients
$D_i$ have dimensions of diffusion constants and are given by
\begin{mathletters}
\begin{eqnarray}
& & D_1=\Gamma_1c_L/\rho_0,\\
& & D_2=\Gamma_2c_L/\rho_0,\\
& & D_3=\Gamma_0K_1/\rho_0,\\
& & D_4=(\Gamma_0c_{11}^y-\Gamma_2K_2)/\rho_0,\\
& & D_5=(\Gamma_0K_2-\Gamma_2c_L)/\rho_0.
\end{eqnarray}
\end{mathletters}

By solving the hydrodynamic equations in the long wavelength limit,
one finds that the decay of density and displacement fluctuations is
governed by two diffusive modes with eigenfrequencies,
\begin{eqnarray}
& &  \omega_\phi=\tilde{v}_3q_x-i\Big[D_3q_x^2+\Big(D_4-
   {\tilde{v}_2D_5\over \tilde{v}_1-\tilde{v}_3}\Big)q_y^2\Big],\\
& & \omega_\rho=\tilde{v}_1q_x+i\Big[D_1q_x^2+\Big(D_2+
   {\tilde{v}_2D_5\over \tilde{v}_1-\tilde{v}_3}\Big)q_y^2\Big].
\end{eqnarray}
For stability we must have $D_4-{\tilde{v}_2D_5\over
\tilde{v}_1-\tilde{v}_3}>0$ and $D_2+{\tilde{v}_2D_5\over
\tilde{v}_1-\tilde{v}_3}>0$.  The diffusion constants $D_1$ and $D_3$
are clearly positive defined.  The first mode describes
long-wavelength deformations of the layers and governs the decay of
displacement fluctuations. The second mode corresponds to the
permeation mode of the driven ``smectic'' and describes the transport
of mass across the layers that can occur in these systems without
destroying the layer periodicity. It is associated with density
fluctuations and it has important physical consequences for the
response of the driven smectic to an additional small driving force
applied normal to the layers (see Sec.\ref{res_corr_functions}).

An important physical quantity that can be measured in both simulations
and experiments is the structure factor of the driven periodic medium.
As the driven smectic has broken translational symmetry in the $y$
direction, normal to the layers and to the external drive, the density field
can be written as
\begin{equation}
\rho({\bf x})=\rho_L({\bf x})
  +\sum_{Q_{ y}} \rho_{Q_{ y}}({\bf x})e^{iQ_{ y}y},
\end{equation}
where $Q_{ y}=n2\pi/a$ are the reciprocal lattice vectors corresponding
to a layer spacing $a$, with $n$ an integer, $\rho_{Q_{ y}}$
are the corresponding Fourier components of the density,
and $\rho_L({\bf x})$ is the smooth (liquid-like) part of the density
field. The smectic structure factor is then given by
\begin{eqnarray}
S(& &{\bf q}_\perp)=S_L({\bf q}_\perp)\nonumber\\
& & +\sum_{Q_{ y}}\int_{{\bf x}}e^{-iq_xx}e^{-i(q_y-Q_{ y})y}
     \langle\rho_{Q_{ y}}({\bf x})\rho_{-Q_{ y}}({\bf 0})\rangle,
\end{eqnarray}
with $S_L({\bf q}_\perp)=\langle|\rho_L({\bf q}_\perp)|^2\rangle$.
The smectic structure function should therefore consists of a broad
liquid-like background $S_L({\bf q}_\perp)$, with superimposed
peaks at the reciprocal lattice vectors $Q_{ y}$, normal to the
direction of mean motion (see Fig.~2(b)).  As discussed in
the Introduction, in a Gaussian theory the correlator of the order
parameters $\rho_{Q_{ y}}({\bf x})$ can be written in terms of the
mean square displacement, according to
\begin{equation}
\langle\rho_{Q_{ y}}({\bf x})\rho^*_{Q_{ y}}({\bf 0})\rangle 
  \approx\rho_1^2e^{-Q_{ y}^2\langle[\phi({\bf x})
        -\phi({\bf 0})]^2\rangle}.
\end{equation}
The disorder-induced transverse mean square displacement is easily
calculated from Eqs.~\ref{smdens}\ and \ref{smdis}\ by treating the
the random force $F_y[{\bf x},\phi]$ as a perturbation, in an analysis
similar to that of Sec.\ref{sec:linearized}.  Conceptually simple but
algebraically tedious calculations show that including the coupling to
the density does not change the resulting decay of the correlation
function. For simplicity we therefore neglect this coupling in
Eq.~(\ref{smdis}) (i.e., let $\delta\rho=0$) and obtain
\begin{equation}
\big[\langle[\phi({\bf x})-\phi({\bf 0})]^2\rangle\big]_{\rm ens.}=
     {(\Delta+g_0)\Gamma_0^2\over 
2\pi\rho_0^2D_4\tilde{v}_3}
     y{\cal F}_0^{(2)}\Big({|x|D_4\over\tilde{v}_3y^2}\Big).
\end{equation}
The scaling function ${\cal F}_0^{(2)}(s)$ is identical to that
obtained in Eq.~(\ref{scalingx}) and it has the asymptotic behavior
${\cal F}_0^{(2)}(0)={\rm constant}$ and ${\cal F}_0^{(2)}(s)\sim
\sqrt{s}$ for $s\gg 1$.  The perturbation theory therefore predicts
that the smectic Bragg peaks at $q_y=Q_{ y}$ decay exponentially with
the system size. In other words, disorder would destroy the transverse
periodicity of the smectic.  In fact, we will see in the following
sections that this result continues to hold non-perturbatively.  The
power-law scaling of transverse Bragg peaks obtained in
simulations of two-dimensional driven vortex lattices\cite{moon}\  is
therefore most likely an artifact of small systems and weak disorder,
and would crossover to a disordered form at longer distances.  We will
return to this point later in Sec.~VII.

\section{Renormalization Group for ``toy'' Smectic}
\label{sec:RGa}

\subsection{Model and MSR formulation}

In this section, we will consider a simplified model for the smectic
phase, in which the hydrodynamic fluctuations of the conserved density
are neglected.  This ``toy'' smectic is then modeled simply by
dropping Eq.~\ref{smdens}\ and setting $\delta\rho = {\rm const.}$,
leaving the single equation of motion
\begin{equation}
  \gamma (\partial_t + v \partial_x)\phi = K_\parallel \partial_x^2 \phi
  + K_\perp \nabla_\perp^2 \phi + F(\phi, {\bf r}) + \eta({\bf r},t),
  \label{toy_smectic}
\end{equation}
Here we have pulled out a factor of $\gamma = \rho_0/\Gamma_0$, let
$K_\parallel \equiv D_3 \rho_0/\Gamma_0$, $K_\perp \equiv D_4
\rho_0/\Gamma_0$, and $F_\perp \rightarrow F$.  For simplicity we assume
longitudinal and transverse elasticity are governed by the
same elastic constant $K_\perp$. We have also added the
random time-dependent ``thermal'' noise $\eta$, satisfying
\begin{equation}
  \langle\eta({\bf r},t)\eta({\bf r'},t')\rangle = 2\gamma T
  \delta({\bf r-r'})\delta(t-t').
\label{thermal_noise}
\end{equation} 
The random force $F$ is characterized by the correlator
\begin{equation}
  \left[ F(\phi,{\bf r}) F(\phi',{\bf r}') \right]_{\rm ens.} =
  \Delta(\phi-\phi') \delta({\bf r - r'}).
\end{equation}
The function $\Delta(\phi)$ is periodic with the smectic lattice
spacing, which we take to be $a = 2\pi$.  It includes both the
equilibrium and non-equilibrium components.  The latter can be viewed
as simply an overall constant contribution to $\Delta(\phi)$.  In
addition to discounting density fluctuations, Eq.~\ref{toy_smectic}\ 
also neglects an allowed KPZ non-linear term of the form
\begin{equation}
  F_{\rm KPZ} = C \partial_x \phi \partial_y \phi,
\end{equation}
which should be added to the right hand side of the equation of
motion.  Simple power-counting shows, however, that, in contrast to a
driven {\it lattice} this term is strongly {\it irrelevant} in a
transverse smectic (see below).

To analyze Eq.~(\ref{toy_smectic}), we use the method of MSR to
transform the stochastic equation of motion into a field theory,
similar to what was already done in the Sec.\ref{mode_elimination} to
perform the single-step coarse-graining.  In this case, the MSR
``partition function'' is
\begin{equation}
  Z = \int [d\hat\phi][d\phi] e^{-S},
\end{equation}
where $S=S_0 + S_1$, with
\begin{equation}
  S_0 = \int_{{\bf r}t} \left\{ \hat\phi_{{\bf r}t} \left[ \gamma(\partial_t +
    v\partial_x) - K_\parallel \partial_x^2 - K_\perp \nabla_\perp^2
  \right] \phi_{{\bf r}t} - \gamma T \hat\phi_{{\bf r}t}^2 \right\},
\end{equation}
and the interaction term
\begin{equation}
  S_1 = -{1 \over 2}\int_{{\bf r},tt'} \hat\phi_{{\bf
      r}t}\hat\phi_{{\bf r}t'} \Delta(\phi_{{\bf r}t} - \phi_{{\bf r}t'}).
\end{equation}
By construction, $Z=1$; non-trivial correlation and response functions
are obtained, however, by inserting appropriate $\hat\phi$ and $\phi$
operators into the functional integrand.

\subsection{Power counting}

Let us first consider under what conditions the random force is a
{\sl relevant} perturbation in the sense of the renormalization group
(RG) using simple power-counting.  To do so, we rescale the
coordinates and fields by a scale factor $b>1$:
\begin{eqnarray}
  {\bf x}_\perp & \rightarrow & b {\bf x}_\perp, \\
  x & \rightarrow & b^{\zeta} x, \\
  t & \rightarrow & b^{z} t, \\
  \hat\phi & \rightarrow & b^{\hat{\chi}} \hat\phi.
\label{rescale_frg}
\end{eqnarray}
Note that, anticipating a periodic fixed point, we will not rescale
$\phi$.  To fix the exponents $\zeta$ and $\hat{\chi}$, we choose to
keep the terms $\gamma v$ and $K_\perp$ fixed in the quadratic action
$S_0$.  This implies $d-3+z+\zeta+\hat{\chi}=0$ and
$d-1+z+\hat{\chi}=0$, or
\begin{eqnarray}
  \zeta & = & 2, \\
  \hat{\chi} & = & -d+1-z.
\end{eqnarray}
Note that because $\phi$ was kept invariant, the temperature $T$
necessarily rescales
\begin{equation}
  T \rightarrow b^{1-d} T. \label{Trescale}
\end{equation}
Clearly this is a somewhat artificial choice for $\Delta=0$, since the
theory retains no memory of the periodicity of $\phi$.  It is rather
natural, however, for $\Delta \neq 0$, and will be returned to below.
The exponent $z$ is more subtle.  Naively, it should be determined by
the condition that $\gamma$ be invariant under rescaling.  Neglecting
the random force, we then obtain
\begin{equation}
  z_{\rm naive} = 2.
\end{equation}
We will see that even for small non-zero $\Delta$, this is actually
very far from correct.

Fortunately, the rescaling of $\Delta(\phi)$ and 
$C$ is in fact independent of
$z$.  The KPZ non-linearity is strongly irrelevant:
\begin{equation}
  C \rightarrow C/b \rightarrow 0,
\end{equation}
while the disorder correlator obeys
\begin{equation}
  \Delta(\phi) \rightarrow b^{3-d} \Delta(\phi).
\end{equation}
We therefore see that for $d > 3$, the random force is irrelevant, and
at long length scales has a negligible effect on the moving smectic.
For $d \leq 3$, however, power-counting is insufficient to determine
the fate of the system.

\subsection{Zero temperature RG}
\label{RGzeroT}

To proceed, we will perform a systematic RG, working perturbatively in
$S_1$, expanding it from the exponential and integrating out ``fast''
modes with large {\sl transverse} momenta.  This is simplest in the
scheme with neither a frequency $\omega$ nor longitudinal momentum
$q_x$ cut-off.  The fields are first decomposed into their slow ($<$)
components with $q_\perp < \Lambda/b$ and fast ($>$) components with
$\Lambda/b < q_\perp < \Lambda$, where $\Lambda$ is a hard transverse
momentum-space cut-off.  That is,
\begin{equation}
  \phi = \phi_< + \phi_>, \qquad \hat\phi = \hat\phi_< + \hat\phi_>
\end{equation}
Then the partition function is
\begin{equation}
  Z = \int [d\hat\phi_<][d\phi_<] e^{-S_0[\phi_<,\hat\phi_<]} \left\langle
    e^{-S_1[\phi_<+\phi_>,\hat\phi_< + \hat\phi_>]} \right\rangle_{>},
\end{equation}
where the brackets denote an average over the fast fields with respect
to the quadratic action $S_0[\phi_>,\hat\phi_>]$.  Relabeling the
surviving slow fields $\phi_< \rightarrow \phi$, $\hat\phi_<
\rightarrow \hat\phi$, the renormalization of the effective action due
to the mode elimination is given by
\begin{equation}
  e^{-S_{\rm eff.}[\phi,\hat\phi]} = e^{-S_0[\phi,\hat\phi]}
  \left\langle e^{-S_1[\phi+\phi_>,\hat\phi + \hat\phi_>]} \right\rangle_>.
\end{equation}

A cumulant expansion gives
\begin{equation}
  \left\langle e^{-S_1} \right\rangle_> = \exp \left\{ -\langle S_1
    \rangle_> + {1 \over 2} \langle S_1^2 \rangle_{>,c} + \cdots \right\},
\end{equation}
where the subscript $c$ indicate the cumulant (connected) correlator.
$ $From Eq.~\ref{Trescale}, it is natural to suspect that much of the
interesting physics is dominated by small (renormalized) temperatures.
With this in mind, we first consider the RG in the extreme $T=0$ case.
Then the only non-vanishing expectation values correspond to {\sl
response} functions:
\begin{eqnarray}
  && G_>({\bf r},t)  =  \langle \phi({\bf r'+r},t'+t) \hat\phi({\bf
    r'},t')\rangle_> \\
  && =  \int_{\Lambda/b}^\Lambda {{d^d{\bf q}_\perp} \over {(2\pi)^{d-1}}} \int
  {{dq_x} \over {2\pi}} \int {{d\omega} \over {2\pi}} {{e^{-i{\bf
          q\cdot r} + i\omega t}} \over {i\gamma(\omega - vq_x) +
      K_\parallel q_x^2 + K_\perp q_\perp^2}}.\nonumber 
\end{eqnarray}
It is instructive to perform some of the integrations above and bring
out the dependence on $x$ and $t$.  We are primarily interested in the
limit $K_\parallel \rightarrow 0$, since $K_\parallel$ is obviously
less relevant than $\gamma v$, which has one less $x$ derivative.  In
that limit these integrals are trivial, and we obtain
\begin{equation}
  G_>({\bf r},t) = \int_{\Lambda/b}^\Lambda {{d^{d-1}{\bf q}_\perp}
    \over {(2\pi)^{d-1}}}{1 \over \gamma} e^{-i{\bf q_\perp\cdot r}}
  e^{-(K_\perp/\gamma) q_\perp^2 t} \theta(t) \delta(x-vt).
\end{equation}
From this we see that the response function is causal and represents
unidirectional propagation along the positive $x$ axis.  Including a
non-zero $K_\parallel$ simply spreads out the $\delta$-function over a
distance $\delta x \propto \sqrt{K_\parallel t}$.  The irrelevance of
$K_\parallel$ is indicated by the smallness of this width relative to
the distance $vt$ propagated in the large $t$ limit.

Working first to $O(\Delta)$, consider the term
\begin{eqnarray}
& & \langle S_1 \rangle_>  -  S_1[\phi,\hat\phi] \nonumber \\
  & \approx & \int_{{\bf r}tt'} \hat\phi_{{\bf r}t}
  [\Delta'(0) + \Delta''(0)(t-t')\partial_t\phi_{{\bf r}t}]
  G_>(0,t-t').
\end{eqnarray}
By symmetry, $\Delta'(0)=0$, and we need to evaluate
\begin{eqnarray}
  \int_\tau \tau G_>({\bf 0},\tau)  & = & \int^>_{\bf q_\perp}
  \int_{\omega\tau} {{\tau e^{i\omega\tau}} \over {i\gamma(\omega-vq_x)
      + K_\parallel q_x^2 + K_\perp q_\perp^2}} \nonumber \\
  & = & {{2\gamma K_\parallel\Lambda^{d-1} C_{d-1} } \over {(\gamma^2
      v^2 + 4K_\parallel K_\perp \Lambda^2)^{3/2}}}dl \nonumber \\
  & = & {\overline{\kappa}\over 1+2\overline{\kappa}}
  {\Lambda^{d-3}C_{d-1}\over v K_\perp} dl,
\end{eqnarray}
where we defined
\begin{equation}
  \overline{\kappa}\equiv {2K_{\parallel} K_\perp\over \gamma^2
    v^2}\Lambda^2\;, \label{tildeKappa}
\end{equation}
chosen the infinitesimal rescaling factor $b=e^{dl}$, and defined
$C_d=S_d/(2\pi)^d$ in terms of the surface area of a d-dimensional
sphere $S_d=2\pi^{d/2}/\Gamma(d/2)$.  Inserting this result above
gives a renormalization of the friction drag coefficient (the inverse
mobility):
\begin{equation}
  \left. \partial_l \gamma \right|_{O(\Delta)} = - 
   {\overline{\kappa}\over 1+2\overline{\kappa}}
  {\Lambda^{d-3}C_{d-1}\over v K_\perp}\Delta''(0).
\end{equation}
Note that we have obtained no renormalization of the spatial gradient
terms.  This is actually a general consequence of taking an ultra-local
(i.e. $\delta$-function correlated) random force.  For such a force,
it is straightforward to show that the static response function
\begin{equation}
  G({\bf q},\omega=0) = {1 \over {-i\gamma v q_x + K_\parallel q_x^2 +
      K_\perp q_\perp^2}}
\end{equation}
is unrenormalized, i.e. exact even when $F$ is included in the
equation of motion.  Thus $\gamma v$, $K_\parallel$, and $K_\perp$
suffer no diagrammatic corrections at any order.

The next step is to examine the renormalization of $\Delta$.  The
first corrections (for $T=0$) arise at $O(\Delta^2)$.  To determine
these, we must compute the next term in the cumulant expansion.  This
is
\end{multicols}
\begin{equation}
  \delta S_2 = - {1 \over 2}\int_{\stackrel{{\bf r}_1 t_1 t'_1}{{\bf
        r}_2 t_2 t'_2}} \left\langle \hat\phi_{{\bf r}_1 t_1}
      \hat\phi_{{\bf r}_1 t'_1} \Delta(\phi_{{\bf r}_1 t_1} -
      \phi_{{\bf r}_1 t'_1}) \hat\phi_{{\bf r}_2 t_2}
      \hat\phi_{{\bf r}_2 t'_2} \Delta(\phi_{{\bf r}_2 t_2} -
      \phi_{{\bf r}_2 t'_2}) \right\rangle_c^> .
\end{equation}

At this point we are aided by a simplifying feature that actually
makes this part of the calculation easier than the equilibrium one.
Since we have a propagating mode along the $x$ axis, in the limit of
zero longitudinal damping, $K_\parallel = 0$, the response function
vanishes unless $x>0$.  This means that two response fields at
different $x$ points cannot be contracted, since this would give the
product $G({\bf x-x'},\cdot)G({\bf x'-x},\cdot)$, which vanishes for
{\sl any} $x$.  Instead, a non-vanishing contribution obtains only
when both $\hat\phi$ fields are taken from the same term.  This gives
\begin{eqnarray}
  \delta S_2 & = & -{1\over 2} \int_{\stackrel{{\bf r}_1 t_1
      t'_1}{{\bf r}_2 t_2 t'_2}} \hat\phi_{{\bf r}_1 t_1}
  \hat\phi_{{\bf r}_1 t'_1} \Delta(\phi_{{\bf r}_2 t_2} \! -\! 
  \phi_{{\bf r}_2 t'_2}) \Delta''(\phi_{{\bf r}_1 t_1} \! -\! 
  \phi_{{\bf r}_1 t'_1})  \nonumber \\
  & & \times \left[ G_>({\bf r}_1 \! -\!  {\bf r}_2,t_1\! -\!
    t_2)G_>({\bf r}_1 \! -\! {\bf r}_2,t_1\! -\! t'_2) \! -\!
    G_>({\bf r}_1 \! -\!  {\bf r}_2,t_1\! -\! t_2) 
    G_>({\bf r}_1 \! -\!  {\bf r}_2,t'_1\! -\! t'_2)\right].
\end{eqnarray}

Each term contains two response functions, which constrain their 
arguments to be small.  To leading (zeroth) order in gradients of 
$\phi$, we may approximate $\phi_{{\bf r}_{2} t_{2}} \approx 
\phi_{{\bf r}_{1},t_{1}}$ and $\phi_{{\bf r}_{2} t'_{2}} \approx 
\phi_{{\bf r}_{1}t_{1}}$ in the first term, while in the second 
$\phi_{{\bf r}_{2} t'_{2}} \approx \phi_{{\bf r}_{1}t'_{1}}$.  This 
leads to the simpler formula
\begin{equation}
  \delta S_{2} = - {I \over 2} \int_{{\bf r}_{1}t_{1}t'_{1}} 
  \hat\phi_{{\bf r}_{1}t_{1}}\hat\phi_{{\bf r}_{1}t'_{1}} 
  \Delta''(\phi_{{\bf r}_{1}t_{1}} - \phi_{{\bf r}_{1}t'_{1}})\left[ 
    \Delta(0) - \Delta(\phi_{{\bf r}_{1}t_{1}} - \phi_{{\bf 
        r}_{1}t'_{1}}) \right],
\end{equation}
where the same integral $I$ occurs in both terms.  In the limit 
$K_\parallel \rightarrow 0$, it becomes
\begin{equation}
  I = \int_{{\bf q}_{\perp}}^{>} \int_{q_{x}} G_{>}({\bf q},\omega=0) 
  G_{>}(-{\bf q},\omega=0) \approx {\Lambda^{d-3}C_{d-1} \over {2\gamma v 
      K_\perp}} dl.
\end{equation}
\begin{multicols}{2}
This gives the second order contribution to the renormalization of 
$\Delta(\phi)$:
\begin{equation}
  \left.\partial_{l}\Delta(\phi)\right|_{O(\Delta^{2})} = 
  -{{\Lambda^{d-3}C_{d-1}} \over {2\gamma v K_\perp}} 
  \Delta''(\phi)\left[\Delta(\phi) - \Delta(0)\right].
\end{equation}

At $T=0$, these are all the necessary mode-elimination contributions.
Combining these results with the scale-changes, we thus arrive at the
zero temperature RG equations
\begin{eqnarray}
  \partial_{l}\Delta(\phi) & = & (3-d)\Delta(\phi) \nonumber \\
  & & -{{\Lambda^{d-3}C_{d-1}} \over {2\gamma v K_\perp}} 
  \Delta''(\phi)\left[\Delta(\phi) - \Delta(0)\right], \\
  \partial_{l}\gamma & = & \left[2-z - 
  {\overline{\kappa}\over 1+2\overline{\kappa}}
  {\Lambda^{d-3}C_{d-1}\over\gamma v K_\perp}\Delta''(0)\right]\gamma,
  \label{gamma_RG} \\
  \partial_{l}K_\parallel & = & -2 K_\parallel, \label{KparallelRG}\\
  \partial_{l}\overline{\kappa} & = & -2 \overline{\kappa}, \label{kappaRG}
\end{eqnarray}
where the flow for the dimensionless parameter $\overline{\kappa}$ is
exact and independent of the choice of the rescaling exponents $z$,
$\zeta$ and $\hat\chi$.
\subsubsection{Behavior in $d=3-\epsilon$ dimensions}

As a first attempt at an analysis of these results, we consider the
problem in $d=3-\epsilon$ dimensions.  Defining the dimensionless
random-force correlator
\begin{equation}
  \Gamma(\phi) \equiv {{\Lambda^{d-3}C_{d-1}} \over {\gamma v
      K_\perp}} \Delta(\phi)
\label{Gamma_define}
\end{equation}
the RG flow equation for this dimensionless random force variance is
given by
\begin{equation}
  \partial_l \Gamma(\phi) = (3-d)\Gamma(\phi) - {1 \over 2}
  \Gamma''(\phi)\left[\Gamma(\phi) -
    \Gamma(0)\right]. \label{zero_temperature_RG}
\end{equation}
For $\epsilon \ll 1$, we thus expect to find a fixed point with
$\Gamma = O(\epsilon)$.  However, evaluating
Eq.~\ref{zero_temperature_RG} at $\phi=0$ demonstrates that (in
contrast to equilibrium case, where there was an additional
stabilizing $-\Gamma'(\phi)^2/2$ term) this is only possible if
$\Gamma(0)^*=0$. Like other functional RG equations at $T=0$,
Eq.~\ref{zero_temperature_RG} leads to a non-analytic force-force
correlator.  This can be seen directly by differentiating twice above
and evaluating at the origin:
\begin{equation}
  \partial_l \Gamma''(0) = (3-d)\Gamma''(0) - {1 \over 2} \left[
    \Gamma''(0) \right]^2.
  \label{curvatureRG}
\end{equation}
Since $\Gamma''(0;l=0) < 0$, this equation leads directly to a
divergence.  In fact, all ``fixed points'' (see below for an
explanation of the quotation marks here) have a slope
discontinuity at the origin.  A little analysis demonstrates that the
large $l$ behavior of $\Gamma$ is
\begin{equation}
  \Gamma(\phi;l) \sim \epsilon \min_n \left\{ (\phi - \pi - 2\pi n)^2 -
    {\pi^2 \over 3} \right\} + \Gamma_0(l), 
  \label{zeroTsolution}
\end{equation}
where $\Gamma_0$ is the zero Fourier component of $\Gamma(\phi)$ and
satisfies
\begin{equation}
  \partial_l \Gamma_0 = \epsilon \Gamma_0 + {{2\pi^2} \over
    3}\epsilon^2 
  \label{static_RG}
\end{equation}
Note that although $\Gamma_0 = -2\pi^2\epsilon/3$ provides a formal
fixed point solution, it is unphysical: for this value, $\Gamma(0) =
0$, which would imply the (positive semidefinite) variance of
$F(\phi,{\bf r})$ vanishes, and hence that $F$ is uniformly zero,
clearly in contradiction with $\Gamma(\phi) \neq 0$.  For any physical
situation $\Gamma_0$ will be {\sl larger}.  For instance, equilibrium
initial conditions require that the force arise as a derivative of a
random potential, and hence that the integral over $\Delta(\phi)$
vanish.  For the solution above, this gives $\Gamma_0 = 0 >
-2\pi^2\epsilon/3$.  Indeed, $\Gamma_0$ can be identified as the
variance of the {\sl static} force calculated using the single-step
coarse graining in Sec.\ref{mode_elimination}.

For physical situations, it is clear from
Eq.~(\ref{static_RG}) that $\Gamma_0$ is a strongly {\sl relevant}
variable which will flow off under the RG.  Luckily, as pointed out by
Narayan and Fisher in their study of CDW depinning\cite{NF}, this does not
really present a problem.  It can be easily shown that the static
random force does not effect the dynamics, by the same argument used
above to demonstrate the exact static response function (i.e. shifting
it away and showing that the distribution of the shifted force is
unmodified).  This is also essentially the same argument used in the
Cardy-Ostlund problem, in which there is also a runaway random force
(which is, in that case however, the gradient of a potential).

The non-analyticity of $\Gamma$ has important consequences for the
dynamics.  Indeed inspection of Eq.~\ref{gamma_RG} shows that it is
problematic: the quantity $\Delta''(0)$ is (minus) infinity at
the fixed point (and indeed becomes infinite at a finite length
scale).  Physically, the non-analyticity of $\Gamma$ is related to the
existence of multiple metastable minima in the effective potential on
the scale $l$ and a concurrent ``sharpness'' of this potential.  At
zero temperature, this sharpness leads to trapping of the phase $\phi$
and hence to a breakdown of the simple assumption of analyticity in
the coarse-grained dynamics.  A detailed analysis\cite{unpublished}\
shows that the divergence of $\Delta''(0)$ can be interpreted as the
signature of a multi-valued force reminiscent of static friction, and
corresponds to the existence of a non-zero transverse critical force
to ``depin'' the smectic\cite{density_note}.  

\subsection{Finite temperature RG}
\label{finite_T_RG}

A simpler means of controlling these singularities is to include
non-zero thermal fluctuations, which act to locally average the
effective potential and effect thermally activated motion between
different metastable states on long time scales.  In hopes of
obtaining a workable dynamics, we are thus led to consider the effects
of a non-zero temperature.  At the same time we must also ask if
temperature is a relevant or irrelevant perturbation around the $T=0$
fixed point considered here.  Naively, from Eq.~\ref{Trescale}, we
would expect $T$ be strongly irrelevant for $d$ near three.  In
equilibrium, the corresponding (power-counting there gives $2-d$)
result is exact; again due to Galilean invariance and the FDT, there
are no diagrammatic renormalizations to $T$\cite{vgdynamics_note}.

Because of the lack of FDT, we will, surprisingly, be led to a
completely different conclusion!    Consider the mode elimination at
a non-zero temperature.  The correlation
function, 
\end{multicols}
\begin{equation}
  C_>({\bf r},t) = \langle \phi_{{\bf r+r'},t+t'} \phi_{{\bf r'} t'}
  \rangle_> = \int^>_{{\bf q}\omega} {{2\gamma T e^{-i{\bf q\cdot r} +
        i \omega t}} \over {\gamma^2(\omega - vq_x)^2 + (K_\parallel
      q_x^2 + K_\perp q_\perp^2)^2}}
\end{equation}
is now non-zero.  This now makes it possible to contract two $\phi$
fields when renormalizing the MSR functional.  It is sufficient to
work only to linear order in $\Delta$ (or $\Gamma$).  Following the
same method as earlier, we find
\begin{equation}
  \delta S_1^{(T)} = \langle S_1 \rangle_>^{(T)} = - {1
    \over 2} \int_{{\bf r}tt'} \! \hat\phi_{{\bf r}t} 
  \hat\phi_{{\bf r}t'} \Delta''(\phi_{{\bf r}t} - \phi_{{\bf r}t'})
  \left[ C_>({\bf 0},0) - C_>({\bf 0},t-t')\right].
  \label{thermal_temp}
\end{equation}
This gives two terms.  The first contributes a renormalization of
$\Delta$.  That correlation function gives
\begin{equation}
  C_>({\bf 0},0) = \int_{{\bf q}_\perp}^> \int_{\omega q_x}
  {{2T\gamma} \over {\gamma^2(\omega - vq_x)^2 + (K_\parallel
      q_x^2 + K_\perp q_\perp^2)^2}} 
   = {T \over {2\sqrt{K_\parallel K_\perp}}}
  {\Lambda^{d-2} C_{d-1}} dl.
\end{equation}
\begin{multicols}{2}
Note that the integral is singular in the $K_\parallel \rightarrow
0$ limit.  Physically, this is because some damping is needed to
control the thermal fluctuations excited in the propagating
longitudinal mode.  Incorporating this piece in the RG equation for
$\Gamma$ gives
\begin{equation}
  \partial_l \Gamma(\phi) = \left[(3-d) + \overline{T} {\partial^2
      \over {\partial\phi^2}} \right] \Gamma(\phi) - {1 \over 2}
      \Gamma''(\phi)\left[\Gamma(\phi) -
        \Gamma(0)\right], \label{finite_temperature_RG}
\end{equation}
where
\begin{equation}
  \overline{T} \equiv {\Lambda^{d-2} C_{d-1}} {T \over
    {2\sqrt{K_\perp K_\parallel}}}.
\label{Tbar_define} 
\end{equation}

The second term in Eq.~\ref{thermal_temp}\ gives a correction to the
temperature.  This is simple to see, since in this term $|t-t'|$ is
kept small, so that $\phi_{{\bf r}t'} \approx \phi_{{\bf r}t}$ (and
likewise for $\hat\phi$).  This gives
\begin{equation}
  \delta S = {1 \over 2} \int_{{\bf r}t} \left(\hat\phi_{{\bf
        r}t}\right)^2 \Delta''(0) \int_{t'} C_>({\bf 0},t-t').
\end{equation}
This integral is finite in the limit $K_\parallel \rightarrow 0$, giving
\begin{eqnarray}
  \int_{t'} C_>({\bf 0},t-t') & = & \int^>_{{\bf q}_\perp} \int_{q_x}
  {{2T\gamma} \over {\gamma^2 v^2 q_x^2 + K_\perp^2 q_\perp^4}}
  \nonumber \\
  & & = {T
    \over {K_\perp v}} \Lambda^{d-3} C_{d-1}dl.
\end{eqnarray}
This is a renormalization of $\gamma T$.  Taking into account the
renormalization of $\gamma$ obtained earlier (Eq.~\ref{gamma_RG}),
one finds
\begin{equation}
\partial_l T = \left[ 1-d - \left( {1 \over 2} - 
{\overline{\kappa}\over 1+ 2\overline{\kappa}}\right)
  \Gamma''(0) \right] T.
\label{tgpp}
\end{equation}
Using the definition of the dimensionless temperature, this becomes 
\begin{equation}
\partial_l \overline{T} = \left[ 2-d - \left( {1 \over 2} - 
{\overline{\kappa}\over 1+ 2\overline{\kappa}}\right)
  \Gamma''(0) \right] \overline{T}. \label{TRG}
\end{equation}

\subsubsection{$d=3-\epsilon$ redux}

Eqs.~\ref{finite_temperature_RG}\ and \ref{TRG}\ complete the modified
set of RG flows at non-zero temperature.  Let us focus again 
on the behavior for $d = 3-\epsilon$, discussed above for $T=0$.
As suspected, the presence of the diffusion-like term in
Eq.~\ref{finite_temperature_RG}\ indeed acts to smooth out the cusp in
$\Gamma(\phi)$ (a simple heuristic argument for this rounding is given
in Ref.~\onlinecite{Balents93}).  To see this, consider the
``adiabatic'' approximation in which Eq.~\ref{finite_temperature_RG}\ 
is solved for fixed non-zero $\overline{T}$, ignoring corrections to
$\partial_l \Gamma$ arising from the scale dependence of
$\overline{T}$.  To carry this out, we search for a solution
\begin{equation}
  \Gamma(\phi,l)=\Gamma(0,l)-(3-d)\tilde\Gamma(\phi,\overline{T}(l)),
\end{equation}
where $\tilde\Gamma(0,l) = 0$.  Evaluating
Eq.\ref{finite_temperature_RG} at $\phi=0$ gives the flow equation for
$\Gamma(0,l)$, which in turn implies an equation for
$\tilde{\Gamma}(\phi)$ in the ``adiabatic'' approximation.
\begin{eqnarray}
  \partial_l\Gamma(0,l) & = &
  (3-d)\Gamma(0)+\overline{T}{\Gamma}''(0), \\ 
  \mu\tilde{\Gamma}''(0) &=& \tilde\Gamma(\phi) + \mu \tilde\Gamma''(\phi)
  + {1 \over 2}\tilde\Gamma''(\phi)\tilde\Gamma(\phi),
  \label{fpeqn}
\end{eqnarray}
where $\mu = \overline{T}/\epsilon \ll 1$.  Multiplying
Eq.~\ref{fpeqn}\ through by $\tilde{\Gamma}'(\phi)$, allows one to
perform one integral, and thereby to solve for $\tilde{\Gamma}(\phi)$
exactly in an implicit form.  One finds
\begin{equation}
  \int_0^{\tilde\Gamma} \! d\hat{\Gamma} \left[ {\pi^2 \over
    |\ln{\mu}|} \ln(1+\hat{\Gamma}/2\mu) -
    \hat{\Gamma}\right]^{-1/2} = 2\phi,
\end{equation}
for $0\leq\phi\leq 2\pi$.  For $|\phi-2\pi n| \gg
\mu\sqrt{|\ln\mu|}$ (with integer $n$), this reduces to the
zero temperature solution in Eq.~\ref{zeroTsolution}.  It contains,
however, a boundary layer near $\phi = 2\pi n$.  Inside this boundary
layer $\Gamma(\phi)$ remains smooth, and a simple computation finds
\begin{equation}
  \Gamma''(0) = -{{\pi^2\epsilon^2} \over {\overline{T}
      \ln(\epsilon/\overline{T})}}. \label{gpp}
\end{equation}
Putting this result back into Eq.~\ref{TRG}\ gives, to leading order
in $\epsilon$,
\begin{equation}
  \partial_l \overline{T} = -\overline{T} + {{\pi^2\epsilon^2} \over
    {2\ln(\epsilon/\overline{T})}}.
\label{Tstar1}
\end{equation}
The new term leads to a fixed point at a non-zero temperature given by
\begin{equation}
  \overline{T}^* \sim {{\pi^2\epsilon^2} \over {2|\ln\epsilon|}}.
  \label{Tstar2}
\end{equation}
Note that in equilibrium, i.e., in $v\rightarrow0$ limit,
$\overline{\kappa}\rightarrow\infty$,
$\overline{\kappa}/(1+2\overline{\kappa})\rightarrow 1/2$, and two
terms appearing inside the parentheses in Eq.~\ref{TRG}| exactly
cancel due to the FDT, preserving the power-counting result and
leaving $\overline{T}$ irrelevant.  Here, for any finite $v$, they do
not, and the noise renormalization (first term) overwhelms the dynamic
renormalization (second term) due to strong irrelevance of
$\overline{\kappa}$ (or equivalently $K_\parallel$), thereby
destabilizing the zero-temperature fixed point! The asymptotic flow of
$\overline{T}$ is then given by
\begin{equation}
\partial_l \overline{T} = \left[ 2-d -{1 \over 2}
  \Gamma''(0) \right] \overline{T}. \label{TRGasymptotic}
\end{equation}

\subsection{Behavior in three dimensions}

In three dimensions, we expect that $\overline{T}$ flows to zero, and
the above analysis is invalid.  To analyze this case, consider first
the behavior of $\Gamma(\phi)$ at $T=0$.  From Eq.~\ref{curvatureRG},
we see that even for $d=3$, $\Gamma''(0)$ diverges at a finite scale,
and the non-analyticity remains.  Indeed, for $T=0$, a solution
(presumably the asymptotic attractor for an arbitrary initial
condition) of Eq.~\ref{zero_temperature_RG}\ is
\begin{eqnarray}
  \Gamma(\phi;l;T=\epsilon=0) & = & a(l) \min_n \bigg\{ (\phi - \pi -
    2\pi n)^2 - {\pi^2 \over 3} \bigg\} \nonumber \\
     & &  + \Gamma_0(l), 
     \label{3done}
\end{eqnarray}
where 
\begin{eqnarray}
  \partial_l a(l) & = & - a^2, \\
  \partial_l \Gamma_0 & = & {{2\pi^2} \over 3}a^2.
\end{eqnarray}
Thus the amplitude of the ``cuspy'' part of the disorder correlator
decays slowly to zero,
\begin{equation}
  a(l) = {{a(0)} \over {1 + l a(0)}}.
  \label{3dtwo}
\end{equation}
Employing again the adiabatic approximation for small $\overline{T}
\neq 0$, one finds 
\begin{equation}
  \Gamma''(0,l;d=3) = {{-\pi^2 [a(l)]^2} \over {\overline{T}(l) \ln
      (a(l)/\overline{T}(l))}}. \label{gpp3d}
\end{equation}
Putting this back into Eq.~\ref{TRG}, the ``fixed point'' for
$\epsilon \neq 0$ now drifts slowly down towards zero.
The correct asymptotic behavior is obtained simply by substituting
$\epsilon \rightarrow a(l)$ in Eq.~\ref{Tstar2}, giving
\begin{equation}
  \overline{T}(l) \sim   {{\pi^2 a^2} \over {2|\ln a|}} \sim {\pi^2
    \over {2l^2 |\ln l|}},
\end{equation}
for large $l$.  Note that this is faster than the usual case of a
marginally irrelevant operator ($\sim 1/l$), but much slower than the
naive power-counting result ($\sim e^{-l}$).

\section{Dynamics of 2+$\epsilon$-dimensional smectic}
\label{sec:RGb}

In this section, we study the toy model for the driven smectic near
{\sl two} dimensions.  Based on an extrapolation of the results of the
previous section (Eq.~\ref{Tstar2}), we expect that in this case the
governing stable fixed point should occur at a renormalized
temperature of order one.  In this regard $d=2$ plays a special role,
since (see Eq.~\ref{TRGasymptotic}) dimensionless temperature
$\overline{T}$ becomes marginal.

For temperatures of $O(1)$, the character of the functional RG
(Eq.~\ref{finite_temperature_RG}) is
substantially changed.  This is because the linear operator
\begin{equation}
  \hat{\cal L} = 3-d + \overline{T} {\partial^2 \over {\partial\phi^2}}
\end{equation}
has a {\sl discrete} spectrum (defined in the space of
$2\pi$-periodic functions),
\begin{equation}
  \hat{\cal L} \cos n\phi = (3-d -\overline{T} n^2) \cos n\phi.
\end{equation}
For $\overline{T} > 3-d$, therefore, all the harmonics but the
constant ($n=0$) are irrelevant, and $\Delta(\phi) \rightarrow
\Delta_0$.  For $(3-d)/4 < \overline{T} < 3-d$, we can study the onset
of non-trivial random-force correlations by truncating the Fourier
expansion of $\Delta$ at $n=1$:
\begin{equation}
  \Delta(\phi) = \Delta_0 + \Delta_1 \cos \phi. \label{truncated}
\end{equation}
Eq.~\ref{truncated}\ displays one distinct advantage over the full
functional RG treatment in the previous section: $\Delta(\phi)$ in
this limit is manifestly {\sl analytic}, so perturbation theory is
uniformly valid.  By contrast, the treatment for $d \approx 3$, while
physically reasonable, is considerably less controlled.  Indeed, from
Eq.~\ref{TRGasymptotic}\ or Eq.~\ref{gpp}, $\Gamma''(0)^*=-2(d-2)$ is
$O(1)$ in this limit, and although $\Gamma(\phi)$ is $O(\epsilon)$, it
is unclear how the singular derivatives might enter into higher order
corrections.

To implement the complimentary, strictly controlled approach for
$d=2+\epsilon$, we simply insert Eq.~\ref{truncated}\ in
Eq.~\ref{finite_temperature_RG}, dropping harmonics with $n \geq 2$.
This gives
\begin{eqnarray}
{d\overline{\Delta}_0(l)\over
dl}&=&(3-d)\overline{\Delta}_0(l)+{1\over4}\overline{\Delta}_1^2\;,
\label{Delta0}\\
{d\overline{\Delta}_1(l)\over
dl}&=&(3-d-\overline{T})\overline{\Delta}_1-{1\over2}\overline{\Delta}_1^2\;,
\label{Delta1}\\
{d\overline{T}(l)\over dl}&=&(2-d +
{1\over2}\overline{\Delta}_1)\overline{T}(l)\label{T}\; ,
\end{eqnarray}
where
$\overline{\Delta}_{0,1}\equiv\Delta_{0,1}C_{d_\perp}\Lambda^{d-3}/
(K_\perp\gamma|v|)$.  We pause to point out that Eq.~\ref{truncated}\ is
equivalent to the physical approximation of the equation of motion,
\begin{eqnarray}
  \gamma (\partial_t + v \partial_x)\phi({\bf r},t)
  &=& (K_\parallel \partial_x^2 + K_\perp \nabla_\perp^2) \phi({\bf
    r},t) + F_0({\bf r}) \nonumber\\
  &+&F_1({\bf r})\cos [ y-\phi({\bf r},t)]+\eta({\bf r},t)\;,\nonumber\\
  \label{eom_singlemode}
\end{eqnarray}
with 
\begin{eqnarray}
  \left[ F_0({\bf r}) F_0({\bf r}') \right]_{\rm ens.} & = & 2 \Delta_0
  \delta^{(d)} ({\bf r-r'}), \\
 \left[ F_1({\bf r}) F_1({\bf r}') \right]_{\rm ens.} & = & 2 \Delta_1
  \delta^{(d)} ({\bf r-r'}).
  \label{FF}
\end{eqnarray}
For completeness, we rederive Eqs.~\ref{Delta0}--\ref{T}\ directly
from Eqs.~\ref{eom_singlemode}--\ref{FF}\ in Appendix E.  Note that
Eq.~\ref{Delta0}\ implies that the random $\phi$-independent drag
correlator $\overline{\Delta}_0$ is always generated by the disorder
and is relevant, even if it is not present in the ``bare'' equation of
motion.  Although $\overline{\Delta}_0$ runs off to infinity, as
discussed in the previous subsection, its effects can luckily be taken
into account exactly, by a simple transformation on the field $\phi$.
The remaining flow equations for $\overline{\Delta}_1(l)$ and
$\overline{T}(l)$ contain three fixed points
\begin{eqnarray}
  \mbox{Gaussian:}\;\;\;\;\;\overline{T}^*=0, \overline{\Delta}_1^*=0\;,\\
  \mbox{Zero Temperature:}\;\;\;\;\;\overline{T}^*=0,
  \overline{\Delta}_1^*=2(3-d)\;,\\ 
  \mbox{Driven
    Smectic:}\;\;\;\;\;\overline{T}^*=5-2d,\overline{\Delta}_1^*=2(d-2)\;. 
  \label{twodeefp}
\end{eqnarray}
Of these, only the Driven Smectic is globally stable.  Fortunately, it
is also perturbative, and hence controlled, for $d$ near $2$, and
indeed becomes exact in the $d\rightarrow 2^+$ limit.  Furthermore, the
Driven Smectic fixed point that is perturbative near $d=2$ appears to
smoothly match onto the finite disorder, finite temperature fixed
point that was obtained at strong coupling by a functional
renormalization group calculation near $d=3$ in Sec.~\ref{finite_T_RG}.

Equations~\ref{Delta1}--\ref{T}\ are also the nonequilibrium analog of the
Cardy-Ostlund fixed line\cite{CO} that describes the 1+1 dimensional vortex
glass state\cite{FFH} and the super-rough phase of a crystal surface
grown on a random substrate\cite{TD,Shapir}. Because of the lack of
fluctuation-dissipation theorem for the driven system considered here,
the Cardy-Ostlund fixed line is destabilized by the nontrivial
renormalization of temperature. This disorder generated
thermal renormalization ($\overline{\Delta}_1$-dependent
term in Eq.\ref{T}) is reminiscent of  the ``shaking'' temperature,
discussed by Koshelev and Vinokur\cite{kv}.  Note, however, that the
``heating'' found here is a multiplicative rather than an additive
effect, as was suggested in Ref.~\onlinecite{kv}.

In $d=2+\epsilon$ dimensions the flow equations have an interesting
spiral structure around the Driven Smectic fixed point, as is
illustrated in Fig.5
\begin{figure}[bth]
{\centering
\setlength{\unitlength}{1mm}
\begin{picture}(150,65)(0,0)
\put(-3,-28){\begin{picture}(150,50)(0,0) 
\includegraphics{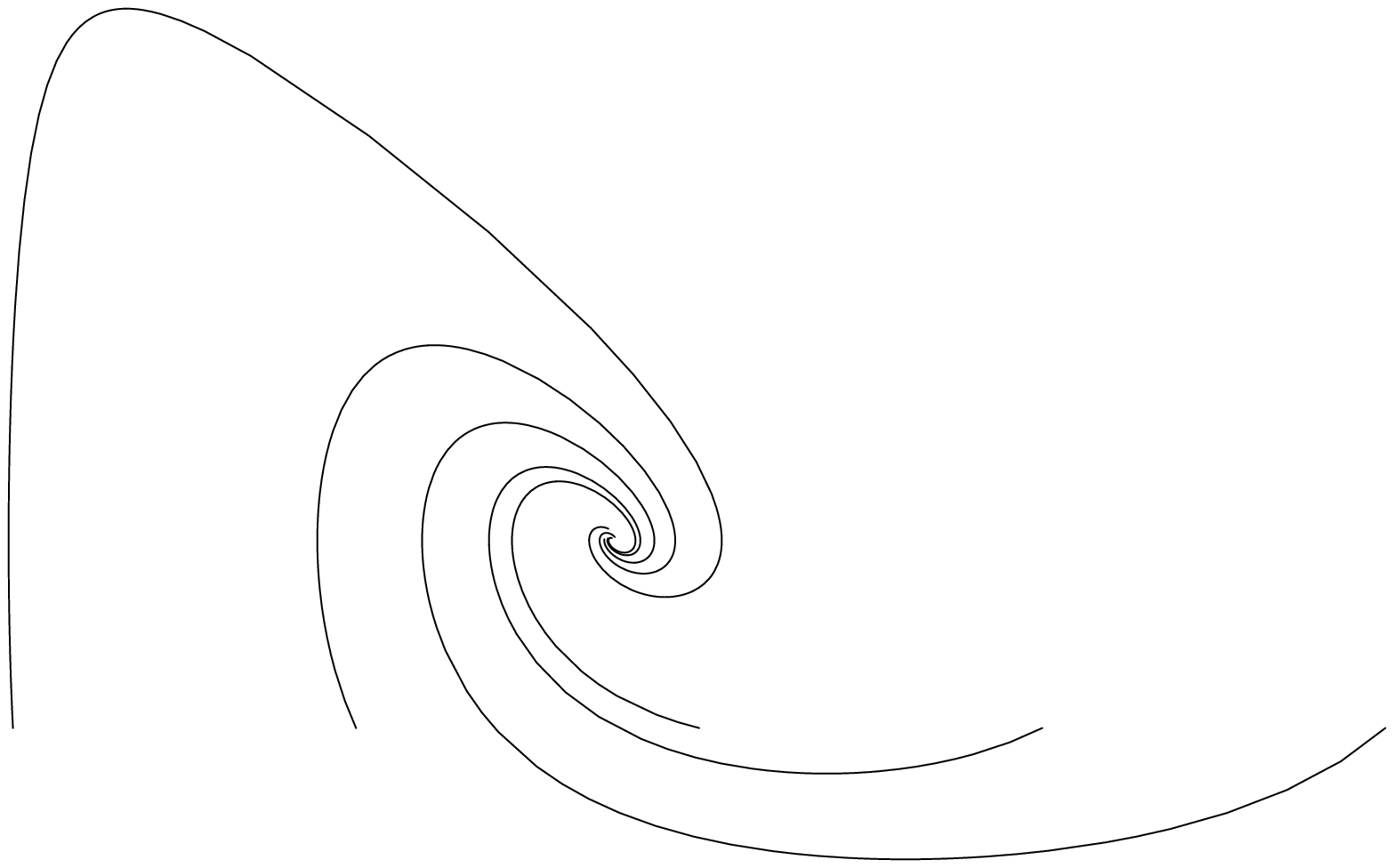}
\end{picture}}
\put(-37,-68){\begin{picture}(150,50)(0,0) 
\includegraphics{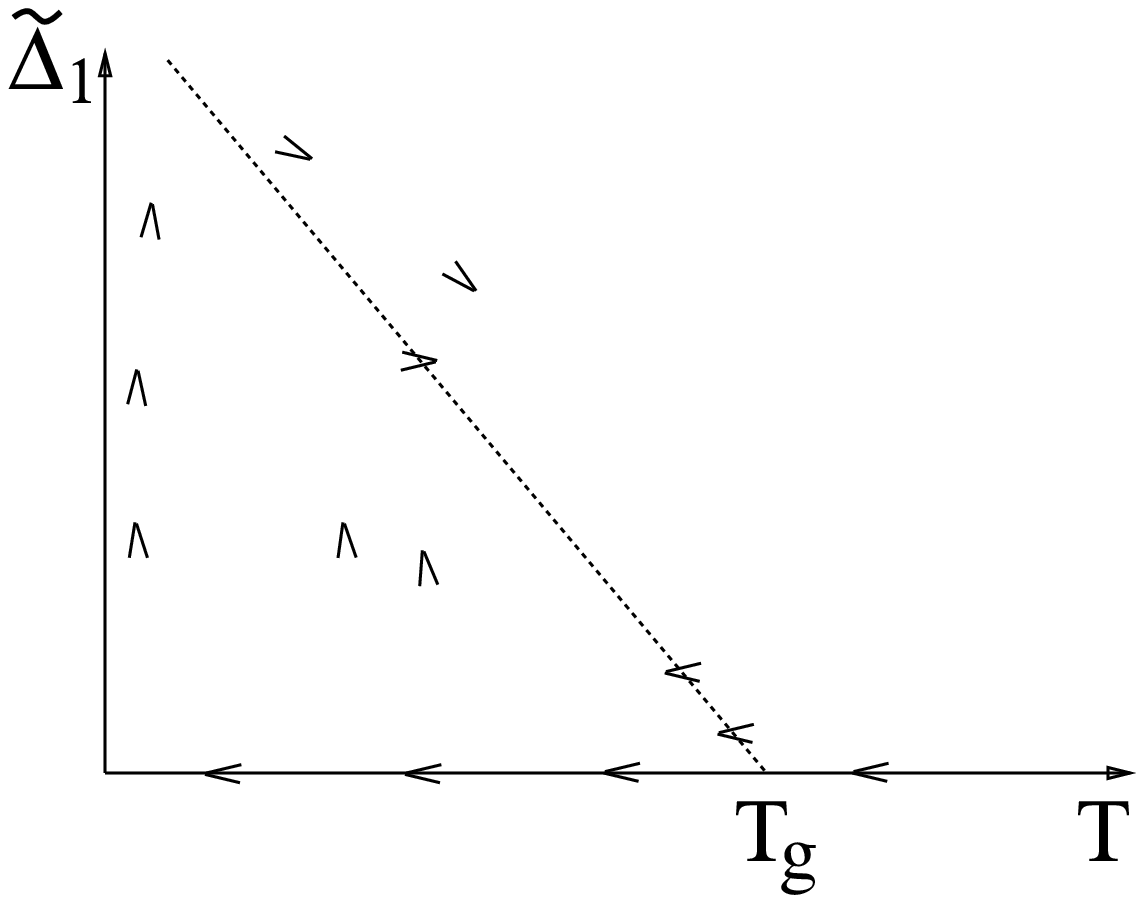}
\end{picture}}
\end{picture}}
Fig.5.:{Renormalization group flow diagram (for $2<d=2+0.3<3$) in the
disorder $\Delta_1$ temperature $T$ plane, for a set of initial
conditions with $\overline{\Delta}_1(0)=0.2$ and $\overline{T}(0)$ ranging
from $0.05$ to $0.85$, in increments of $0.2$}
\label{flow_d}
\end{figure}

The structure of the renormalization group flow in the physically
interesting case of $d=2$ is displayed in Fig.6.
\begin{figure}[bth]
{\centering
\setlength{\unitlength}{1mm}
\begin{picture}(150,65)(0,0)
\put(-3,-28){\begin{picture}(150,50)(0,0) 
\includegraphics{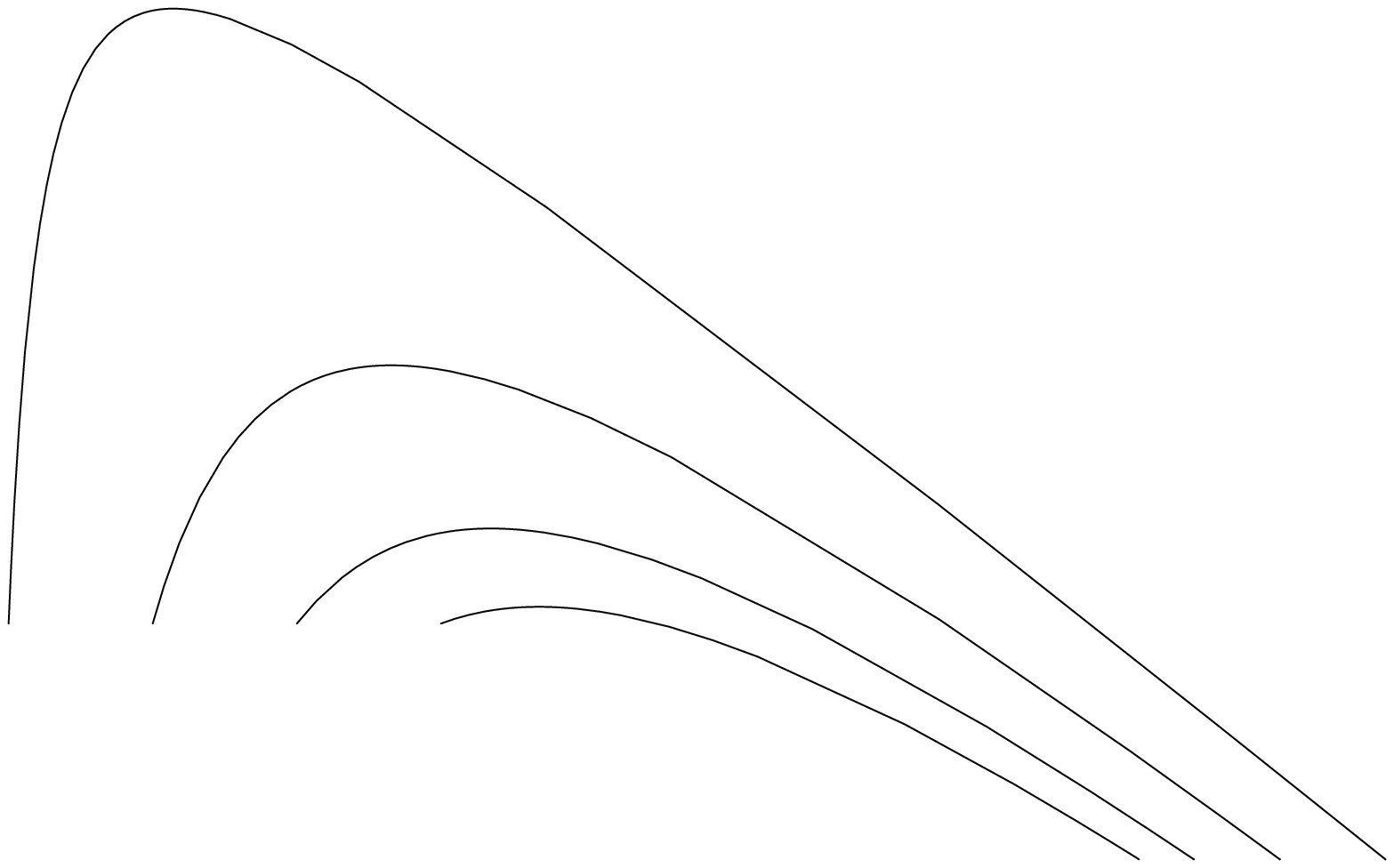}
\end{picture}}
\put(-37,-68){\begin{picture}(150,50)(0,0) 
\includegraphics{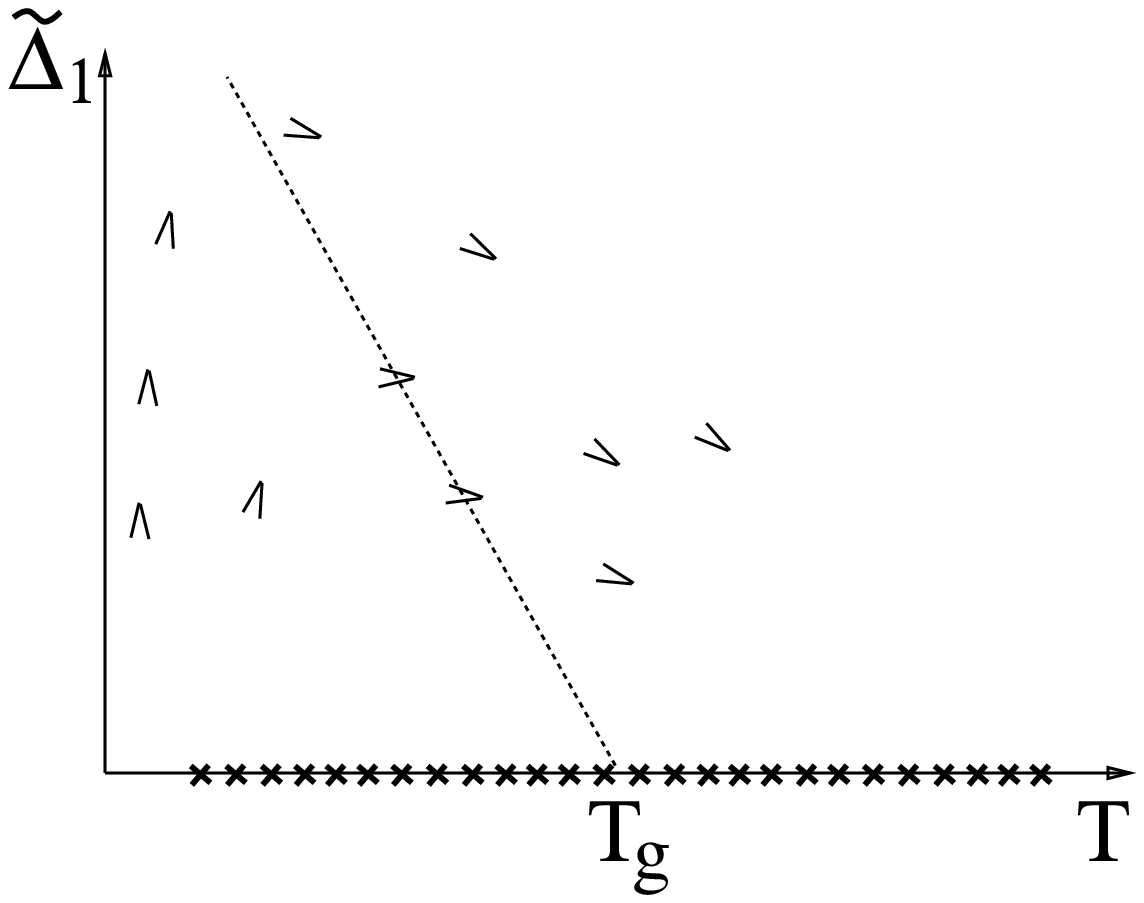}
\end{picture}}
\end{picture}}

{Fig.6: Renormalization group flow diagram (for $d=2$) in the
disorder $\Delta_1$ temperature $T$ plane, for a set of initial
conditions with $\overline{\Delta}_1(0)=0.2$ and $\overline{T}(0)$ ranging
from $0.05$ to $0.65$, in increments of $0.2$}
\label{flow_2}
\end{figure}
In $d=2$ the Driven Smectic fixed point moves down to zero disorder
and merges into the zero-disorder fixed line. Despite the absence of a
globally stable finite disorder fixed point in $d=2$, we expect
nontrivial observable effects associated with the interesting RG flows
displayed in Fig.6. Qualitatively, a moving lattice at temperature $T$
with disorder $\Delta_1$ behaves at long times and length scales as a
thermal moving smectic with an effective disorder-enhanced temperature
$T_{\rm eff}(T,\Delta_1)$.

To determine $T_{\rm eff.}$, we take advantage of the solvability of
Eqs.~\ref{Delta1}--\ref{T}\ in two dimensions.  Dividing
Eq.~\ref{Delta1}\ by Eq.~\ref{T}\ gives
\begin{equation}
  {d\overline{\Delta}_1(\overline{T})\over d
    \overline{T}}=2{1-\overline{T}\over\overline{T}} 
  -{\overline{\Delta}_1\over\overline{T}}\;.
  \label{Delta1T}
\end{equation}
This has the exact solution
\begin{equation}
  \overline{\Delta}_1^{\rm eff} = 2-\overline{T}_{\rm
    eff} + f/\overline{T}_{\rm eff}\;,
\label{Delta1Tsolution}
\end{equation}
where $f$ is given by the initial conditions $\overline{\Delta}_1(l=0)
= \overline{\Delta}_1$, $\overline{T}(l=0) = \overline{T}$ as
\begin{equation}
  f = \overline{T}\left[ \overline{\Delta}_1 - 2 + \overline{T}\right].
\end{equation}
%
%
The asymptotic effective temperature is determined from
Eq.~\ref{Delta1Tsolution}\ by setting $\overline{\Delta}_1^{\rm eff} =
0$, which gives
\begin{equation}
  \overline{T}_{\rm eff} = 1 + \sqrt{1+\overline{T}(\overline{\Delta}_1 +
    \overline{T} - 2) }. \label{Tbareff}
\end{equation}
This effective temperature has the following meaning: asymptotically,
the {\sl connected} correlation and response functions of a moving
smectic at temperature $\overline{T}$ in the presence of disorder
$\overline{\Delta}_1$ are given by the noninteracting, disorder-free
functions with $T$ replaced by $T_{\rm
  eff}(\overline{T},\overline{\Delta}_1)$. Of course the details of 
the approach to the zero disorder fixed line and to this effective
temperature will determine subdominant $q$-dependent corrections to
the correlation and the response functions.

\section{Nonequilibrium Response and Correlation Functions}
\label{res_corr_functions}

We now turn our attention to the dynamic response and correlation
functions of a moving smectic.  We will consider only mean
(disorder-averaged) properties here.  The mean response function is
defined by
\begin{eqnarray}
  R({\bf r},t) & = & \left[ {{\delta \langle \phi({\bf r+r'},t+t')\rangle}
      \over {\delta \eta({\bf r}',t')}} \right]_{\rm ens.} \nonumber \\
  & = & \left[ \langle \phi({\bf r+r'},t+t') \hat\phi({\bf r}',t')
    \rangle\right]_{\rm ens.} .
\end{eqnarray}
We similarly define a mean correlation function,
\begin{equation}
  C({\bf r},t) = \left[ \left\langle \left(\phi({\bf r+r'},t+t')
      -\phi({\bf r'},t')\right)^2 \right\rangle \right]_{\rm ens.}.
\end{equation}

Given the results of the renormalization group analysis of the previous
two sections, the long length and time asymptotics of these functions
can be computed using standard matching techniques.  For the
momentum-shell regularization used here, this matching is
most directly done in momentum and frequency space.  Consider first
the correlation function, which for transverse momentum ${\bf
  k}_\perp$ and rescaling factor $1<b\leq |\Lambda/{\bf k_\perp}|$ satisfies
the relation 
\begin{equation}
  C({\bf k},\omega; \{ \lambda_i \}) = 
  b^{d_\perp+\zeta+z}
  C({\bf k}_\perp b, k_x b^\zeta, \omega b^z; \{ \lambda_i(b)
  \}). \label{relationC}
\end{equation}
Here $\{ \lambda_i(b) \}$ denotes the set of running coupling
constants at scale $b$.  The prefactor on the right hand side arises
from the (conventional) definition of the Fourier transform,
\begin{equation}
  C({\bf r},t) = \int_{{\bf k}\omega} 2\left[1 - e^{-i{\bf k\cdot
      r}+i\omega t}\right] C({\bf k},\omega),
\end{equation}
and the dimensionlessness of $\phi({\bf r},t)$.

To calculate the correlators at long length and time scales, we will
choose $b=\Lambda/|{\bf k_\perp}| \gg 1$ and evaluate the
right-hand-side of Eq.~\ref{relationC}.  This is simple because the
rescaled correlator is evaluated at a large rescaled transverse
wavevector $|{\bf k_\perp}|b = \Lambda$ equal to the uv cut-off, at which
fluctuations are small, and therefore can be taken into account
perturbatively, without encountering any infa-red divergences.
However, the computation of the rescaled propagator requires a
knowledge of the running couplings $\{ \lambda_i(b) \}$.

These flows have been studied in Sections IV-V.  Throughout, unless
explicitly indicated otherwise, all running couplings {\it without}
arguments refer to the bare couplings, i.e.,
$\lambda_i\equiv\lambda_i(b=1)$. It is convenient to choose
$z=\zeta=2$ and $\hat{\chi}=-d-1$. With this choice, $K_\perp$ and
$\gamma v$ are invariant, i.e.
\begin{eqnarray}
  K_\perp(b) & = & K_\perp, \\
  \gamma v(b) & = & \gamma v.
\end{eqnarray}
The remaining parameters behave nontrivially.  The rescaled
temperature and longitudinal elastic modulus follow from
Eq.\ref{Tbar_define}\ and Eq.~\ref{kappaRG}, respectively:
\begin{eqnarray}
  T(b) & \;\; \widetilde{\scriptstyle b \gg 1}\;\; & {2(K_\perp
    K_\parallel)^{1/2}\over\Lambda^{d-2} 
    C_{d-1}} \overline{T}(b)\;b^{-1},
  \label{Tb}\\
  \overline{\kappa}(b) & = & \overline{\kappa}\;b^{-2}. \label{kappaRGb}
\end{eqnarray}
In $d=2$, the dimensionless temperature flows to the fixed half-line
(and is hence non-universal), while in $d=3$, it flows logarithmically
to zero, i.e.
\begin{equation}
  \overline{T}(b=e^l) \;\; \widetilde{\scriptstyle b \gg 1}\;\;
  \cases{\overline{T}_{\rm eff} > 1, & $d=2$ \cr
    {\pi^2/{2l^2|\ln l|}} & $d=3$ \cr}.
\end{equation}
For $2<d<3$ it flows to a universal value (see Eqs.~\ref{Tstar2} and
\ref{twodeefp}), which is sadly of only formal interest since the
$2.5$-dimensional smectic is currently experimentally inaccessible.

The mobility is more complicated, but using Eq.~\ref{gamma_RG}\ and
Eq.~\ref{kappaRGb}, we obtain
\begin{equation}
  \gamma(b)=\gamma e^{\Phi(b)}\;,\label{gamma_Phi} 
\end{equation}
with 
\begin{equation}
  \Phi(b)=-\int_0^{\log{b}} d l\; {\overline{\kappa}\;e^{-2l}
  \over 1+2\;\overline{\kappa}\;e^{-2l}}
  \Gamma''(0,l)\;.\label{Phi}
\end{equation}
%
The integrand in Eq.~\ref{Phi}\ is exponentially suppressed at large
$l$, so that $\Phi(b)$ has a finite limit as $b \rightarrow \infty$.
This implies the finite renormalization
\begin{equation}
  \gamma(b) \;\;\widetilde{\scriptstyle b\rightarrow\infty} \;\; \gamma_R =
  \gamma e^{\sigma K_\parallel K_\perp \Lambda^2/(\gamma v)^2}.
  \label{renormalizedGamma}
\end{equation}
Because, for large $v$ (small $\overline\kappa$) the integral in
Eq.~\ref{Phi}\ is dominated by small $l$ (due to the exponential
behavior of $\kappa(l)$), the constant $\sigma$ is highly
nonuniversal.  Above two dimensions at sufficiently high velocities
and temperatures ($\overline{\kappa}<< 1$ and $\overline{T} \gg a_0^2$
in 3d), $\overline{T}$ flows rapidly to a unique fixed-point, and it
becomes parameter-independent, with $\sigma = 2(d-2)$.  For $d=2$,
$\sigma$ depends upon the bare disorder strength even for
$\overline{\kappa}<< 1$, due to the semi-infinite fixed line with
$\overline{T}_{\rm eff}>1$.  More interesting is the extreme
low-temperature limit, in which (see Eqs.~\ref{gpp},\ref{gpp3d})
$\Gamma''(0)$ becomes singular.  In 3d, this limit gives
\begin{equation}
  \sigma \approx {{\pi^2 a_0^2} \over {\overline{T}\ln(a_0/\overline{T})}},
  \qquad \overline{T} \ll a_0^2. \label{lowTsigma}
\end{equation}
The corresponding low-temperature regime in two dimensions is outside
the limits of the controlled RG, but we expect a result similar to
Eq.~\ref{lowTsigma}\ to hold with, however, $a_0$ of order one.  Note
that in all cases $\gamma_R$ is strongly enhanced as the velocity of
the moving smectic is lowered.

The remaining flow parameter is the force-force correlator
$\Delta(\phi)$.  Again, we quote here the results only directly in
$d=2,3$.

In two dimensions, the correlator is well-described asymptotically in
the single-harmonic approximation (Eq.~\ref{truncated}).  From
Eq.~\ref{Delta0}, $\overline{\Delta}_0$ grows linearly with $b$,
\begin{equation}
  \overline{\Delta}_0(b) \;\;\widetilde{\scriptstyle
    b\rightarrow\infty} \;\;   \tilde{\Delta}\; b,
\end{equation}
where $\tilde{\Delta} > \Delta_0$, and for $\overline{\Delta}_1 \ll 1$
  and $\overline{T}>1/2$
  can be estimated by
\begin{equation}
  \tilde{\Delta} \approx \Delta_0 + {1 \over {4\pi\Lambda
      K_\perp\gamma v}} {{\Delta_1^2} \over {2\overline{T}-1}}.
\end{equation}
As can be easily seen from Eq.\ref{Delta1}, in $d=2$, the first
harmonic flows asymptotically to zero according to
\begin{equation}
  \Delta_1(b) \;\;\widetilde{\scriptstyle
    b\rightarrow\infty} \;\; {\tilde{\Delta}_1}
    b^{-(\overline{T}_{\rm\small eff} -1)}.
\end{equation}

In three dimensions, we have instead logarithmic flows, and simple
manipulations of Eqs.~\ref{3done}--\ref{3dtwo}\ give
\begin{eqnarray}
  \Delta(\phi,l) & \;\;\widetilde{\scriptstyle
    b\rightarrow\infty} \;\; & 2\pi\gamma v K_\perp {a_0 \over
    {1+la_0}} \min_n \left\{ (\phi - \pi - 2\pi
    n)^2 - {\pi^2} \right\} \nonumber \\
  & & + \tilde\Delta,
\end{eqnarray}
where
\begin{equation}
  \tilde\Delta =  2\pi\gamma v K_\perp\left[\Gamma_0 + {2\pi^2 \over
      3} a_0\right].
\end{equation}

We are now in a position to evaluate the right-hand-side of
Eq.~\ref{relationC}.  Setting $b=e^l = \Lambda/|{\bf k_\perp}|$ and using the
above results, we obtain
\begin{equation}
  C({\bf k},\omega) = \cases{{\cal D}^{-1}({\bf k},\omega) \left[ 4\pi
      K \gamma_R \overline{T}_{\rm eff} + \tilde\Delta
      \delta(\omega)\right],\; & $d=2$ \cr
    {\cal D}^{-1}({\bf k},\omega) \left[{1\over k_\perp} 
        {{4\pi^3 a^2 K\gamma_R}
        \over {|\ln a|}} + \tilde\Delta \delta(\omega)\right],\; &
    $d=3$ \cr}, \label{MFR1}
\end{equation}
where 
\begin{equation}
  {\cal D}({\bf k},\omega) = (\gamma_R \omega - \gamma v k_x)^2 +
  (K_\parallel k_x^2 + K_\perp k_\perp^2)^2, \label{MFR2}
\end{equation}
we have defined $K= \sqrt{K_\perp K_\parallel}$ and
from Eq.~\ref{3dtwo}, the three-dimensional logarithmic coupling
constant is 
\begin{equation}
  a = {{a_0} \over {1+a_0 \ln(\Lambda/k_\perp)}}.
  \label{MFR3}
\end{equation}

Eqs.~\ref{MFR1}--\ref{MFR3}\ have simple physical significance.  The
first term in each of the square brackets in Eq.~\ref{MFR1}\
represents time-dependent ``thermal'' fluctuations.  The second,
$\delta$--function term represents a static, time-independent
distortion of the smectic, and is in fact identical in form to the
perturbative expressions of Section IV.  This is roughly because at
the uv momentum cutoff (where the rescaled correlator is evaluated),
the displacement fluctuations are strongly suppressed, and dominated
by the value of the random force at origin $\phi=0$ (as in the naive
Larkin approximation).

To further explore the implications of Eqs.~\ref{MFR1}--\ref{MFR3}, we
now discuss the corresponding expressions in real space and time.
First consider the long-time limit $C_{\rm EA}({\bf r}) = \lim_{t
  \rightarrow \infty} C({\bf r},t)$, 
\begin{equation}
  C_{\rm EA}({\bf r}) = \int_{\bf k} {{2\tilde\Delta\left[1-e^{-i{\bf
            k\cdot r}}\right]} \over {(\gamma v k_x)^2 + (K_\parallel
      k_x^2 + K_\perp k_\perp^2)^2}}. 
\end{equation}
This correlator is analogous to the Edwards-Anderson correlator in a
spin-glass, and represents a static but random conformation of the
$\phi$ field that persists at long times.  Performing the above
integration, we find two limits.  For $d<4$ and $x<K_\parallel/\gamma
v$, the static roughness {\it scales} isotropically (similarly to the
equilibrium case), with {\it finite} anisotropy arising due to the
difference between $K_\perp$ and $K_\parallel$
\begin{equation}
  C_{\rm EA}({\bf r})={\tilde{\Delta}\over K} \left({K_\parallel \over
      K_\perp}\right)^{(d-2)/4} r_\perp^{4-d} 
  f^{(1)}_\Delta\left({|x| K_\perp^{1/2}\over r_\perp
      K_\parallel^{1/2}}\right) \;,
\end{equation}
where the scaling function obeys
$f_\Delta^{(1)}(x\rightarrow0)\rightarrow\mbox{constant}$ and
$f_\Delta^{(1)}(x\rightarrow\infty)\rightarrow x^{4-d}$.  For $d\leq
3$, in the asymptotic limit $x>K_\parallel/\gamma v$, $C_\Delta({\bf
r})$ becomes infinitely anisotropic owing to the difference between
the convective behavior along the direction of motion ($x$) and
diffusive transport transverse to it (along ${\bf r}_\perp$).  We find
\begin{equation}
  \label{C_Delta}
C_{\rm EA}({\bf r})=\tilde{\Delta}{r_\perp^{3-d}\over\gamma v K_\perp} 
  f_\Delta^{(2)}\left({|x| K_\perp\over r_\perp^2 \gamma v}\right)\;,
\end{equation}
where the asymptotics of the scaling function are
\begin{equation}
\label{f_Delta2}
f_\Delta^{(1)}(x)\propto
\left\{ \begin{array}{lr}
\mbox{constant}, & \mbox{for}\; x\rightarrow 0 \\
x^{(3-d)/2}, & \mbox{for}\; x\rightarrow\infty
\end{array} \right.\;.
\end{equation}
In three dimensions, the power-law in Eq.~\ref{C_Delta}\ should be
replaced by the logarithm $(r_\perp)^0 \rightarrow \ln |\Lambda r_\perp|$.

The other important physical correlator measures the thermal
fluctuations around the static distortion measured by $C_{\rm EA}$.
The thermal correlator is naturally defined as
\begin{equation}
  C_{\rm T}({\bf r},t) \equiv C({\bf r},t) - C_{\rm EA}({\bf r}).
\end{equation}
For simplicity, we consider only the equal-time thermal fluctuations;
corresponding results for non-equal times are easily obtained from
Eqs.~\ref{MFR1}--\ref{MFR3}.  In two dimensions, the thermal
correlator is logarithmic,
\begin{equation}
  C_{\rm T}({\bf r},0) \sim 2 \overline{T}_{\rm eff} \ln \sqrt{ \aleph x^2
    + \aleph^{-1}y^2},   
\end{equation}
where $\aleph= \sqrt{K_\perp/K_\parallel}$.  Note that the (in 2d
nonuniversal) coefficient of the logarithm above is proportional to
the effective temperature, and therefore {\sl bounded} below!  This is
a consequence of the roughly semicircular RG flows in Fig.~6.  It is
interesting to further note that by extending above calculations to
$2+\epsilon$ dimensions, using results of Sec.\ref{sec:RGb}, we find
that the logarithmic growth of thermal correlation function is infact
super-universal (i.e., independent of $d$) for $2<d<3$, with the
prefactor $\overline{T}_{\mbox {eff}}$ above replaced by a {\it
universal} fixed point value of $\overline{T}^*=1-2\epsilon$
(Eq.\ref{twodeefp}).

In three dimensions, the logarithmic decay of $a$ in Eq.~\ref{MFR3}\
renders the Fourier transform of the thermal term in Eq.~\ref{MFR1}\
non-divergent.  The thermal correlator thus {\sl saturates} at long
distances in $d=3$,
\begin{equation}
  C_{\rm T}({\bf r},0) \stackrel{\longrightarrow}{\scriptstyle
    |r|\rightarrow\infty} C_{\rm T,0} \propto {a_0 \over {|\ln a_0|}}. 
\end{equation}

Comparing the static (EA) and thermal correlators above, we see that
at {\sl any} fixed time, the static contribution to $C({\bf r},t)$
dominates at long distances, i.e.
\begin{equation}
  C({\bf r},t) \;\;\widetilde{\scriptstyle
    |r|\rightarrow\infty} \;\; C_{\rm EA}({\bf r}).
\end{equation}
Thus the naive perturbative results of section IV are essentially
correct for the long-distance properties of the correlation
functions.  In particular, {\sl the structure function in the smectic
phase displays power-law smectic Bragg peaks (translational QLRO) in three
dimensions, with fully rounded smectic peaks (translation SRO) in two
dimensions}. 

Similar analysis of the response function $R({\bf k},\omega)$ shows
that to the leading order in disorder it is given by
\begin{equation}
  R({\bf k},\omega) = {1 \over {i(\gamma_R(k_\perp)\omega-\gamma v
      k_x) + K_\parallel k_x^2 + K_\perp k_\perp^2}}\;,
\end{equation}
and is identical to the response function in the linearized theory of
Sec.\ref{sec:linearized}, but with a drag coefficient $\gamma$
multiplying $\omega$ (but {\it not} $k_x$) replaced by the disorder
enhanced renormalized $\gamma_R(k_\perp)$,
\begin{eqnarray}
  \gamma_R(k_\perp )&=& \gamma e^{\Phi(\Lambda/k_\perp)}\;,
  \label{gamma_PhiR1}\\
  &\;\;\widetilde{\scriptstyle
    k_\perp \ll \Lambda} \;\; & \gamma_R. \label{gamma_PhiR2}
\end{eqnarray}
Since $\gamma_R$ is finite, the renormalized response function $R({\bf
k},\omega)$ implies an {\it analytic} response to a uniform transverse
external force, and a {\it finite linear} mobility $\gamma_R^{-1}$ in
the limit of vanishing transverse force.

We now turn our attention to the {\it nonlinear} dynamic response to a
transverse force $f_\perp$.  A true calculation of the non-linear
response to $f_\perp$ is beyond the current capabilities of these
methods.  A relatively simple scaling argument suffices, however, to
obtain a rough picture of the dynamics.  Consider introducing an
explicit force into the MSR functional.  Since it couples directly to
$\phi$, simple
rescaling and use of the relations between the scaling exponents
$\hat{\chi}$, $z$ and $\zeta$ leads to a strong growth of the rescaled
force $f_\perp(b)$ under the RG,
\begin{equation}
f_\perp(b)=f_\perp b^2\;.
\label{fperp}
\end{equation}
When this rescaled force becomes large, of the order of the typical
depinning force $f_c\sim \Lambda^{d/2}\sqrt{\Delta''(0)}$, it becomes
a strong perturbation and the damping coefficient $\gamma$ should
cease to renormalize. Choosing $f_\perp b^2 = f_c$ defines a
($f_\perp$-dependent) rescaling factor $b$ at which to evaluate
$\gamma$ via Eqs.~\ref{gamma_Phi}--\ref{Phi}.  This gives an effective
drag coefficient
\begin{eqnarray}
\gamma_{\rm eff}(f_\perp) &=&\gamma_R e^{-\nu (f_\perp/f_c)}\;,\;\;\;\mbox{for}\; f_\perp<f_c\;,\\
&=&\gamma\;,\;\;\;\mbox{for}\; f_\perp>f_c\;,
\label{gamma_eff}
\end{eqnarray}
where $\nu \equiv \sigma K_\parallel K_\perp\Lambda^2/(\gamma v)^2$. Combining
this result with the definition of $\gamma$
\begin{equation}
v_\perp(f_\perp)={f_\perp\over\gamma_{\rm eff}(f_\perp)}\;,
\end{equation}
we obtain a nonlinear response which interpolates between two different linear
responses at small and large $f_\perp$, schematically displayed in
Fig.7.  We remark in passing that this simple argument neglects
possible renormalizations of $f_\perp$ by the random force.  This
occurs at zero temperature, and allows for the possibility of a true
critical transverse force in that case.  A careful treatment of the
finite-temperature crossover to the above (generic) behavior is still an
open problem.  
\begin{figure}[bth]
{\centering
\setlength{\unitlength}{1mm}
\begin{picture}(150,65)(0,0)
\put(-25,-60){\begin{picture}(150,50)(0,0) 
\includegraphics{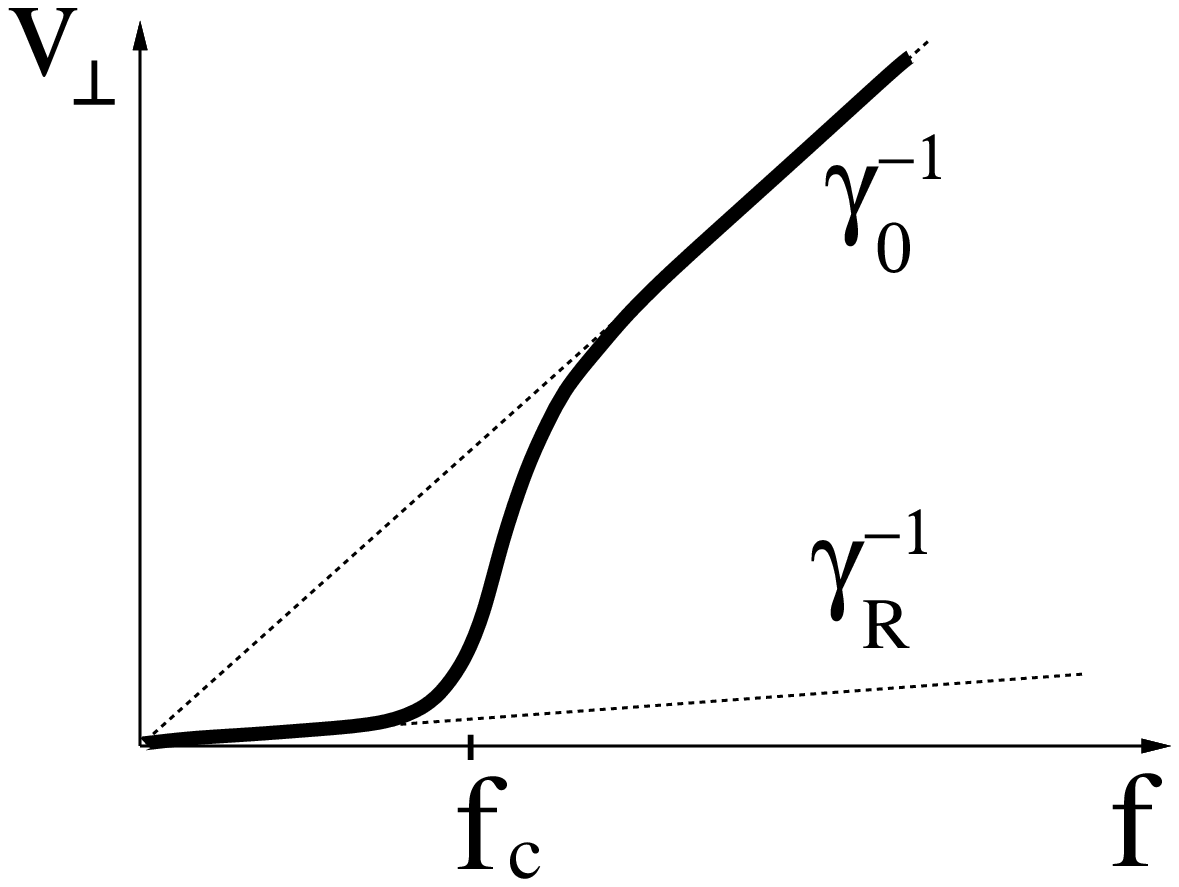}
\end{picture}}
\end{picture}}

{Fig.7: Schematic of the nonlinear transverse response for the moving
  smectic (neglecting the permeation mode).  Including the permeation
  mode simply adds a linear function of slope $\mu_{\rm perm.}$ to the
  above plot.}
\label{mobility}
\end{figure}

The finite-temperature ``IV'' (velocity-force) curve gives an
interpolation between two linear regimes, above and below $f_c$.  For
small $\gamma v$, the renormalized mobility $\gamma_R^{-1}$ is quite
small.  In this case the IV has a non-linear feature similar to that
of a threshold, but nevertheless completely analytic.  In addition,
the permeation mode provides a second channel for
transport, which further enhances the linear mobility.  A possible
(though highly speculative) scenario at zero temperature is a small
linear mobility via the permeation of a small concentration of
fluctuating defects ``activated'' through local chaotic dynamics,
superimposed upon a sharp threshold for motion of the smectic density
wave. However, we again caution (as in the Introduction), that zero
temperature dynamics may be highly nonuniversal, with qualitatively
different behavior occurring in different systems.

\section{Discussion}

\subsection{Relation to previous work}

Much of the recent interest in driven solids stems from the work of
Koshelev and Vinokur (KV),\cite{kv}\ who applied perturbation theory to
compute the mean-square displacement of the driven solid and argued
that the driven system might exhibit a nonequilibrium phase transition
from a moving liquid to an ordered moving solid.\cite{schmid_note}\ 
The argument was based on the notion that at large velocities the
effect of pinning could be described by an effective ``shaking''
temperature $T_{sh}\sim1/v^2$.  With this assumption, KV recast the
problem into equilibrium form.  This suggested only thermal roughness
for the driven lattice, so that the mean square displacement would
always be bounded in $d>2$ (with translational QLRO in $d=2$),
arguing indeed for the stability of the driven lattice for $d \geq 2$.  

Experiments on flux lattices in type-II superconductors have indeed
shown evidence for current-induced ordering of the vortex array.
This evidence has been obtained both indirectly through transport
experiments \cite{shobo,ak} and by directly probing the structure of a
driven vortex lattice by neutron scattering \cite{yaron} and
decoration experiments \cite{duarte,marche}.  Numerical simulations of
driven vortex arrays in two dimensions also provide clear indication
of ordering of the sliding lattice at large
drives\cite{jensen,shi,kv,faleski}. We discuss these experiments in
more detail below.

The notion that non-equilibrium effects might play an important role
was addressed by Balents and Fisher\cite{bf}\ in the simpler context
of CDWs.  They classified the possible phases of driven CDWs and showed
that the KV predictions were violated in this case.  In fact, certain
static components of the quenched disorder persist in a coarse-grained
model even at very large velocity.  This static random force arises
physically from spatial inhomogeneities in the impurity distribution
and represents a sort of random drag. Therefore noise in the
nonequilibrium steady state of the sliding CDW never mimics thermal
noise, which is uncorrelated in time.  As a result these authors
concluded that the analogous moving solid phase of a CDW is stable in
$3d$ at large velocities, but does not exhibit the true long-ranged
translational order of an equilibrium $3d$ crystal. Rather it exhibits
the algebraic decay of correlations characteristic of a $2d$
equilibrium crystal (QLRO).  In $d=2$ in contrast the moving CDW
appears unstable to the proliferation of phase slips.

A similar reinvestigation of the driven lattice was recently
undertaken by Giamarchi and Le Doussal (GL)\cite{gl}.  These
authors focused on the behavior of fluctuations in the direction
transverse to that of the driving force.  They pointed out that
periodic static component of the pinning force persist in the
transverse direction at large drives and suggested that motion in the
driven solid occurs along elastically coupled channels parallel to the
direction of motion. These authors assumed that disorder-induced
displacements in the direction parallel to the driving force always
remain bounded and neglected all fluctuations in this direction.  They
also suggested that the signature of this anisotropic sliding state,
the anisotropic moving Bragg Glass (BG), would be the existence of a finite
transverse critical force at zero temperature.

Considerable elements of GL's work are retained in our calculations.
The static periodic transverse force is also non-vanishing in our
model, and leads to glassy channel-like motion at very low
temperatures.  There are, however, several key differences from GLs
work.  First, we give a {\sl model independent} characterization of
the possible phases.  Second, we {\sl derive} our equations of motion
from a microscopic dynamics.  These have a truly {\it non}-equilibrium
form, which cannot be obtained simply by a Galilean boost of the
equilibrium equations of motion.  Third, we argue that the most stable
driven phase is the {\it transverse} smectic, with {\it short-range}
rather than long-range {\it longitudinal} order.  Fourth, we point out
the importance of the {\it permeation mode} in the transverse smectic,
which implies a {\it non-zero transverse linear response at any finite
temperature}.  Lastly, we show that the transverse displacements
themselves are much less glassy than in equilibrium, owing to the
strong enhancement of thermal noise caused by the breakdown of the
fluctuation dissipation theorem.

Recent simulations in $d=2$ have confirmed the anisotropic channel
structure of the sliding state.\cite{moon,ryu}\  These simulations
also indicate that dislocations with Burger's vectors parallel to
the flow are unbound, so that the intra-channel order is liquid-like.
The driven array thus indeed appears more consistent with a transverse
smectic phase than the moving Bragg glass.  

All the aforementioned analytical treatments neglect KPZ-type
nonlinearities ($C$ coefficients in Eq.~\ref{maineq}).  These are
perturbatively irrelevant in the smectic state, as argued above.  They
probably play a role in the behavior of longitudinal correlations, and
may well determine the length scale at which longitudinal dislocations
appear in the moving lattice, i.e., the scale at which a driven
lattice is unstable to the transverse smectic.  Such effects have
recently been studied in a simpler model relevant for CDW motion,
which contains {\sl only} longitudinal degrees of freedom.\cite{chen}\

\subsection{Experiments}

As discussed in the Introduction, there are many physical realizations
of dirty driven periodic media. Among these, the magnetic flux lattice
in type-II superconductors is perhaps the system that has been most
studied experimentally in recent years and where our predictions can
most easily be tested.

The large majority of experimental work has focused on the nonlinear
transport properties of these systems for driving forces near the
zero-temperature depinning threshold. Our work, in contrast, focuses
on the properties of the sliding state well above threshold, where the
velocity-force characteristic approaches a linear form.  In this
regime one is rather interested in the positional and temporal order
of the sliding medium.

The positional correlations in a current-driven magnetic flux-lattice
can be studied directly both by numerical simulations and experiments.
Numerical work is very useful as it can provide both direct real space
images of the sliding lattice as well as quantitative structural
information like the structure factor. Recent simulations of
two-dimensional flux lattices ($d_t =2$, $d_l=0$) by Moon et
al.\cite{moon}\ are in agreement with our finding that the periodicity
of the driven flux lattice {\it along} the direction of motion is
absent, i.e., the sliding lattice is a smectic. Real space images of
the driven lattice show that motion occurs along channels that are
aligned with the direction of the driving force and periodically
spaced in the transverse direction.  Phase slips, however, occur at
the channel boundaries, indicating that the channels are uncorrelated
and the longitudinal structure is liquid-like.  The structure factors
obtained from these simulations show sharp algebraically-divergent
peaks at the reciprocal lattice vectors ${\bf Q}$ normal to the
external drive.  The peaks at the other reciprocal lattice vectors
have a very small intensity that decays exponentially with system
size.  Similar results have also been obtained by other
authors.\cite{ryu,spencer}\ One detailed aspect of the results of
Ref.~\onlinecite{moon}\ which {\sl does not} agree with our
theoretical expectations is the algebraic decay of the smectic Bragg
peaks.  Our theory predicts in fact (possibly stretched) exponential
decay, due to the linear displacement growth in Eq.~\ref{C_Delta}.  We
believe the observed power-law structure factor scaling is likely a
crossover phenomenon, perhaps enhanced by the dispersive elastic
moduli due to the short-scale logarithmic character of the
inter-vortex interaction.

One of the first direct experimental evidence of the ordering of the
sliding flux lattice at large velocities was obtained some time ago by
neutron diffraction\cite{yaron}\  These experiments have not,
however, been able to quantitatively determine the structural
properties of the driven state. More recently, the channel structure
of the driven flux lattice was observed directly by decorating the
current-driven flux array in $NbSe_2$.\cite{marche}\  By
digitizing the decoration images Pardo et al. have also very recently
obtained the structure factor of the driven array that indeed has the
transverse peaks characteristic of a smectic \cite{pardo}.

Temporal order in the driven medium should manifest itself in narrow
band noise (NBN) and mode-locking phenomena.  Both of these have been
studied extensively in the context of charge density waves, but have
not been observed in flux lattices, indicating that the driven flux
lattice  lacks long-range temporal order.  The spectrum of
voltage fluctuations can be probed by applying a dc current $I$ that
yields a driving force $f\sim I$ on the flux lattice.  The local
field induced by flux motion is given by
\begin{equation}
E_i({\bf r},t)=-{\phi_0\over c}\epsilon_{ij}\partial_t\tilde{\phi}_j
  \Big[\rho_0+ \sum_{{\bf Q}}
   \rho_{{\bf Q},0}e^{i{\bf Q}\cdot
  ({\bf x}-\bbox{\tilde{\phi}})}\Big],
\end{equation}
with $\phi_0$ the flux quantum. In a moving solid we expect
$\bbox{\tilde{\phi}}={\bf v}t+\bbox{\phi}$. In a perfect lattice the
local voltage contains therefore oscillatory components at the
frequencies $\omega_{\bf Q} = {\bf Q}\cdot{\bf v}$.  The Fourier
spectrum of voltage fluctuations, defined as,
\begin{equation}
  S(\omega)=\int_{{\bf r},{\bf r'}}\int_t e^{i\omega t}
  \langle E_x({\bf r},t)E_x({\bf r'},0)\rangle,
\end{equation}
will then contain, in addition to a dc component, a sharp fundamental
peak at $\omega_1={\bf Q}_{1}\cdot{\bf v}$, with $Q_1\sim 2\pi/a$, 
and smaller peaks at all the higher harmonics.  This
type of spectrum, usually referred to as NBN, has indeed been observed
in sliding CDWs and is generally regarded as the signature of the
absence of appreciable phase slips in the system.\cite{fleming}\  In
contrast, in a liquid phase we expect the Fourier spectrum to have a
Lorentzian form, centered at the frequency $\omega_1$.  A broad power
spectrum of voltage fluctuations, with an $\omega^{-\alpha}$ decay at
large frequencies, known as broad band noise (BBN), is observed in
CDWs when macroscopic velocity inhomogeneities arising from phase
slips are present in the driven system.\cite{shoboBBN}\  NBN has not
been observed in current-driven flux lattices. This is consistent with
our finding that correlations along the direction of motion are always
liquid-like, indicating longitudinal phase slips are present in the
system.

Systems with temporal LRO should also exhibit {\it complete}
mode-locking.  Mode locking is an interference effect that can occur
in a periodic medium driven by both a dc and an ac force. Keeping
fixed the amplitude and the frequency of the ac force and varying the
amplitude of the dc component, one observes steps in the dc response,
known as Shapiro steps.  These steps arise from mode locking of the
frequency of the applied ac force with the internal oscillation
frequencies $\omega_{{\bf Q}}$ of the periodic medium.  {\it Complete}
mode locking (steps in the dc response that remain constant over some
finite range of dc bias) to an arbitrarily weak external ac drive has
again been observed in CDWs.\cite{monceau}\ Assuming, as our
calculations suggest, that the driven flux lattice has only
short-range longitudinal order (i.e. SRO for wavevectors with non-zero
$\omega_{\bf Q}$), we would expect at best {\it incomplete} mode
locking above a non-zero (perhaps large) threshold ac drive.  For
$(2+1)$-dimensional flux lattices again no complete mode locking is
expected in the smectic state, as the driven lattice only has
quasi-Bragg peaks along the direction in wavevector space
perpendicular to the velocity.  If a longitudinally ordered phase
(``Bragg glass'') were stable in 3d, it ought to exhibit mode locking
and NBN; the available experimental evidence seems not to support this
possibility.

Finally, it has been suggested that a sliding flux lattice will
exhibit a finite threshold force for response to an additional driving
force $f_\perp$ transverse to the mean velocity\cite{gl,moon}\ and no
transverse linear response at zero temperature.  The behavior in a
purely dissipative (overdamped) model at zero temperature is somewhat
non-universal, and indeed, such a transverse critical force is
certainly likely {\sl in many possible phases, including both the
  smectic and Bragg glass (BG)}.  At {\it finite} temperature, a sharper
distinction can be drawn.  A naive extrapolation of the Bragg glass
theory to finite temperature would predict an {\sl exponentially
  small, non-linear} transverse response at finite temperatures,
\begin{equation}
  v_\perp^{\rm BG} \sim \exp \left[ - \left( {f_0 \over f_\perp}
    \right)^{\Upsilon}
    \right],\label{nonlinearIVa} 
\end{equation} 
with $f_0^\Upsilon \propto 1/T$ at low temperatures and small
$f_\perp$.  Eq.~\ref{nonlinearIVa} relies, however, upon two
assumptions: (1) density fluctuations (the permeation mode) can be
neglected and (2) the $T=0$ fixed point is stable.  Our works shows
that both these assumptions are incorrect and invalidates
Eq.~\ref{nonlinearIVa}.  As discussed, in Section VI, we then expect a
velocity-force curve with the crossover behavior shown in Fig.~7.  For
low transverse driving forces there are two small (but nonzero) {\sl
linear} components to the mobility, i.e.
\begin{equation}
  v_\perp^{\rm sm.} \sim \left(\mu_{\rm
      dw.} + \mu_{\rm perm.}(T) \right) f_\perp.
\end{equation} 
The mobility of the density wave, $\mu_{\rm dw.}$, has been estimated
in Sec.VII.  It is activated at low temperatures, and also suppressed
for small sliding (longitudinal) velocities, with the form
\begin{equation}
  \mu_{\rm dw.} =\gamma_R^{-1}\approx \gamma^{-1} e^{-\sigma K_\parallel K_\perp
    \Lambda^2/(\gamma v)^2},
\end{equation}
where $\sigma \propto 1/T$ at very low temperatures (more details are
given in the discussion following Eq.~\ref{renormalizedGamma}).  The
second term is a mobility associated with the permeation mode:
\begin{equation}
  \mu_{\rm perm.}(T) \sim \mu_{0} e^{-U_d/T},
\end{equation}
for $T \ll U_d$.  Although we have not analytically derived the above
exponential form of the mobility, it seems extremely likely on
physical grounds.  Such behavior arises from an activated
concentration of mobile defects (vacancies or interstitials in 2d,
kinks and/or vacancy/interstitial lines in 3d) with finite ``energy''
cost $U_d$.  These flow linearly in response to a driving force with
mobility $\mu_0$.  Additional non-linear behavior will be superimposed
upon this linear term, but it is subdominant at small $f_\perp$.  At
low temperatures these non-linearities sharpen to a threshold-like
feature around a finite $f_{\perp c}$, so that the identification of
an experimental system as either a BG or a smectic by the transverse
``threshold'' or critical current may be misleading.

\subsection{Open questions}

We conclude with a summary of some of the many remaining open
questions.  Eqs.~\ref{maineq}\ provide for the first time a complete
set of hydrodynamic equations to describe nonequilibrium states of
driven periodic media.  A systematic RG (or other) treatment of them
is, however, daunting.  Aside from the obvious algebraic complexity,
the structure of the equations also raises some interesting conceptual
issues.  Naively, since first derivative terms are present in all the
variables, all spatial coordinates seem to scale isotropically, with
dynamical exponent $z=1$.  This power counting, however, is
inconsistent with direct perturbative (and more sophisticated)
calculations of physical quantities.  Preliminary investigations of
this problem suggests novel scaling without unique anisotropy or
dynamical exponents.\cite{unpublished}\ A more modest goal, requiring
only straightforward (if rather tedious) calculations is to extend the
RG treatment of the smectic to include the permeation mode, i.e., the
coupling to density fluctuations, as described by
Eqs.~\ref{smdens}--\ref{smdis}.

A systematic treatment of topological defects in disordered lattices is also
lacking.  Some recent progress along these lines was made recently for
equilibrium elastic glasses in Ref.\onlinecite{DSFrecent} -- it remains to
be seen whether this work can be extended to driven systems.  When
dislocations {\sl are} present, many interesting questions remain at
zero temperature \cite{svglide}.  
A possible plastic depinning transition and its
properties -- the associated scaling, noise, and hysteresis -- are not
presently understood.

Finally, an extremely interesting line of inquiry is to explore how
some of the techniques used here might be extended to inertial models
appropriate for friction and lubrication.  Such systems presumably
are controlled not only by the physics described here, but
also by nontrivial mechanisms of dissipation and possibly chaotic
dynamics.

\acknowledgements

\vskip .2in It is a pleasure to acknowledge discussions with Matthew
Fisher.  L. Radzihovsky acknowledges discussions with John Toner.
L. Balents was supported by the NSF PHY94--07194.  M.C. Marchetti was
supported by the NSF through Grant DMR--9419257.  L. Radzihovsky
acknowledges support by the NSF CAREER award, through Grant
DMR--9625111, and partial support by the A.P. Sloan Foundation.
\vspace{-.2in}

\section*{Appendix A}

In this appendix we carry out the mode-elimination needed to
coarse-grain the driven lattice in a {\em periodic} potential.  Expanding
the $e^{-\tilde{S}_1}$ within the brackets in Eq.~\ref{effectiveS}\ to first
order gives
\begin{eqnarray}
  \delta S_{\rm eff.}^{(1)} & = & - \sum_{\bf x} \int_{{\bf z}t}
  \bigg\langle \left[\hat{u}_{<i}({\bf r},t) + \hat{u}_{>i}({\bf
    r},t) \right] \nonumber \\
& & \,\,\times \tilde{F}_i[{\bf x} + {\bf u}_<({\bf r},t) + {\bf
  u}_>({\bf r},t) + {\bf v} t,{\bf
    z}] \bigg\rangle_{0>} .
\end{eqnarray}
This average can be evaluated by expanding the force ${\bf \tilde{F}}$ in
${\bf u}_>$ and averaging over the fast modes term by term.  However,
in the absence of thermal noise, the only non-zero fast correlator is
the response function, which vanishes by causality at equal times.
Since all the fields in the first order term are at equal time, all
the terms involving {\sl any} ${\bf u}_>$ or $\hat{\bf u}_>$ fields
are zero, and this average simply gives
\begin{equation}
  \delta S_{\rm eff.}^{(1)} = - \sum_{\bf x} \int_{{\bf z}t}
  \hat{u}_{<i}({\bf r},t) \tilde{F}_i[{\bf x} + {\bf u}_<({\bf r},t) +
  {\bf v} t,{\bf z}].
\end{equation}
To leading order, then, the force is unrenormalized.  

The first non-trivial correction arises at second order.  This effect
is actually physically transparent.  At first order the fast modes
simply respond linearly to the adiabatic motion of the slow modes.
However, because of the mode coupling, this response is fed back and
felt {\sl at second order} by the slow modes again.  It is this
feedback that corrects the motion of the slow degrees of freedom, and
is perturbatively estimated in the next correction.  This is given by 
\begin{eqnarray}
  \delta S_{\rm eff.}^{(2)} & = & - {1 \over 2} \sum_{\bf x x'}
  \int_{{\bf zz'}tt'} \bigg\langle \hat{u}_i \hat{u}'_j \nonumber \\
  & & \,\,\times \tilde{F}_i[{\bf
    x} + {\bf u} + {\bf v} t,{\bf z}] \tilde{F}_j[{\bf
    x}' + {\bf u}' + {\bf v} t',{\bf z'}]
  \bigg\rangle_{0>}^c, 
\end{eqnarray}
where we have indicated the arguments of ${\bf u}$ and $\hat{\bf u}$
by the presence or absence of a prime, and furthermore suppressed the
mode decomposition.  The superscript $c$ indicates a cumulant, or
disconnected average, meaning that half of the square of $\delta
S_{\rm eff.}^{(2)}$ is subtracted off, as demanded by the logarithm in
Eq.~(\ref{effectiveS}).  As we saw earlier, only non-equal-time
response functions can survive the average.  This can occur {\sl only}
via the first order expansion of the displacement field out of one of
the force terms, to be contracted against the response field
at the non-equal time.  Taking into account the two ways in which this
can be achieved gives the result
\begin{eqnarray}
  \delta S_{\rm eff.}^{(2)} & = & - \sum_{\bf x x'}
  \int_{{\bf zz'}tt'} \hat{u}_{<i} \partial_k \tilde{F}_i[{\bf
    x} + {\bf u}_< + {\bf v} t,{\bf z}] \nonumber \\
  & & \times \tilde{F}_j[{\bf
    x}' + {\bf u}'_< + {\bf v} t',{\bf z'}] G_{jk}({\bf
    r-r'},t-t'). 
  \label{SecondOrder}
\end{eqnarray}
Here the response function is (in matrix form)
\begin{equation}  
G({\bf q},\omega) = \bigg[ i\omega I + K({\bf q}) \bigg]^{-1}
\theta(|q| - \Lambda).
\end{equation}
The $\theta$-function indicates that only a partial mode elimination
has been performed, so that the slow modes remain as dynamical
variables in the coarse-grained theory.  Generally, however, the
expressions obtained in this section have smooth limits as $\Lambda
\rightarrow 0$, and are well approximated therefore by using the
response function in the full Brillouin zone.

Using the Fourier decomposition, Eq.~\ref{Fourier}, the second order
contribution to the effective action can then be cast into the form
\begin{equation}
  \delta S_{\rm eff.}^{(2)} = - \sum_{\bf x} \int_{{\bf z}t}
  \hat{u}_{<i}  \delta f_i[{\bf u_<},{\bf r},t],
\end{equation}
where $\delta f_i$ may be interpreted as an additional effective force
in the coarse-grained equation of motion for the slow modes ${\bf
  u}_<$.  The dominant terms in $\delta f_i$ are those which do not
oscillate as ${\bf x}$ is varied.  Keeping only these, it takes
the form
\end{multicols}
\begin{equation}
  \delta f_i[{\bf u_<},{\bf r},t] = \sum_{\bf x'} \int_{{\bf
      z'}t'} \sum_{\bf Q} i Q_{ i}Q_{ j}Q_{ k}
  e^{i{\bf Q \cdot(x-x' + u-u' +v(t-t'))}} |U_{\bf
    Q}|^2 G_{jk}({\bf r-r'},t-t').
\end{equation}
This is greatly simplified by gradient expanding the difference in
displacement fields
\begin{equation}
  {\bf u-u'} \approx ({\bf r-r'})^\alpha\partial_\alpha {\bf u} +
  (t-t')\partial_t {\bf u} - {1 \over 2} {\bf (r-r')^\alpha
    (r-r')^\beta} \partial_\alpha\partial_\beta {\bf u}.
  \label{gradient_expand}
\end{equation}
We then obtain an expansion for the force corrections,
\begin{equation}
  \delta f_i = \sum_{\bf x'} \int_{{\bf z'}t'} \sum_{\bf
    Q} i Q_{ i}Q_{ j}Q_{ k} |U_{\bf
    Q}|^2 e^{i{\bf Q}\cdot({\bf x'} + {\bf v} t')}
  G_{jk}({\bf r'},t') {\cal F}[{\bf u}_<],
\end{equation}
where
\begin{eqnarray}
  {\cal F}[{\bf u}] & = & 1 + iQ_{ l}\left[ x^{'m} \partial_m 
    + t'\partial_t  - {1 \over 2} (x^{'m}
    x^{'n}\partial_m\partial_n  +  z^{'a} z^{'b} \partial_a
    \partial_b )\right] u^l, \nonumber \\
  & & - {1 \over 2} Q_{ l} Q_{ m} r^{'\alpha}
    r^{'\beta} \partial_\alpha u^l \partial_\beta u^m + \cdots.
\end{eqnarray}
Including these corrections into the effective action gives
Eq.~\ref{eom}\ in the main text.

\begin{multicols}{2}

\section*{Appendix B}

Here we present the details of mode elimination for a lattice driven
over a {\em disordered} potential.  Expanding $S_1$ from the
exponential ($e^{-\tilde{S}}$) and averaging it perturbatively over
the modes outside of the cut-off gives, to linear order in $S_1$,
\begin{equation}
  \langle S_1 \rangle_> = \delta S_{\rm eff.}^{(1a)} + \delta S_{\rm
    eff.}^{(1b)}, 
\end{equation}
where the first term is, as before
\begin{equation}
  \delta S_{\rm eff.}^{(1a)} = S_1[{\bf u} \rightarrow {\bf u}_<],
\end{equation}
and simply returns the uncorrected bare random force.  The next
correction is 
\begin{eqnarray}
  \delta S_{\rm eff.}^{(1b)} & = & \sum_{\bf x,x'}\int_{{\bf
      zz'}tt'} \hat{u}^i({\bf r},t)  G_{jk}({\bf r-r'},t-t') \nonumber \\
  & & \hspace{-0.5in} \times
  \partial_i\partial_j\partial_k \tilde\Gamma[{\bf r-r'+u(r,t)-u(r',t')
    +v}(t-t')].
\end{eqnarray}

Using Eq.~(\ref{gradient_expand}), this becomes
\begin{eqnarray}
  \delta S_{\rm eff.}^{(1b)} & & = \sum_{\bf x,x'}\int_{{\bf
      zz'}tt'} \hat{u}^i \bigg\{ \tilde\Gamma_{ijkl}[{\bf x'} - {\bf
    v}_0 t',{\bf z'}]\\
  & & \times \bigg[ r^{'\alpha}\partial_\alpha u_l +
  t'\partial_t u_l - {1 \over 2}
  r^{'\alpha}r^{'\beta}\partial_\alpha\partial_\beta u_l \bigg]
  \nonumber \\
  & & \!\!\! + {1 \over 2} \tilde\Gamma_{ijklm}[{\bf x'}+{\bf v} t',{\bf
    z'}] r^{'\alpha}r^{'\beta} \partial_\alpha u_l \partial_\beta u_m
  \bigg\} G_{jk}({\bf r'},t'),\nonumber
\end{eqnarray}
where we have abbreviated $\Gamma_{ij\cdots} =
\partial_i\partial_j\cdots \Gamma$.  This correction to the effective
action again represents gradient terms in the coarse-grained equation
of motion, of the form of Eq.~(\ref{eom}).  Extracting these
coefficients, we obtain the formulae quoted in Section II.B.2 for the
derivative coefficients $\gamma,A,B$ and $C$.

It remains to consider the renormalization of the force term itself.
This vanished in the case of the periodic force, due to the (assumed)
incommensurability of the lattice and the substrate.  The random
potential, however, has Fourier components commensurate with the
moving lattice, which thereby generates such a renormalization.  To
evaluate it, however, we must go to higher order in the disorder
variance $\tilde\Gamma$.  In particular, we consider the second correction term
\begin{equation}
  \delta S_{\rm eff.}^{(2)} = -{1\over 2}\left\langle S_1^2
    \right\rangle_{0>}^{C}.
\end{equation}
\end{multicols}
This is explicitly
\begin{eqnarray}
  \delta S_{\rm eff.}^{(2)} & = & -{1 \over 8}\,\int
  \!\!\!\!\!\!\!\sum_{1-4} \bigg\langle \hat{u}_1^i \hat{u}_2^j
  \tilde\Gamma_{ij}[{\bf x}_{1}\! - \!{\bf x}_{2}\! +\! {\bf u}_1\! -\!
    {\bf u}_2 \!+\! {\bf
    v} (t_1\! -\! t_2),{\bf z}_1\! -\! {\bf z}_2] \nonumber \\ & & \times
  \hat{u}_3^k \hat{u}_4^l 
  \tilde\Gamma_{kl}[{\bf x}_{3}\! -\! {\bf x}_{4} \! +\! {\bf
    u}_3\! -\! {\bf u}_4 \! + \! {\bf
    v} (t_3-t_4),{\bf z}_3\! -\! {\bf z}_4] \bigg\rangle_{0>}^{C},
\end{eqnarray}
where we introduced the obvious abbreviation for the four lattice sums
and longitudinal space and time integrals.  This expectation value
contains several terms, depending upon the number of $\hat{\bf u}$
fields which are contracted to give response functions.  Terms with
one contraction leave 3 response fields, which represents the
generation of a skewness to the distribution of the random force, and
can be neglected in what follows.  Forming three contractions leaves a
single response field, which will give higher-order corrections to the
coefficients determined above, and can thus also be neglected (for
weak disorder).  Forming two contractions leaves two response fields,
which is of the proper form to renormalize the random force.

To obtain these terms, the $\tilde\Gamma$'s must be expanded to second order
in the fast fields ${\bf u}_>$.  This gives
\begin{eqnarray}
  \delta S_{\rm eff.}^{(2)} & = & -{1 \over 8}\,\int
  \!\!\!\!\!\!\!\sum_{1-4} \Bigg[ \tilde\Gamma_{ijm}(12)
  \tilde\Gamma_{kln}(34) \left\langle \hat{u}_1^i \hat{u}_2^j
  (u_{1>}^m-u_{2>}^m) \hat{u}_3^k \hat{u}_4^l
  (u_{3>}^n-u_{4>}^n)\right\rangle _{>0}^c \nonumber \\
& & + \tilde\Gamma_{ijmn}(12) \tilde\Gamma_{kl}(34) \left\langle
  \hat{u}_1^i \hat{u}_2^j 
  (u_{1>}^m-u_{2>}^m) (u_{1>}^n-u_{2>}^n)\hat{u}_3^k \hat{u}_4^l
  \right\rangle _{>0}^c \Bigg],
\end{eqnarray}
where we have now also abbreviated the arguments of the force
correlators.  The contractions inside the angular brackets can still
be performed in several ways.  Each such choice gives rise to a
separate term containing two response functions and a combination of
slow fields at different space-time points.  To determine the desired
correction to the random force correlator, we keep only the leading
term in a gradient expansion of the slow fields (i.e. zeroth order in
the gradients).  Lengthy but simple calculation gives
\begin{equation}
  \delta S_{\rm eff.}^{(2)} = {1 \over 2} \sum_{\bf x,x' }
  \int_{{\bf zz'}tt'} \hat{u}^i({\bf r},t)
  \hat{u}^j({\bf r}',t') \delta\Gamma_{ij}[{\bf
    x-x'+{\bf u_<-u'_<} +{\bf v}}(t-t'),{\bf z-z'}],
\end{equation}
where the renormalization of the force-force correlator is
\begin{eqnarray}
  && \!\!\!\!\!\!\delta \tilde\Gamma_{ij}[{\bf x},{\bf z}]  = 
  \nonumber \\
  & & -\int
  \!\!\!\!\!\!\!\sum_{12} \tilde\Gamma_{ikm}[{\bf x}\! \! -\! {\bf
    x}_{1}\!\! - \! {\bf
    v} t_1,{\bf z}\!\! - \! {\bf z_1}] \tilde\Gamma_{jln}[{\bf x} \!\! +
  \! {\bf x}_{2} \!\! +\! {\bf v} t_2,{\bf z}\!\!+{\bf z}_2]
  G_{kn}(1) G_{lm}(2) \nonumber \\
  & & +2\int
  \!\!\!\!\!\!\!\sum_{12} \tilde\Gamma_{ikm}[{\bf x}\! \! -\! {\bf
    x}_{1}\!\! - \! {\bf
    v} t_1,{\bf z}\!\! - \! {\bf z_1}] \tilde\Gamma_{jln}[{\bf x}_{2}
  \!\! +\! {\bf v} t_2,{\bf z}_2] 
  G_{kn}(1) G_{lm}(2-1) \nonumber \\
  & & + \tilde\Gamma_{ijmn}[{\bf x},{\bf z}] \int
  \!\!\!\!\!\!\!\sum_{12} \tilde\Gamma_{kl}[{\bf x}_{1}\!\!-\!{\bf
    x}_{2}\! +\! {\bf v}(t_1\! -\! t_2),{\bf z}_1\!\! -\! {\bf
    z}_2] G_{km}(1) G_{ln}(2) \nonumber \\
  & & - \tilde\Gamma_{ijmn}[{\bf x},{\bf z}] \int
  \!\!\!\!\!\!\!\sum_{12} \tilde\Gamma_{kl}[{\bf x} \! + \! {\bf
    x}_{1}\!\!-\!{\bf x}_{2}\! +\! {\bf v}(t_1\! -\! t_2),
  {\bf z}\! + \! {\bf z}_1\!\! -\! {\bf
    z}_2] G_{km}(1) G_{ln}(2) \nonumber \\
  & & + \int
  \!\!\!\!\!\!\!\sum_{12} \tilde\Gamma_{ilmn}[{\bf
    x}_{1}\!\! + \! {\bf
    v} t_1,{\bf z_1}] \tilde\Gamma_{kj}[{\bf x} \!\! +
  \! {\bf x}_{2} \!\! +\! {\bf v} t_2,{\bf z}\!\!+{\bf z}_2]
  G_{lm}(1) G_{kn}(2) \nonumber \\
  & & - \int
  \!\!\!\!\!\!\!\sum_{12} \tilde\Gamma_{ilmn}[{\bf
    x}_1\!\! + \! {\bf
    v} t_1,{\bf z_1}] \tilde\Gamma_{kj}[{\bf x} \!\! +
  \! {\bf x}_{2} \!\! +\! {\bf v} t_2,{\bf z}\!\!+{\bf z}_2]
  G_{lm}(1) G_{kn}(2-1).
  \label{toodamnlong}
\end{eqnarray}
Integrating this horrendous equation over ${\bf r}$, one obtains the
formula for $g_{ij}$ quoted in the text.

\begin{multicols}{2}
\section*{Appendix C}

Here we estimate the correlator of the nonequilibrium 
part of the static random force, $g_{ij}$, given in Eq.~\ref{gij_result}\,
in the limit of large sliding velocity $v$.
First we note that as $g_{ij}$ is a symmetric tensor,
it can be written as
\begin{equation}
g_{ij}=g_0 (\delta_{ij}-\hat{v}_{i}\hat{v}_{j})
       +g_1\hat{v}_{i}\hat{v}_{j},
\end{equation}
where ${\bf \hat{v}}={\bf v}/v$ and
\begin{eqnarray}
g_0+g_1= & & \int_{\bf q}q^2q_kq_lq_mq_n|\Gamma({\bf q})|^2
   G_{km}({\bf q},{\bf q}_t\cdot{\bf v})\nonumber\\
 & & \times\big[G_{ln}({\bf q},-{\bf q}_t\cdot{\bf v})
      -G_{ln}({\bf q},{\bf q}_t\cdot{\bf v})\big],
\end{eqnarray}
and
\begin{eqnarray}
g_1=& & \int_{\bf q}({\bf q}_t\cdot{\bf \hat{v}})^2q_kq_lq_mq_n
     |\Gamma({\bf q})|^2
   G_{km}({\bf q},{\bf q}_t\cdot{\bf v})\nonumber\\
 & & \times\big[G_{ln}({\bf q},-{\bf q}_t\cdot{\bf v})
      -G_{ln}({\bf q},{\bf q}_t\cdot{\bf v})\big].
\end{eqnarray}

The physical case of interest here is the one where the range $\xi$
of the pinning potential is small compared to the lattice constant,
$a$. The Fourier transform of the variance of the random potential,
$\Gamma({\bf q})$ is then a very broad function on the scale
of the Brillouin zone. To proceed, we exploit the periodicity
of the elastic propagators in reciprocal space, by letting 
${\bf q}_t={\bf k}+{\bf Q}$, where ${\bf k}$
only spans the first Brillouin zone. The integral over ${\bf q}_t$
is then replaced by an integral over the first Brillouin zone and 
a sum over all reciprocal lattice vectors. Focusing for concreteness
on the evaluation of $g_1$ ($g_0$ can be evaluated by a similar
procedure), we obtain
\end{multicols}
\begin{eqnarray}
g_1=\sum_{{\bf Q}\not= 0} & & \int_{{\bf k}}\int_{{\bf q}_z}
   \big[({\bf k}+{\bf Q})\cdot{\bf \hat{v}}\big]^2
   (k+Q)_k(k+Q)_l(k+Q)_m
   (k+Q)_n |\Gamma({\bf k}+{\bf Q})|^2
\nonumber\\
  & &\times G_{km}({\bf k},q_z,({\bf k}+{\bf Q})\cdot{\bf v})
    \big[G_{ln}({\bf k},q_z,-({\bf k}+{\bf Q})\cdot{\bf v})
      -G_{ln}({\bf k},q_z,({\bf k}+{\bf Q})\cdot{\bf v})\big].
\label{g0}
\end{eqnarray}
We now split the reciprocal lattice vector sum in Eq.~(\ref{g0}) in two
parts by separating out the terms with ${\bf Q}\cdot{\bf v}=0$
and write
\begin{equation}
g_1=g_1^{(1)}+g_1^{(2)}.
\label{splitg0}
\end{equation}
The term $g_1^{(1)}$ denotes the contribution to $g_1$ from the sum
over reciprocal lattice vectors satisfying 
${\bf Q}\cdot{\bf v}\not=0$. In this term we neglect everywhere
$k$ compared to $Q$ and obtain
\begin{eqnarray}
g_1^{(1)}\approx \sum_{{\bf Q}\cdot{\bf v}\not=0}& &
  ({\bf Q}\cdot{\bf \hat{v}})^2
  Q_{ k}Q_{ l}Q_{ m}Q_{ n}|\Gamma({\bf Q})|^2
   \int_{{\bf k},q_z}  
   G_{km}({\bf k},q_z,{\bf Q}\cdot{\bf v})\nonumber\\
 & &   \big[G_{ln}({\bf k},q_z,-{\bf Q}\cdot{\bf v})
      -G_{ln}({\bf k},q_z,{\bf Q}\cdot{\bf v})\big].
\label{g0one}
\end{eqnarray}
\begin{multicols}{2}
In the limit $v>>2\pi\gamma/(ac)$, with $c$ a typical elastic constant,
we can now approximate the elastic propagators in Eq.~(\ref{g0one}) 
by neglecting
the in-plane elastic matrix compared to the frequency 
$\gamma{\bf Q}\cdot{\bf v}$, i.e.,
\begin{equation}
G_{km}({\bf k},q_z,{\bf Q}\cdot{\bf v})\approx
   {\delta_{km}\over c_{44}q_z^2+i\gamma{\bf Q}\cdot{\bf v}}
\label{propiso}
\end{equation}
By inserting Eq.~(\ref{propiso}) into Eq.~(\ref{g0one}),
we obtain
\begin{eqnarray}
g_1^{(1)}\approx 2\rho_0\sum_{{\bf Q}\cdot{\bf v}\not=0}& &
    ({\bf Q}\cdot{\bf \hat{v}}_0)^2 Q^4
     |\Gamma(Q)|^2 \nonumber\\
   & & \times\int_{q_z}
    {(\gamma{\bf Q}\cdot{\bf v})^2\over
    [(c_{44}q_z^2)^2+(\gamma{\bf Q}\cdot{\bf v})^2]^2}.
\end{eqnarray}
For $d_l=0$ this gives
\begin{eqnarray}
g_1^{(1)} & & \approx {2\rho_0\over \gamma^2v^2}\sum_{{\bf Q}\cdot{\bf 
v}_0\not=0}
        Q^4|\Gamma(Q)|^2 \nonumber\\
& & \approx \Big({\Gamma(Q=0)\over\gamma v}\Big)^2{1\over\xi^6}
    \approx\Big({\Delta\over\gamma va}\Big)^2,
\end{eqnarray}
where $\Delta\approx\Gamma(Q=0)a/\xi$ is the variance of the equilibrium part
of the static pinning force defined in Eq.~\ref{delta_eq}.
For $d_l=1$ the $q_z$-integral is easily performed, with
the result,
\begin{eqnarray}
g_1^{(1)}& &\approx {3\rho_0\over 8\sqrt{2c_{44}}}{1\over(\gamma v)^{3/2}}
       \sum_{{\bf Q}\cdot{\bf v}\not=0}
        |Q_{ x}|Q^4|\Gamma(Q)|^2\nonumber\\
 & & \approx {\Delta^2\over (a\gamma v)^{3/2}\sqrt{c_{44}a\xi^2}}.
\end{eqnarray}

In the contribution $g_1^{(2)}$ from the ${\bf Q}\cdot{\bf v}=0$ of
the reciprocal lattice vector sum the integral over ${\bf k}$ is
dominated by $k$ near the center of the Brillouin zone. In this term
we can therefore approximate the elastic propagators by their long
wavelength form, given in Eqs.~\ref{propl}\ and \ref{propt}\ . After
some lengthy algebra, one can show that in the limit of large sliding
velocities, $v>>2\pi\gamma/(ac_{66})$, $g_1^{(2)}\sim (\xi/a)g_1^{(1)}$ 
and is therefore negligible for short-ranged pinning potentials.

The evaluation of $g_0$ can be performed by the same method, with the result
\begin{eqnarray}
& & g_0\sim g_1{a\over\xi},\hskip 0.2in d_l=0,\nonumber\\
& & g_0\sim g_1\ln(a/\xi),\hskip 0.2in d_l=1.
\end{eqnarray}

In summary, we find that at large sliding velocities both  components
$g_0$ and $g_1$ of the correlator
of the nonequilibrium part of the static random force have the same 
asymptotic dependence on $v$ and the disorder strength $\Delta$, with
\begin{equation}
g_{0,1}\approx {\Delta^2\over v^{(2d_t-d_l)/2}}. 
\end{equation}

\section*{Appendix D}

Here we examine the predictions of the  perturbation theory described
in Section III for the real-space decay of positional correlations in a
$d_t$-dimensional lattice of magnetic flux lines ($d_l=1$), 
driven in the $x$ direction. To obtain the mean-square displacement
in real space, we need to evaluate integrals of the form,
\begin{equation}
B({\bf r})=2\int^{'}_{{\bf q}_t,q_z}
     {1-\cos({\bf q}\cdot{\bf r}) \over
     (\gamma vq_x)^2+[cq_t^2+c_{44}q^2_z]^2},
\end{equation}
where $c$ stands for either $c_{66}$ or $c_{11}+c_{66}$
and the prime denotes a cutoff at $|q_\perp|=\Lambda$.
The integrals over $q_x$ and $q_z$ are easily performed. Letting
$u=q_\perp x_\perp$, one obtains,
\begin{equation}
B({\bf r})={|y|^{2-d_t}\over \sqrt{c c_{44}}\gamma v}
         {\cal F}_1^{(d_t)}(s,\zeta,\Lambda|y|),
\end{equation}
where $s=|x|c/(v\gamma y^2)$ and
$\zeta=\sqrt{c/c_{44}}|z|/|y|$. The scaling function 
${\cal F}_1^{(d_t)}(s,\zeta,\Lambda|y|)$ is given by
\begin{eqnarray}
{\cal F}_1^{(d_t)}& &(s,\zeta,\Lambda|x_\perp|)=
  \int{d{\bf \hat{u}}\over(2\pi)^{d_t -1}} \int_0^{\Lambda y}
   u^{d_t-3}
  \bigg\{1-\cos(\hat{\bf y}\cdot{\bf u})\nonumber\\
  & & \times\Big[\cosh(\zeta u)
  -{1\over 2}e^{-\zeta u}\Phi\Big(\sqrt{s} u-{\zeta\over 
2\sqrt{s}}\Big)\nonumber\\
 & & -{1\over 2}e^{\zeta u}\Phi\Big(\sqrt{s} u+{\zeta\over 2\sqrt{s}}\Big)
  \Big]\bigg\}.
\end{eqnarray}
Here ${\bf u}$ is a $(d_t-1)$-dimensional vector, with
${\bf \hat{u}}={\bf u}/u$, and $\Phi(x)$
is the error function. We are interested in the asymptotic behavior of the 
scaling function
for $d_t =2$. For $\zeta=0$ we find
\begin{eqnarray}
& & {\cal F}^{(2)}_1(s\rightarrow 0,0,\Lambda|y|)\sim \ln(\Lambda 
|y|),\nonumber\\
& & {\cal F}^{(2)}_1(s>>1,0,\Lambda|y|)\sim 
       \ln\Big({c\Lambda^2|x|\over\gamma v}\Big).
\end{eqnarray}
For $\zeta>>1$, or $|z|>>|y|\sqrt{c_{44}/c}$, 
\begin{eqnarray}
& & {\cal F}^{(2)}_1(s\rightarrow 0,0,\Lambda|y|)\sim 
    \ln\Big({|z|\over\Lambda}\sqrt{c\over c_{44}}\Big),\nonumber\\
& & {\cal F}^{(2)}_1(s>>1,0,\Lambda|y|)\sim 
    \ln\Big({|x|c\over v\gamma y^2}\Big).
\end{eqnarray}
The scaling of the mean square displacement is therefore anisotropic,
but logarithmic in all directions.

\section*{Appendix E}

In this appendix we outline the details of the $2+\epsilon$ RG
calculation for the single Fourier mode driven smectic model
defined by Eq.\ref{eom_singlemode} in Sec.\ref{sec:RGb} of the main
text. It is convenient to employ the Martin-Siggia-Rose (MSR)
formalism\cite{MSR}. In this formalism one studies the dynamic
generating functional $Z$ which is a trace over the displacements
$\phi({\bf r},t)$, with the constraint that $\phi({\bf r},t)$
satisfies the equation of motion Eq.\ref{eom_singlemode}, imposed via a
functional $\delta$-function as an integral over a response field
$\hat{\phi}({\bf r},t)$. Averaging over the noise $\eta({\bf r},t)$
and the quenched random force $F_p[\phi({\bf r},t),{\bf r}]$, the
problem can be recast in the form of a dynamical field theory,
\begin{equation}
Z = \int [d\hat{\phi}\;d \phi] e^{-S_0[\hat{\phi},\phi]-
S_1[\hat{\phi},\phi]}\;,
\end{equation}
where in addition to the standard quadratic part of the action $S_0$
\begin{eqnarray}
\hspace{-.1in}S_0 &=& \int_{{\bf r},t}\bigg[\hat{\phi}({\bf r},t)
\left\{\gamma(\partial_t + v\partial_x)
-(K_\parallel\partial_x^2 + K_\perp\nabla_\perp^2)\right\}\phi({\bf 
r},t)\nonumber\\
&-&\gamma T\hat{\phi}({\bf r},t)^2\bigg]\;,
\label{S0}
\end{eqnarray}
there is a contribution $S_1$ due to disorder
\begin{equation}
S_1=-{1\over2}\int_{{\bf r},t,t'}
\hat{\phi}({\bf r},t)\hat{\phi}({\bf r},t')
\Delta_1\cos[q_0(\phi({\bf r},t)-\phi({\bf r},t'))]\;.
\label{Sp}
\end{equation}
In the above, after averaging over disorder we have kept only the most
relevant lowest Fourier component of the random force correlation
function. We have also deformed the
functional integral contour over the response field $\hat{\phi}$ to the
imaginary axis.

We employ the standard momentum shell renormalization group
transformation\cite{Wilson}, by writing the displacement field as
$\phi({\bf r},t)=\phi^<({\bf r},t)+\phi^>({\bf r},t)$, integrating
perturbatively in $\Delta_1$ the high wavevector field $\phi^>({\bf r},t)$
nonvanishing for $\Lambda e^{-l}<q_\perp<\Lambda$ (with no cutoff
on the momentum along the direction of motion $q_x$ and on $\omega$),
and rescaling the lengths, the time, and long wavelength part of the
fields with
\begin{mathletters}
\begin{eqnarray}
{\bf x}_\perp&=&{\bf x'}_\perp e^{l}\;,\label{rescalings_f}\\
x&=&x'e^{\zeta l}\;,\\
t&=&t'e^{zl}\;,\\
\phi^<({\bf r},t)&=&e^{\chi l}\phi({\bf r'},t')\;,\\
\hat{\phi}^<({\bf r},t)&=&e^{\hat{\chi}l}\hat{\phi}({\bf r'},t')\;,
\label{rescalings_l}
\end{eqnarray}
\end{mathletters}
so as to restore the ultraviolet cutoff back to $\Lambda$.  Because
the random-force correlator $\Delta_1$ term is a periodic function of
$\phi$, it is convenient (but not necessary) to take the arbitrary field
dimension $\chi=0$, thereby preserving the period $2\pi/q_0$ under the
renormalization group transformation. Under this transformation the
resulting effective free energy functional can be restored into its
original form Eqs.\ref{S0}--\ref{Sp} with the effective
$l$-dependent couplings. To zeroth order we obtain:
\begin{mathletters}
\begin{eqnarray}
{d\gamma(l)\over dl}=(d_\perp+\zeta+\hat{\chi})\gamma(l)
\label{a}\\ 
{d\gamma v(l)\over dl}=(d_\perp+z+\hat{\chi})\gamma v(l)\label{b}\\ 
{d K_\perp(l)\over
dl}=(d_\perp+\zeta+z-2+\hat{\chi})K_\perp(l)
\label{c}\\ 
{d K_\parallel(l)\over dl}=(d_\perp-\zeta+z+\hat{\chi})K_\parallel(l)
\\
{d\Delta_1(l)\over dl}=(d_\perp+\zeta+2z+2\hat{\chi})\Delta_1(l)\\ 
{d T\gamma(l)\over dl}=(d_\perp+\zeta+z+2\hat{\chi})T\gamma(l)
\end{eqnarray}
\end{mathletters}

As in the calculation of Sec.\ref{sec:RGa} statistical symmetry under an
arbitrary time-independent shift of the displacement field $\phi({\bf
r},t)\rightarrow \phi({\bf r},t)+f({\bf r})$ (time-translational
invariance) guarantees that $\gamma v$, $K_\perp$ and $K_\parallel$ do not
acquire any graphical corrections, i.e. their flow equations above are
{\em exact}. Imposing this requirement at the tree level on the first
two coefficients, and using Eq.\ref{a} we obtain $\zeta=2$ and
$\hat{\chi}=-d_\perp-z$.

More generally but equivalently we look at the dimensionless coupling
constants
\begin{eqnarray}
\overline{T}&\equiv&{T\over 2(K_\perp K_\parallel)^{1/2}}
C_{d-1}\Lambda^{d-2}\label{overlineT}\\
\overline{\Delta}_1&\equiv&{\Delta_1\over K_\perp\gamma|v|}
C_{d-1}\Lambda^{d-3}\label{overlineDelta}\;,
\end{eqnarray}
which have tree-level flow equations
\begin{eqnarray}
{d\overline{T}\over dl}&=&(2-d)\overline{T}(l)\\
{d\overline{\Delta}_1\over dl}&=&(3-d)\overline{\Delta}_1(l)\;,
\end{eqnarray}
whose flow is independent of the arbitrary choice of rescaling
exponents appearing in Eqs.\ref{rescalings_f}-\ref{rescalings_l}. 

We now proceed to higher order in $\Delta_1$, perturbatively
integrating the short length modes $\hat{\phi}^>({\bf r},t)$ and
$\phi^>({\bf r},t)$
\begin{eqnarray}
Z&=&\int[d\hat{\phi}^<d \phi^<]\;e^{-S_0[\hat{\phi}^<,\phi^<]}
\int[d\hat{\phi}^>d \phi^>]\;e^{-S_0[\hat{\phi}^>,\phi^>]}\nonumber\\
&\times&\left[1-S_1[\hat{\phi},\phi]+{1\over2}S_1[\hat{\phi},\phi]^2
+\ldots\right]\;,\\
&\equiv&\int[d\hat{\phi}^<d \phi^<]\;e^{-S_0[\hat{\phi}^<,\phi^<]-
\delta S[\hat{\phi}^<,\phi^<]}\;,
\label{deltaS}
\end{eqnarray}
where the graphical correction $\delta S$ to the action (dropping an
unimportant constant) is
\begin{equation}
\delta S[\hat{\phi}^<,\phi^<]=\langle
S_1[\hat{\phi},\phi]\rangle_>-{1\over2}\langle
S_1[\hat{\phi},\phi]^2\rangle_>^c+\ldots\;,
\end{equation}
where the subscript ``c'' means cumulant average, and the averages are
performed with the quadratic action with correlation and response
functions, $C({\bf q},\omega)V\equiv\langle \phi({\bf
q},\omega)\phi(-{\bf q},-\omega)\rangle$, $G({\bf
q},\omega)V\equiv\langle \phi({\bf q},\omega)\hat{\phi}(-{\bf
q},-\omega)\rangle$, respectively given by
\begin{eqnarray}
C({\bf q},\omega)&=&{2T\gamma\over \gamma^2(\omega - v
q_x)^2+(K_\perp q_\perp^2+K_\parallel q_x^2)^2}\;,
\label{G11}\\
G({\bf q},\omega)&=&{1\over i\gamma(\omega-v q_x)+
(K_\perp q_\perp^2+K_\parallel q_x^2)}\;,
\label{G12}
\end{eqnarray}
which can be read off from Eq.\ref{S0}. Although naively one would
expect from Eq.\ref{T} that as in the equilibrium problem for $d>2$
temperature is an irrelevant variable, as we demonstrate below
(consistent with the functional RG treatment of the Sec.\ref{finite_T_RG}
this zero temperature fixed point is destabilized by the finite
velocity motion. We will therefore work at a finite temperature.

The first order correction $\langle S_1[\hat{\phi},\phi]\rangle_>$
contributes to the renormalization of $\gamma$, $T\gamma$ and $\Delta_1$,
which we designate as $\delta\gamma^{(1)}$, $\delta(T\gamma)^{(1)}$,
and $\delta\Delta^{(1)}_1$ and illustrate graphically in
Fig.~8. 
\begin{figure}[bth]
{\centering
\setlength{\unitlength}{1mm}
\begin{picture}(50,20)(0,0)
\put(-13,-50){\begin{picture}(150,0)(0,0)
\includegraphics{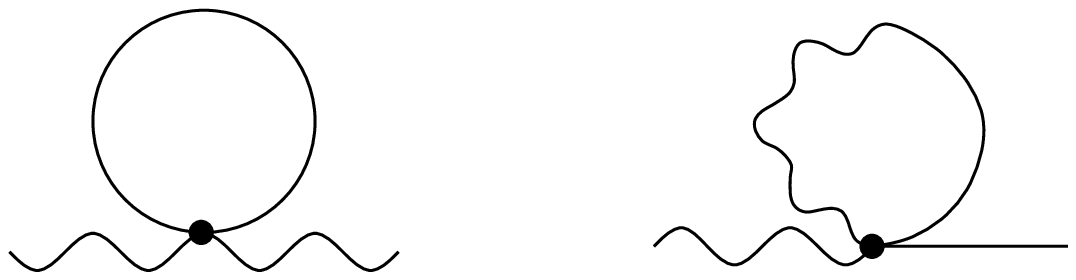}
\end{picture}}
\end{picture}}

\noindent{Fig.~8: Two diagrams that contribute to the renormalization of
$\Delta_1$, $T\gamma$, and $\gamma$. The full line corresponds to the
correlator $C^>$, the full-wiggle line is the response function
$G^>$, wiggly line is the $\hat{\phi}$ field and the vertex is the
$S_1$ nonlinearity. The first diagram, is proportional to $T$, and is
the graphical correction to $\Delta_1$ and $T\gamma$, while the second
one, survives even at zero temperature and renormalizes $\gamma$.}
\label{fig1}
\end{figure}
Expanding to quadratic order in the short-scale fields
$\hat{\phi}^>$ and $\phi^>$ and averaging we obtain
\begin{eqnarray}
\langle\hspace{-.15in} &&S_1\rangle_>= S_1[\hat{\phi}_<,\phi_<]\nonumber\\
&+&{{\Delta}_1\over 2}\int_{{\bf r},t,t'} 
\bigg[{q_0^2\over2}\hat{\phi}^<_{{\bf
r},t}\hat{\phi}^<_{{\bf r},t'} \cos[q_0(\phi^<_{{\bf r},t}-\phi^<_{{\bf
r},t'})] \langle(\phi^>_{{\bf r},t}-\phi^>_{{\bf
r},t'})^2\rangle_>\nonumber\\ 
&+&2q_0\hat{\phi}^<_{{\bf r},t}
\sin[q_0(\phi^<_{{\bf r},t}-\phi^<_{{\bf r},t'})]
\langle \phi^>_{{\bf r},t} \hat{\phi}^>_{{\bf r},t'}\rangle_>\bigg]\;,\\
&\approx&\big(1-q_0^2 C^>({\bf r= 0},t=0)\big)
S_1[\hat{\phi}_<,\phi_<]\nonumber\\
&-&{1\over2}q_0^2\Delta_1\int_{{\bf r},t}
\hat{\phi}^<_{{\bf r},t}\hat{\phi}^<_{{\bf r},t}\int_{\delta t}
C^>({\bf r=0},\delta t)\nonumber\\ 
&+&q_0^2\Delta_1\int_{{\bf r},t}
\hat{\phi}^<_{{\bf r},t}\partial_t \phi^<_{{\bf r},t}
\int_{\delta t} \delta t G^>({\bf r=0},\delta t)\;,
\label{Sp1}
\end{eqnarray}
where in above we used causality (selecting the discretization with
$\theta(0)\equiv0$) and took advantage of the fact that the
correlations functions are short-range in time to perform small time
gradient expansion. Performing above integrals over $\delta t$ (best
evaluated in Fourier $\omega$ space) and noting that the terms in last
part of the Eq.\ref{Sp1} renormalize $\Delta_1$, $T\gamma$ and $\gamma$,
respectively, we find
\begin{eqnarray}
\delta\Delta_1^{(1)}&=&-\Delta_1 q_0^2 C^>({\bf r=0},\delta 
t=0)\nonumber\\
&=&-\Delta_1{T q_0^2\over2(K_\parallel K_\perp)^{1/2}}C_{d-1}\Lambda^{d-2}dl
\;,
\label{Delta1_appE}\\
\delta(T\gamma)^{(1)}&=&{1\over2}\Delta_1 q_0^2 C^>({\bf r=0},\omega=0)\nonumber\\ 
&=&\Delta_1{T q_0^2\over 2 K_\perp |v|}C_{d-1}\Lambda^{d-3}dl\;
\label{Tgamma1}\\
\delta\gamma^{(1)}&=&\Delta_1 q_0^2 i \partial_\omega G^>({\bf
r=0},\omega=0)\nonumber\\
&=&\Delta_1{2K_\parallel q_0^2\over \gamma^2|v|^3}C_{d-1}\Lambda^{d-1}dl\;.
\label{gamma1}
\end{eqnarray}

The calculation to second order in $\Delta_1$ given by $\delta
S_2\equiv-{1\over2}\langle S_1^2\rangle_>^c$ can of course be done
directly, however, it is convenient to utilize the functional
renormalization group calculation of Sec.\ref{RGzeroT}. There we found
for an arbitrary force correlation function $\Delta(\phi)$
\begin{eqnarray}
\delta S_2&=&{1\over2}\int_{{\bf r},t,t'}\hspace{-.15in} 
\hat{\phi}^<_{{\bf r},t}\hat{\phi}^<_{{\bf r},t'}
\Delta''(\phi^<_{{\bf r},t}- \phi^<_{{\bf r},t'})\nonumber\\
&&\hspace{.2in}\times\left(\Delta(\phi^<_{{\bf r},t}- 
\phi^<_{{\bf r},t'})-\Delta(0)\right)
{C_{d-1}\Lambda^{d-3}\over 2 K_\perp\gamma|v|}dl\;,
\label{S2i}
\end{eqnarray}
which when applied to the lowest harmonic $\Delta(\phi)=\Delta_1\cos[q_0
\phi]$ gives
\begin{equation}
\delta S_2=\Delta_1^2 q_0^2\int_{{\bf r},t,t'}\hspace{-.15in} 
\hat{\phi}^<_{{\bf r},t}\hat{\phi}^<_{{\bf r},t'}
\bigg[\cos[\phi^<_{{\bf r},t}- \phi^<_{{\bf r},t'}]-{1\over2}\bigg]
{C_{d-1}\Lambda^{d-3}\over 4 K_\perp\gamma|v|}dl\;,
\label{S2ii}
\end{equation}
The diagram that leads to the above contribution to $\delta S_2$,
renormalizing the 0th and 1st harmonics, i.e. $\Delta_0$, and
$\Delta_1$, respectively, is illustrated in Fig.9
\begin{figure}[bth]
{\centering
\setlength{\unitlength}{1mm}
\begin{picture}(50,30)(0,0)
\put(-20,-72){\begin{picture}(150,0)(0,0)
\includegraphics{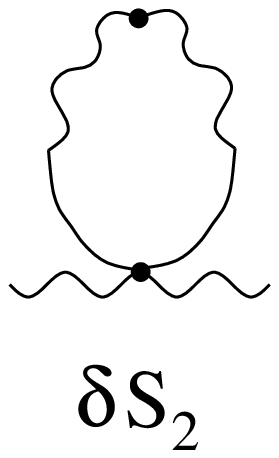}
\end{picture}}
\end{picture}}

{Fig.~9: Diagram that contributes to the renormalization of
$\Delta(u)$, to second order in $\Delta(u)$, i.e. renormalizes the 0th
and 1st harmonics, $\Delta_0$ and $\Delta_1$, respectively with the
same notation as in Fig.8.}
\label{fig2}
\end{figure}
We therefore have to second order in $\Delta_1$ 
\begin{eqnarray}
\delta\Delta_0^{(2)}&=&{\Delta_1^2 q_0^2 C_{d-1}\Lambda^{d-3}
\over 4 K_\perp\gamma|v|}dl\;\\
\delta\Delta_1^{(2)}&=&-{\Delta_1^2 q_0^2 C_{d-1}\Lambda^{d-3}
\over 2 K_\perp\gamma|v|}dl\;,
\label{Delta2}
\end{eqnarray}
where $\Delta_0$ is the zeroth harmonic of the force-force correlator,
i.e. the $\phi$-independent correlator of the random drag. Combining
Eqs.\ref{Delta2} with the zeroth (trivial dimensional rescaling) and
first order results of Eqs.\ref{b},\ref{Delta1_appE}--\ref{gamma1} and
rewriting the flow equations for the dimensionless couplings
$\overline{\Delta}_0$, $\overline{\Delta}_1$, and $\overline{T}$ defined by
Eqs.\ref{overlineT}--\ref{overlineDelta}, we obtain the RG flow equations
quoted in the main text, Sec.\ref{sec:RGb}, Eqs.\ref{Delta0}--\ref{T}.

\section{Appendix F}

In this appendix we discuss the hydrodynamic equations for the driven
smectic of lines (illustrated in Fig.~4) that may be
obtained in a three-dimensional superconductor ($d_t=2$, $d_l=1$). 
This smectic of lines has qualitatively new properties as compared to the
smectic of point particles described in Section IV.
As discussed by Marchetti and Nelson \cite{mcmdrn}, in an isotropic
flux-line liquid the conserved variables associated with hydrodynamic modes
are the density field $\rho({\bf r})$ and the two components of a tangent
field density $\bbox\tau({\bf r})=(\tau_x,\tau_y)$, 
describing the instantaneous
bending of the lines away from the direction of the external field
($z$ direction). Since flux lines cannot start or stop inside the sample, 
the density and the two components of the tilt field are not independent 
dynamical variables, but are related by a continuity equation in the time-like variable $z$,
\begin{equation}
\partial_z\delta\rho+\bbox{\nabla}_t\cdot\bbox{\tau}=0.
\label{cont_z}
\end{equation}
This is simply the condition of no magnetic monopoles.  The line
smectic retains some degree of periodicity along the transverse
direction $y$. This broken symmetry is described by the layer
displacement $\phi_y=\phi$. In addition both the conserved density and
one component of the conserved tilt density are associated with
independent hydrodynamic modes since they are not slaved to the layer
displacement field.  Having identified the relevant hydrodynamic
variables for the line smectic as a one-dimensional layer displacement
$\phi$, a density $\rho$ and a tilt density $\bbox{\tau}$, related by
Eq.~\ref{cont_z}, we now proceed to construct the phenomenological
hydrodynamic equations for the line smectic and to study the spectrum
of the hydrodynamic modes of this system.

The continuum hydrodynamic free energy for the overdamped line smectic
is given by
\begin{eqnarray}
\label{fsmec_lines}
{\cal F}_{ls}={1\over 2}\int_{{\bf r}}\Big\{& &
  c_L\Big({\delta\rho\over\rho_0}\Big)^2
  +c_{44}\Big({\bbox{\tau}\over\rho_0}\Big)^2
  +c_{11}^y (\partial_y\phi)^2 \nonumber\\
& &  +K_1^x(\partial_x\phi)^2+K_1^z(\partial_z\phi)^2
  +2 K_2(\partial_y\phi){\delta\rho\over\rho_0}\Big\},\nonumber\\
\end{eqnarray}
where $\delta\rho=\rho-\rho_0$, with $\rho_0$ the equilibrium
density. Here $c_L$ and $c_{44}$ are the smectic bulk and tilt moduli,
respectively,
$c_{11}^y$ is the in-layer compressibility,
$K_1^x$ and $K_1^z$ are layer bending stiffnesses. The coupling constant $K_2$ 
has dimensions of an elastic constant.
The hydrodynamic equations of the driven smectic contain additional
nonequilibrium terms, as compared to their equilibrium counterpart.
The nonequilibrium terms can be constructed by preserving two important 
symmetries of the driven system, the invariance under inversions
about the direction of the external drive ($y\rightarrow -y$,
$\phi\rightarrow -\phi$) and the broken
translational invariance in the $y$ directions 
($y\rightarrow y+a$, $\phi\rightarrow \phi+a$).

Density and tilt density conservation requires the density and the tilt
field to satisfy continuity equations \cite{mcmdrn},
\begin{eqnarray}
& & \partial_t\delta\rho+\bbox{\nabla}\cdot{\bf j}=0,\label{dens_cont}\\
& & \partial_t\tau_i+\partial_jJ_{ij}=\partial_zj_i,\label{tilt_cont},
\end{eqnarray}
where ${\bf j}$ is the number current density and $J_{ij}$ is the
antisymmetric tilt flux tensor. The density and tilt density fields
are also related by the ``continuity'' equation \ref{cont_z}.  The
equation for the layer displacement has the same structure as that for
the two-dimensional lattice, Eq.~\ref{smecphi}, and is repeated here
for completeness
\begin{equation}
\label{smecphi_line}
(\partial_t+v\partial_x)\phi={j_y\over \rho}
  -{\Gamma_0\over\rho_0}{\delta {\cal F}_{ls}\over\delta\phi},
\end{equation}
with $\Gamma_0$ a kinetic coefficient.
We will not discuss here the role of disorder on the hydrodynamics
of the line smectic. Therefore we have not included any pinning
force in the equations of motion.
The hydrodynamic equations need to be supplemented by constitutive 
relations for the current
flux ${\bf j}$ and the tilt flux $J_{ij}$. For simplicity we only 
consider here local hydrodynamics, but the nonlocality of the elastic 
constants that is often important in flux-line systems can be trivially
incorporated. The constitutive equations for the {\it driven}
line smectic contain, however, new nonequilibrium terms not present
in their equilibrium counterpart discussed in Ref. \onlinecite{mcmdrn}.
The two components of the current density are given by
\begin{eqnarray}
& & j_x=(v+v_1)\delta\rho+\rho_0v_2\partial_y\phi
   -a_1\partial_z\tau_x\nonumber\\
  & & \hspace{0.8in} -\rho_0\Gamma_1\bigg(\partial_x{\delta {\cal F}_{ls}
         \over\delta\rho}
   -\partial_z{\delta {\cal F}_{ls}\over\delta\tau_x}\bigg),
 \label{currentx_line}\\
& & j_y=\rho_0v_3\partial_x\phi -a_2\partial_z\tau_y
   -\rho_0\Gamma_2\bigg(\partial_y{\delta {\cal F}_{ls}
      \over\delta\rho}
       -\partial_z{\delta {\cal F}_{ls}\over\delta\tau_y}\bigg).
\label{currenty_line}
\end{eqnarray}
The antisymmetric tilt flux tensor is written as
\begin{equation}
J_{ij}=\epsilon_{ij}\Big[v_4{\bf \hat{z}}\cdot({\bf \hat{v}}_0\times
  \bbox{\tau}) +\rho_0\Gamma_\tau{\bf \hat{z}}\cdot\Big(
  \bbox{\nabla}_t\times{\delta{\cal F}_{ls}\over\delta\bbox{\tau}}\Big)\Big]
\end{equation}
All the parameters $v_i$ and $a_i$ entering the nonequilibrium terms
are proportional to the mean velocity $v$.
Since the longitudinal part of the tilt vector can be eliminated
in favor of the density using 
Eq.~\ref{cont_z}, it is convenient to work in Fourier space.
We introduce longitudinal and transverse components of the tilt vector
as
\begin{equation}
\bbox{\tau}({\bf q})={\bf \hat{q}}_t\tau_l({\bf q})+
  ({\bf \hat{z}}\times{\bf \hat{q}}_t)\tau_T({\bf q}),
\end{equation}
with ${\bf \hat{q}}_t={\bf q}_t/q_t$. Then 
$\tau_L={\bf \hat{q}}_t\cdot\bbox{\tau}$ and
$\tau_T=({\bf \hat{z}}\times{\bf \hat{q}}_t)\cdot\bbox{\tau}$.
By inserting the constitutive equation for the fluxes in 
Eqs. (\ref{dens_cont}), (\ref{tilt_cont}) and (\ref{smecphi_line}), 
we obtain,
\end{multicols}
\begin{equation}
\Big[\partial_t-i\tilde{v}_1q_x+ D_1q_x^2+D_2q_y^2
   -\big(\hat{q}_x^2D_6+\hat{q}_y^2D_7\big)q_z^2 \Big]\delta\rho =
    \rho_0\tilde{v}_2q_xq_y\phi 
    +(D_6-D_7)\hat{q}_x\hat{q}_yq_zq_t\tau_T,
\label{smdens_line}
\end{equation}
\begin{equation}
\Big[\partial_t- i\tilde{v}_3q_x+D_3q_x^2+D_4q_y^2+D_8q_z^2\Big]\phi= 
   -iq_y\big(D_5-{q_z^2\over q_t^2}D_7\big){\delta\rho\over\rho_0}
    -iq_z\hat{q}_x{D_7\over\rho_0}\tau_T
\label{smdis_line}
\end{equation}
\begin{eqnarray}
\Big[\partial_t+iv_4q_x+D_9q_t^2-\big(D_6\hat{q}_x^2+ & &D_7\hat{q}_y^2\big)
    q_z^2\Big]\tau_T =
    \rho_0q_zq_t(v_2\hat{q}_y^2-v_3\hat{q}_x^2)\phi\nonumber\\
& &  + i(\tilde{v}_1+v_4)\hat{q}_yq_z\delta\rho
     +q_z^2(D_6-D_7)\hat{q}_x\hat{q}_y{q_z\over q_t}\delta\rho.
\label{tilt_transv}
\end{eqnarray}
\begin{multicols}{2}
Finally, the longitudinal part of the tilt density is simply 
related to the density,
\begin{equation}
\tau_L=-{q_z\over q_t}\delta\rho
\label{constraint}
\end{equation}
The ``velocities'' $\tilde{v}_1$, $\tilde{v}_2$, $\tilde{v}_3$ have been
defined as
\begin{mathletters}
\begin{eqnarray}
& & \tilde{v}_1=v+v_1,\\
& & \tilde{v}_2=v_2+v_3,\\
& & \tilde{v}_3=v-v_3
\end{eqnarray}
\end{mathletters}
The coefficients
$D_i$ have dimensions of diffusion constants and are given by
\begin{mathletters}
\begin{eqnarray}
& & D_1=\Gamma_1c_L/\rho_0,\\
& & D_2=\Gamma_2c_L/\rho_0,\\
& & D_3=\Gamma_0K_1/\rho_0,\\
& & D_4=(\Gamma_0c_{11}^y-\Gamma_2K_2)/\rho_0,\\
& & D_5=(\Gamma_0K_2-\Gamma_2c_L)/\rho_0,\\
& & D_6=a_1-\Gamma_1c_{44}/\rho_0,\\
& & D_7=a_2-\Gamma_2c_{44}/\rho_0,\\
& & D_8=\Gamma_0 K_1^z/\rho_0,\\
& & D_9=\Gamma_\tau c_{44}/\rho_0.
\end{eqnarray}
\end{mathletters}

By solving the hydrodynamic equations in the long wavelength limit, 
we can find the hydrodynamic eigenfrequencies that govern
the relaxation of density, tilt and displacement fluctuations.
All the modes are propagating at finite velocities and are given by
\begin{eqnarray}
& & \omega_\rho=\tilde{v}_1q_x+i\Big[D_1q_x^2+\Big(D_2+
   {\tilde{v}_2D_5\over \tilde{v}_1-\tilde{v}_3}\Big)q_y^2
    -D_6q_z^2\Big],\\
& &  \omega_\phi=\tilde{v}_3q_x+i\Big[D_3q_x^2 +\Big(D_4-
   {\tilde{v}_2D_5\over \tilde{v}_1-\tilde{v}_3}\Big)q_y^2\nonumber \\
    & & \hspace{1.in}+\Big(D_8-{v_3D_7\over \tilde{v}_3+v_4}\Big)q_z^2\Big],\\
& & \omega_\tau=-v_4q_x+i\Big[D_9q_t^2-\Big(D_6+
    {v_3 D_7\over v_4+\tilde{v}_3}\Big)q_z^2\Big].
\end{eqnarray}
For stability, in addition to the conditions stated for the two-dimensional
smectic, we must have $D_6<0$, $D_8-{v_3D_7\over \tilde{v}_3+v_4}>0$,
$D_9>0$ and $D_6+{v_3 D_7\over v_4+\tilde{v}_3}<0$.
The first mode corresponds
to the permeation mode of smectic liquid crystals and describes the transport
of mass across the layers that can occur in these systems
without destroying the layer 
periodicity. The second mode describes long-wavelength deformations 
of the layers and
governs the decay of displacement fluctuations.  
Finally, the third mode governs the relaxation of tilt fluctuations,
that, like density, can occur both in and out of the layers,
while preserving the line smectic periodicity.

\end{multicols}


\begin{references}
\bibitem{lowestFourier_note} It can be easily seen from the functional
RG flow equation for $\Gamma(\phi)$, Eq.\ref{finite_temperature_RG}, that
at the Gaussian fixed point, the disorder eigenvalue for the $Q_n=n
q_0$ Fourier component of the random potential is $\lambda_n=3-d -
\overline{T} n^2 q_0^2$. This therefore implies that near a finite
temperature fixed point one needs to keep only the most relevant
lowest Fourier component $Q_1=q_0$ of the disorder potential.
\bibitem{hexatic_note} Since a driving force necessary selects an
  axis, thereby breaking rotational symmetry explicitly, a hexatic
  phase has no distinction from the liquid (orientational ordering
  {\sl can} play a role in the dynamics as the applied force tends to
  zero, changing the nature of the response -- this dynamics is an
  important open problem).
\end{references}
\end{document}